\documentclass[traditabstract,longauth]{aa}

\pdfoutput=1

\usepackage[nonamebreak]{natbib}
\usepackage[stable]{footmisc}
\usepackage{comment}
\usepackage{amsmath}
\usepackage{txfonts}
\usepackage{placeins}
\usepackage{natbib}
\bibpunct{(}{)}{;}{a}{}{,} 
\usepackage{graphicx}
\usepackage{ifthen}
\usepackage[table,usenames,dvipsnames]{xcolor}
\definecolor{linkcolor}{rgb}{0.6,0,0}
\definecolor{citecolor}{rgb}{0,0,0.75}
\definecolor{urlcolor}{rgb}{0.12,0.46,0.7}
\usepackage[breaklinks, colorlinks, urlcolor=urlcolor, linkcolor=linkcolor,citecolor=citecolor,pdfencoding=auto]{hyperref}
\hypersetup{linktocpage}
\usepackage{subfigure}
\usepackage[utf8]{inputenc}
\usepackage{enumitem}

\setlist[enumerate]{itemindent=\dimexpr\labelwidth+\labelsep\relax,leftmargin=0pt}

\newcommand{\bin}{\mathcal{B}}
\newcommand{\elp}{L}
\newcommand{\pliklite}{{\tt plik\_lite}}

\providecommand{\sorthelp}[1]{}

\newcommand{\isdraft}[1]{}\newcommand{\wantcomments}[1]{}

\newcommand{\JC}[1]{{\wantcomments{\color{purple} JC: #1}}}
\newcommand{\AL}[1]{{\wantcomments{\color{red} AL: #1}}}
\newcommand{\AC}[1]{{\wantcomments{\color{ForestGreen} AC: #1}}}

\newcommand{\JCrev}[1]{{#1}} 
\newcommand{\ALrev}[1]{{#1}}
\newcommand{\ACrev}[1]{{#1}}
\newcommand{\JCrrev}[1]{{#1}} 

\newcommand{\Nzero}{\ensuremath{N^{(0)}}}
\newcommand{\None}{\ensuremath{N^{(1)}}}

\newcommand{\clppfid}{C_{L}^{\phi\phi,\ {\rm fid}}}
\newcommand{\clkkfid}{C_{L}^{\kappa\kappa,\ {\rm fid}}}

\newcommand{\hn}{\hat{\boldsymbol{n}}}
\newcommand{\Res}{\textrm{Res}}
\newcommand{\Cov}{\textrm{Cov}}
\newcommand{\beq}{\begin{equation}}
\newcommand{\enq}{\end{equation}}

\newcommand{\FFP}{FFP10}

\let\origla\la

\renewcommand{\la}{\langle}
\newcommand{\ra}{\rangle}

\newcommand{\fidprefix}{_{\rm fid}} 
\newcommand{\bfprefix}{_{\rm b.f.}} 

\newcommand{\threeonesig}[4]{
\begin{equation}
\left.
 \begin{aligned}
#1 \\ #2 \\ #3
 \end{aligned}
\ \right\} \ \ \mbox{\parbox{5cm}{68\%, #4}}
\end{equation}
}

\newenvironment{unindentedlist}{
 \begin{list}{{$\bullet$}}{
  \setlength\partopsep{0pt}
  \setlength\parskip{0pt}
  \setlength\parsep{0pt}
  \setlength\topsep{0pt}
  \setlength\itemsep{0pt}
  \setlength{\itemindent}{\leftmargin}
  \setlength{\leftmargin}{0pt}
 }
}{
 \end{list}
}

\newcommand{\smica}{{\tt SMICA}}
\newcommand{\GNILC}{{\tt GNILC}}

\newcommand{\av}[1]{\left\langle {#1} \right\rangle}
\newcommand{\MCNzero}{\textrm{MC-}N^{(0)}}
\newcommand{\RDNzero}{\textrm{RD-}N^{(0)}}
\newcommand{\RDNone}[0]{\textrm{RD-}N^{(1)}}
\newcommand{\xpRDNone}[0]{\textrm{xpRD-}N^{(1)}}
\newcommand{\MCNone}[0]{\textrm{MC-}N^{(1)}}

\newcommand{\mksym}[1]{\ifmmode {\rm #1}\else #1\fi}
\newcommand{\dataplus}{\allowbreak{+}}

\newcommand{\lensing}{\mksym{lensing}}

\newcommand{\TT}{\mksym{TT}}
\newcommand{\TTTEEE}{\mksym{TT,TE,EE}}
\newcommand{\planckTTonly}{\planck\ \TT}
\newcommand{\planckTTTEEEonly}{\planck\ \TTTEEE}
\newcommand{\lowTEB}{\mksym{lowE}}
\newcommand{\planckTT}{\planckTTonly\dataplus\lowTEB}
\newcommand{\planckall}{\planckTTTEEEonly\dataplus\lowTEB}

\newcommand{\planckalllensing}{\planckall\dataplus\lensing}

\newcommand{\As}{A_{\rm s}}

\newcommand{\ns}{n_{\rm s}}
\newcommand{\lcdm}{{$\rm{\Lambda CDM}$}}

\newcommand{\Alens}{A_{\rm L}}

\newcommand{\thetaMC}{\theta_{\rm MC}}

\newcommand{\zre}{z_{\text{re}}}

\newcommand{\sumnu}{\sum m_\nu}

\providecommand{\Planck}{\textit{Planck}}
\providecommand{\planck}{\Planck}

\providecommand{\alt}{\lea}

\providecommand{\text}[1]{\rm{#1}}

\newcommand{\Mpc}{\text{Mpc}}

\newcommand{\Hunit}{~\text{km}~\text{s}^{-1} \Mpc^{-1}}

\providecommand{\muK}{\mu\rm{K}}
\newcommand{\muKarcmin}{\,\muK-{\rm arcmin}}
\newcommand{\muKa}{\muK\textrm{-arcmin}}

\newcommand{\lmax}{l_{\text{max}}}

\newcommand{\eV}{\,\text{eV}}

\providecommand{\Omb}{\Omega_{\mathrm{b}}}
\providecommand{\Omc}{\Omega_{\mathrm{c}}}
\providecommand{\Omm}{\Omega_{\mathrm{m}}}

\newcommand{\lensit}{\textsc{\tt LensIt}}

\newcommand{\LENSPIX}{{\tt LensPix}}
\providecommand{\CAMB}{{\tt camb}}
\providecommand{\COSMOMC}{{\tt CosmoMC}}

\providecommand{\CLASS}{{\tt class}}

\providecommand{\LCDM}{{$\rm{\Lambda CDM}$}}

\providecommand{\HALOFIT}{{\tt halofit}}

\newcommand{\begm}{\begin{pmatrix}}
\newcommand{\enm}{\end{pmatrix}}

\newcommand\ba{\begin{eqnarray}}
\newcommand\ea{\end{eqnarray}}
\newcommand\bea{\begin{eqnarray}}
\newcommand\eea{\end{eqnarray}}

\newcommand\be{\begin{equation}}
\newcommand\ee{\end{equation}}

\newcommand{\vgrad}{{\boldsymbol{\nabla}}}

\newcommand{\vtheta}{\mathbf{\theta}}

\providecommand{\var}{\text{var}}
\providecommand{\cov}{\text{cov}}

\newcommand{\boldvec}[1]{{\mbox{\boldmath{$#1$}}}}

\newcommand{\vl}{\boldvec{\ell}}

\newcommand{\clo}{\mathcal{O}}

\def\setsymbol#1#2{\expandafter\def\csname #1\endcsname{#2}}
\def\getsymbol#1{\csname #1\endcsname}

\def\Planck{\textit{Planck}}

\newbox\tablebox    \newdimen\tablewidth
\def\leaderfil{\leaders\hbox to 5pt{\hss.\hss}\hfil}
\def\endPlancktable{\tablewidth=\columnwidth 
    $$\hss\copy\tablebox\hss$$
    \vskip-\lastskip\vskip -2pt}
\def\endPlancktablewide{\tablewidth=\textwidth 
    $$\hss\copy\tablebox\hss$$
    \vskip-\lastskip\vskip -2pt}
\def\tablenote#1 #2\par{\begingroup \parindent=0.8em
    \abovedisplayshortskip=0pt\belowdisplayshortskip=0pt
    \noindent
    $$\hss\vbox{\hsize\tablewidth \hangindent=\parindent \hangafter=1 \noindent
    \hbox to \parindent{$^#1$\hss}\strut#2\strut\par}\hss$$
    \endgroup}
\def\doubleline{\vskip 3pt\hrule \vskip 1.5pt \hrule \vskip 5pt}

\def\L2{\ifmmode L_2\else $L_2$\fi}

\def\DeltaT{\ifmmode \Delta T\else $\Delta T$\fi}
\def\deltat{\ifmmode \Delta t\else $\Delta t$\fi}
\def\fknee{\ifmmode f_{\rm knee}\else $f_{\rm knee}$\fi}
\def\Fmax{\ifmmode F_{\rm max}\else $F_{\rm max}$\fi}
\def\solar{\ifmmode{\rm M}_{\mathord\odot}\else${\rm M}_{\mathord\odot}$\fi}
\def\Msolar{\ifmmode{\rm M}_{\mathord\odot}\else${\rm M}_{\mathord\odot}$\fi}
\def\Lsolar{\ifmmode{\rm L}_{\mathord\odot}\else${\rm L}_{\mathord\odot}$\fi}
\def\inv{\ifmmode^{-1}\else$^{-1}$\fi}
\def\mo{\ifmmode^{-1}\else$^{-1}$\fi}
\def\sup#1{\ifmmode ^{\rm #1}\else $^{\rm #1}$\fi}
\def\expo#1{\ifmmode \times 10^{#1}\else $\times 10^{#1}$\fi}
\def\,{\thinspace}
\def\lsim{\mathrel{\raise .4ex\hbox{\rlap{$<$}\lower 1.2ex\hbox{$\sim$}}}}
\def\gsim{\mathrel{\raise .4ex\hbox{\rlap{$>$}\lower 1.2ex\hbox{$\sim$}}}}

\def\simprop{\mathrel{\raise .4ex\hbox{\rlap{$\propto$}\lower 1.2ex\hbox{$\sim$}}}}
\def\deg{\ifmmode^\circ\else$^\circ$\fi}
\def\pdeg{\ifmmode $\setbox0=\hbox{$^{\circ}$}\rlap{\hskip.11\wd0 .}$^{\circ}
          \else \setbox0=\hbox{$^{\circ}$}\rlap{\hskip.11\wd0 .}$^{\circ}$\fi}
\def\arcs{\ifmmode {^{\scriptstyle\prime\prime}}
          \else $^{\scriptstyle\prime\prime}$\fi}
\def\arcm{\ifmmode {^{\scriptstyle\prime}}
          \else $^{\scriptstyle\prime}$\fi}
\newdimen\sa  \newdimen\sb
\def\parcs{\sa=.07em \sb=.03em
     \ifmmode \hbox{\rlap{.}}^{\scriptstyle\prime\kern -\sb\prime}\hbox{\kern -\sa}
     \else \rlap{.}$^{\scriptstyle\prime\kern -\sb\prime}$\kern -\sa\fi}
\def\parcm{\sa=.08em \sb=.03em
     \ifmmode \hbox{\rlap{.}\kern\sa}^{\scriptstyle\prime}\hbox{\kern-\sb}
     \else \rlap{.}\kern\sa$^{\scriptstyle\prime}$\kern-\sb\fi}
\def\ra[#1 #2 #3.#4]{#1\sup{h}#2\sup{m}#3\sup{s}\llap.#4}
\def\dec[#1 #2 #3.#4]{#1\deg#2\arcm#3\arcs\llap.#4}
\def\deco[#1 #2 #3]{#1\deg#2\arcm#3\arcs}
\def\rra[#1 #2]{#1\sup{h}#2\sup{m}}

\def\dots{\relax\ifmmode \ldots\else $\ldots$\fi}
\def\WHzsr{\ifmmode $W\,Hz\mo\,sr\mo$\else W\,Hz\mo\,sr\mo\fi}
\def\mHz{\ifmmode $\,mHz$\else \,mHz\fi}
\def\GHz{\ifmmode $\,GHz$\else \,GHz\fi}
\def\mKs{\ifmmode $\,mK\,s$^{1/2}\else \,mK\,s$^{1/2}$\fi}
\def\muKs{\ifmmode \,\mu$K\,s$^{1/2}\else \,$\mu$K\,s$^{1/2}$\fi}
\def\muKRJs{\ifmmode \,\mu$K$_{\rm RJ}$\,s$^{1/2}\else \,$\mu$K$_{\rm RJ}$\,s$^{1/2}$\fi}
\def\muKHz{\ifmmode \,\mu$K\,Hz$^{-1/2}\else \,$\mu$K\,Hz$^{-1/2}$\fi}
\def\MJysr{\ifmmode \,$MJy\,sr\mo$\else \,MJy\,sr\mo\fi}
\def\MJysrmK{\ifmmode \,$MJy\,sr\mo$\,mK$_{\rm CMB}\mo\else \,MJy\,sr\mo\,mK$_{\rm CMB}\mo$\fi}
\def\microns{\ifmmode \,\mu$m$\else \,$\mu$m\fi}

\def\muK{\ifmmode \,\mu$K$\else \,$\mu$\hbox{K}\fi}
\def\microK{\ifmmode \,\mu$K$\else \,$\mu$\hbox{K}\fi}
\def\muW{\ifmmode \,\mu$W$\else \,$\mu$\hbox{W}\fi}
\def\kms{\ifmmode $\,km\,s$^{-1}\else \,km\,s$^{-1}$\fi}
\def\kmsMpc{\ifmmode $\,\kms\,Mpc\mo$\else \,\kms\,Mpc\mo\fi}
\providecommand{\sorthelp}[1]{}

\newcommand{\PLA}{\url{https://pla.esac.esa.int}}

\newcommand{\paramsIII}{\citetalias{planck2016-l06}}
\defcitealias{planck2013-p11}{PCP13}
\defcitealias{planck2014-a15}{PCP15}
\defcitealias{planck2016-l06}{PCP18}

\defcitealias{pb2015}{BKP}

\defcitealias{planck2013-p08}{PPL13}
\defcitealias{planck2014-a13}{PPL15}
\defcitealias{planck2016-l05}{PPL18}

\defcitealias{planck2016-l08}{PL2018}
\defcitealias{planck2014-a17}{PL2015}
\defcitealias{planck2013-p12}{PL2013}

\newcommand{\PlanckLensTwo}{\citetalias{planck2014-a17}}
\newcommand{\PlanckLensOne}{\citetalias{planck2013-p12}}

\renewcommand{\alt}{\origla}

\pdfmapfile{+txfonts.map}

\begin{document}
\author{\small
Planck Collaboration: N.~Aghanim\inst{50}
\and
Y.~Akrami\inst{52, 54}
\and
M.~Ashdown\inst{61, 5}
\and
J.~Aumont\inst{88}
\and
C.~Baccigalupi\inst{72}
\and
M.~Ballardini\inst{19, 38}
\and
A.~J.~Banday\inst{88, 8}
\and
R.~B.~Barreiro\inst{56}
\and
N.~Bartolo\inst{27, 57}
\and
S.~Basak\inst{79}
% ...
}
\institute{\small
AIM, CEA, CNRS, Universit\'{e} Paris-Saclay, F-91191 Gif sur Yvette, France. AIM, Universit\'{e} Paris Diderot, Sorbonne Paris Cit\'{e}, F-91191 Gif sur Yvette, France.\goodbreak
\and
APC, AstroParticule et Cosmologie, Universit\'{e} Paris Diderot, CNRS/IN2P3, CEA/lrfu, Observatoire de Paris, Sorbonne Paris Cit\'{e}, 10, rue Alice Domon et L\'{e}onie Duquet, 75205 Paris Cedex 13, France\goodbreak
\and
African Institute for Mathematical Sciences, 6-8 Melrose Road, Muizenberg, Cape Town, South Africa\goodbreak
\and
Aix Marseille Univ, CNRS, CNES, LAM, Marseille, France\goodbreak
\and
Astrophysics Group, Cavendish Laboratory, University of Cambridge, J J Thomson Avenue, Cambridge CB3 0HE, U.K.\goodbreak
\and
Astrophysics \& Cosmology Research Unit, School of Mathematics, Statistics \& Computer Science, University of KwaZulu-Natal, Westville Campus, Private Bag X54001, Durban 4000, South Africa\goodbreak
\and
CITA, University of Toronto, 60 St. George St., Toronto, ON M5S 3H8, Canada\goodbreak
\and
CNRS, IRAP, 9 Av. colonel Roche, BP 44346, F-31028 Toulouse cedex 4, France\goodbreak
\and
Cahill Center for Astronomy and Astrophysics, California Institute of Technology, Pasadena CA,  91125, USA\goodbreak
\and
California Institute of Technology, Pasadena, California, U.S.A.\goodbreak
\and
Centre for Theoretical Cosmology, DAMTP, University of Cambridge, Wilberforce Road, Cambridge CB3 0WA, U.K.\goodbreak
\and
Computational Cosmology Center, Lawrence Berkeley National Laboratory, Berkeley, California, U.S.A.\goodbreak
\and
D\'{e}partement de Physique Th\'{e}orique, Universit\'{e} de Gen\`{e}ve, 24, Quai E. Ansermet,1211 Gen\`{e}ve 4, Switzerland\goodbreak
\and
Departamento de Astrof\'{i}sica, Universidad de La Laguna (ULL), E-38206 La Laguna, Tenerife, Spain\goodbreak
\and
Departamento de F\'{\i}sica, Universidad de Oviedo, C/ Federico Garc\'{\i}a Lorca, 18 , Oviedo, Spain\goodbreak
\and
Department of Astrophysics/IMAPP, Radboud University, P.O. Box 9010, 6500 GL Nijmegen, The Netherlands\goodbreak
\and
Department of Mathematics, University of Stellenbosch, Stellenbosch 7602, South Africa\goodbreak
\and
Department of Physics \& Astronomy, University of British Columbia, 6224 Agricultural Road, Vancouver, British Columbia, Canada\goodbreak
\and
Department of Physics \& Astronomy, University of the Western Cape, Cape Town 7535, South Africa\goodbreak
\and
Department of Physics and Astronomy, University of Sussex, Brighton BN1 9QH, U.K.\goodbreak
\and
Department of Physics, Gustaf H\"{a}llstr\"{o}min katu 2a, University of Helsinki, Helsinki, Finland\goodbreak
\and
Department of Physics, Princeton University, Princeton, New Jersey, U.S.A.\goodbreak
\and
Department of Physics, University of California, Berkeley, California, U.S.A.\goodbreak
\and
Department of Physics, University of California, One Shields Avenue, Davis, California, U.S.A.\goodbreak
\and
Department of Physics, University of California, Santa Barbara, California, U.S.A.\goodbreak
\and
Department of Physics, University of Illinois at Urbana-Champaign, 1110 West Green Street, Urbana, Illinois, U.S.A.\goodbreak
\and
Dipartimento di Fisica e Astronomia G. Galilei, Universit\`{a} degli Studi di Padova, via Marzolo 8, 35131 Padova, Italy\goodbreak
\and
Dipartimento di Fisica e Scienze della Terra, Universit\`{a} di Ferrara, Via Saragat 1, 44122 Ferrara, Italy\goodbreak
\and
Dipartimento di Fisica, Universit\`{a} La Sapienza, P. le A. Moro 2, Roma, Italy\goodbreak
\and
Dipartimento di Fisica, Universit\`{a} degli Studi di Milano, Via Celoria, 16, Milano, Italy\goodbreak
\and
Dipartimento di Fisica, Universit\`{a} degli Studi di Trieste, via A. Valerio 2, Trieste, Italy\goodbreak
\and
Dipartimento di Fisica, Universit\`{a} di Roma Tor Vergata, Via della Ricerca Scientifica, 1, Roma, Italy\goodbreak
\and
European Space Agency, ESAC, Planck Science Office, Camino bajo del Castillo, s/n, Urbanizaci\'{o}n Villafranca del Castillo, Villanueva de la Ca\~{n}ada, Madrid, Spain\goodbreak
\and
European Space Agency, ESTEC, Keplerlaan 1, 2201 AZ Noordwijk, The Netherlands\goodbreak
\and
Gran Sasso Science Institute, INFN, viale F. Crispi 7, 67100 L'Aquila, Italy\goodbreak
\and
Haverford College Astronomy Department, 370 Lancaster Avenue, Haverford, Pennsylvania, U.S.A.\goodbreak
\and
Helsinki Institute of Physics, Gustaf H\"{a}llstr\"{o}min katu 2, University of Helsinki, Helsinki, Finland\goodbreak
\and
INAF - OAS Bologna, Istituto Nazionale di Astrofisica - Osservatorio di Astrofisica e Scienza dello Spazio di Bologna, Area della Ricerca del CNR, Via Gobetti 101, 40129, Bologna, Italy\goodbreak
\and
INAF - Osservatorio Astronomico di Padova, Vicolo dell'Osservatorio 5, Padova, Italy\goodbreak
\and
INAF - Osservatorio Astronomico di Trieste, Via G.B. Tiepolo 11, Trieste, Italy\goodbreak
\and
INAF, Istituto di Radioastronomia, Via Piero Gobetti 101, I-40129 Bologna, Italy\goodbreak
\and
INAF/IASF Milano, Via E. Bassini 15, Milano, Italy\goodbreak
\and
INFN - CNAF, viale Berti Pichat 6/2, 40127 Bologna, Italy\goodbreak
\and
INFN, Sezione di Bologna, viale Berti Pichat 6/2, 40127 Bologna, Italy\goodbreak
\and
INFN, Sezione di Ferrara, Via Saragat 1, 44122 Ferrara, Italy\goodbreak
\and
INFN, Sezione di Milano, Via Celoria 16, Milano, Italy\goodbreak
\and
INFN, Sezione di Roma 1, Universit\`{a} di Roma Sapienza, Piazzale Aldo Moro 2, 00185, Roma, Italy\goodbreak
\and
INFN, Sezione di Roma 2, Universit\`{a} di Roma Tor Vergata, Via della Ricerca Scientifica, 1, Roma, Italy\goodbreak
\and
Imperial College London, Astrophysics group, Blackett Laboratory, Prince Consort Road, London, SW7 2AZ, U.K.\goodbreak
\and
Institut d'Astrophysique Spatiale, CNRS, Univ. Paris-Sud, Universit\'{e} Paris-Saclay, B\^{a}t. 121, 91405 Orsay cedex, France\goodbreak
\and
Institut d'Astrophysique de Paris, CNRS (UMR7095), 98 bis Boulevard Arago, F-75014, Paris, France\goodbreak
\and
Institute Lorentz, Leiden University, PO Box 9506, Leiden 2300 RA, The Netherlands\goodbreak
\and
Institute of Astronomy, University of Cambridge, Madingley Road, Cambridge CB3 0HA, U.K.\goodbreak
\and
Institute of Theoretical Astrophysics, University of Oslo, Blindern, Oslo, Norway\goodbreak
\and
Instituto de Astrof\'{\i}sica de Canarias, C/V\'{\i}a L\'{a}ctea s/n, La Laguna, Tenerife, Spain\goodbreak
\and
Instituto de F\'{\i}sica de Cantabria (CSIC-Universidad de Cantabria), Avda. de los Castros s/n, Santander, Spain\goodbreak
\and
Istituto Nazionale di Fisica Nucleare, Sezione di Padova, via Marzolo 8, I-35131 Padova, Italy\goodbreak
\and
Jet Propulsion Laboratory, California Institute of Technology, 4800 Oak Grove Drive, Pasadena, California, U.S.A.\goodbreak
\and
Jodrell Bank Centre for Astrophysics, Alan Turing Building, School of Physics and Astronomy, The University of Manchester, Oxford Road, Manchester, M13 9PL, U.K.\goodbreak
\and
Kavli Institute for Cosmological Physics, University of Chicago, Chicago, IL 60637, USA\goodbreak
\and
Kavli Institute for Cosmology Cambridge, Madingley Road, Cambridge, CB3 0HA, U.K.\goodbreak
\and
LERMA, CNRS, Observatoire de Paris, 61 Avenue de l'Observatoire, Paris, France\goodbreak
\and
LERMA/LRA, Observatoire de Paris, PSL Research University, CNRS, Ecole Normale Sup\'erieure, 75005 Paris, France\goodbreak
\and
Laboratoire AIM, CEA - Universit\'{e} Paris-Saclay, 91191 Gif-sur-Yvette, France\goodbreak
\and
Laboratoire de Physique Subatomique et Cosmologie, Universit\'{e} Grenoble-Alpes, CNRS/IN2P3, 53, rue des Martyrs, 38026 Grenoble Cedex, France\goodbreak
\and
Laboratoire de Physique Th\'{e}orique, Universit\'{e} Paris-Sud 11 \& CNRS, B\^{a}timent 210, 91405 Orsay, France\goodbreak
\and
Low Temperature Laboratory, Department of Applied Physics, Aalto University, Espoo, FI-00076 AALTO, Finland\goodbreak
\and
Max-Planck-Institut f\"{u}r Astrophysik, Karl-Schwarzschild-Str. 1, 85741 Garching, Germany\goodbreak
\and
Mullard Space Science Laboratory, University College London, Surrey RH5 6NT, U.K.\goodbreak
\and
NAOC-UKZN Computational Astrophysics Centre (NUCAC), University of KwaZulu-Natal, Durban 4000, South Africa\goodbreak
\and
Nicolaus Copernicus Astronomical Center, Polish Academy of Sciences, Bartycka 18, 00-716 Warsaw, Poland\goodbreak
\and
SISSA, Astrophysics Sector, via Bonomea 265, 34136, Trieste, Italy\goodbreak
\and
San Diego Supercomputer Center, University of California, San Diego, 9500 Gilman Drive, La Jolla, CA 92093, USA\goodbreak
\and
School of Chemistry and Physics, University of KwaZulu-Natal, Westville Campus, Private Bag X54001, Durban, 4000, South Africa\goodbreak
\and
School of Physical Sciences, National Institute of Science Education and Research, HBNI, Jatni-752050, Odissa, India\goodbreak
\and
School of Physics and Astronomy, Cardiff University, Queens Buildings, The Parade, Cardiff, CF24 3AA, U.K.\goodbreak
\and
School of Physics and Astronomy, Sun Yat-sen University, 2 Daxue Rd, Tangjia, Zhuhai, China\goodbreak
\and
School of Physics and Astronomy, University of Nottingham, Nottingham NG7 2RD, U.K.\goodbreak
\and
School of Physics, Indian Institute of Science Education and Research Thiruvananthapuram, Maruthamala PO, Vithura, Thiruvananthapuram 695551, Kerala, India\goodbreak
\and
School of Physics, The University of New South Wales, Sydney NSW 2052, Australia\goodbreak
\and
Simon Fraser University, Department of Physics, 8888 University Drive, Burnaby BC, Canada\goodbreak
\and
Sorbonne Universit\'{e}-UPMC, UMR7095, Institut d'Astrophysique de Paris, 98 bis Boulevard Arago, F-75014, Paris, France\goodbreak
\and
Space Research Institute (IKI), Russian Academy of Sciences, Profsoyuznaya Str, 84/32, Moscow, 117997, Russia\goodbreak
\and
Space Science Data Center - Agenzia Spaziale Italiana, Via del Politecnico snc, 00133, Roma, Italy\goodbreak
\and
Space Sciences Laboratory, University of California, Berkeley, California, U.S.A.\goodbreak
\and
The Oskar Klein Centre for Cosmoparticle Physics, Department of Physics, Stockholm University, AlbaNova, SE-106 91 Stockholm, Sweden\goodbreak
\and
UPMC Univ Paris 06, UMR7095, 98 bis Boulevard Arago, F-75014, Paris, France\goodbreak
\and
Universit\'{e} de Toulouse, UPS-OMP, IRAP, F-31028 Toulouse cedex 4, France\goodbreak
% ...
}

\title{\textit{Planck} 2018 results. VIII. Gravitational lensing}

\abstract{
We present measurements of the cosmic microwave background (CMB) lensing potential using the final $\textit{Planck}$ 2018 temperature and polarization data. Using polarization maps filtered to account for the noise anisotropy, we increase the significance of the detection of lensing in the polarization maps from $5\,\sigma$ to $9\,\sigma$. Combined with temperature, lensing is detected at $40\,\sigma$. We present an extensive set of tests of the robustness of the lensing-potential power spectrum, and construct a minimum-variance estimator likelihood over lensing multipoles $8 \le L \le 400$ (extending the range to lower $L$ compared to 2015), which we use to constrain cosmological parameters. We find good consistency between lensing constraints and the results from the $\textit{Planck}$ CMB power spectra within the $\rm{\Lambda CDM}$ model. Combined with baryon density and other weak priors, the lensing analysis alone constrains $\sigma_8 \Omega_{\rm m}^{0.25} = 0.589\pm 0.020$ ($1\,\sigma$ errors). Also combining with baryon acoustic oscillation (BAO) data, we find tight individual parameter constraints, $\sigma_8 = 0.811\pm 0.019$, $H_0 =67.9_{-1.3}^{+1.2}\,\text{km}\,\text{s}^{-1}\,\rm{Mpc}^{-1}$, and $\Omega_{\rm m} = 0.303^{+0.016}_{-0.018}$. Combining with $\textit{Planck}$ CMB power spectrum data, we measure $\sigma_8$ to better than $1\,\%$ precision, finding $\sigma_8=0.811\pm 0.006$. CMB lensing reconstruction data are complementary to galaxy lensing data at lower redshift, having a different degeneracy direction in $\sigma_8$--$\Omega_{\rm m}$ space; we find consistency with the lensing results from the Dark Energy Survey, and give combined lensing-only parameter constraints that are tighter than joint results using galaxy clustering. Using the $\textit{Planck}$ cosmic infrared background (CIB) maps as an additional tracer of high-redshift matter, we make a combined $\textit{Planck}$-only estimate of the lensing potential over $60\,\%$ of the sky with considerably more small-scale signal. We additionally demonstrate delensing of the $\textit{Planck}$ power spectra using the joint and individual lensing potential estimates, detecting a maximum removal of $40\,\%$ of the lensing-induced power in all spectra. The improvement in the sharpening of the acoustic peaks by including both CIB and the quadratic lensing reconstruction is detected at high significance.
\phantom{Dummy text to make abstract longer to Contents page split is in more sensible position
}
}
\renewcommand{\abstractname}{} 
\keywords{gravitational lensing: weak -- cosmological parameters --
 cosmic background radiation -- large-scale structure of Universe --
 cosmology: observations}

\date{Draft compiled \today}

\titlerunning{\textit{Planck} 2018 lensing}
\authorrunning{Planck Collaboration}
\maketitle
\clearpage

\tableofcontents

\section{Introduction}
\label{sec:intro}

Gravitational lensing distorts our view of the last-scattering surface, generating new non-Gaussian signals and $B$-mode polarization, as well as smoothing the shape of the observed power spectra. The large distance to recombination means that each photon is effectively independently lensed many times, boosting the signal compared to other second- and higher-order effects. The sharply-defined acoustic scale in the unlensed CMB perturbation power also makes small magnification and shear distortions easily detectable, allowing us to use observations of the lensed sky to reconstruct the lensing deflections and hence learn about the large-scale structure and geometry of the Universe between recombination and today~\citep{1987A&A...184....1B,Hu:2001kj,Lewis:2006fu}. In this paper we present the final \planck\footnote{\Planck\ (\url{https://www.esa.int/Planck}) is a project of the
European Space Agency (ESA) with instruments provided by two scientific
consortia funded by ESA member states and led by Principal Investigators
from France and Italy, telescope reflectors provided through a collaboration
between ESA and a scientific consortium led and funded by Denmark, and
additional contributions from NASA (USA).} lensing reconstruction analysis, giving the most significant detection of lensing to date over 70\,\% of the sky. We also give new results for polarization-based reconstructions, a combination of \Planck's lensing reconstruction and measurements of the cosmic infrared background (CIB), and delensing of the CMB temperature and polarization fields.

In the \planck\ 2013 analysis \citep[hereafter PL2013]{planck2013-p12} we produced the first nearly full-sky lensing reconstruction based on the nominal-mission temperature data. In the 2015 analysis \citep[hereafter PL2015]{planck2014-a17} this was updated to include the full-mission temperature data, as well as polarization, along with a variety of analysis improvements.
The final full-mission analysis presented here uses essentially the same data as \PlanckLensTwo: most of the map-level improvements discussed by~\citet{planck2014-a10} and~\citet{planck2016-l03} are focussed on large scales (especially the low-$\ell$ polarization), which have almost no impact on lensing reconstruction (the lensing analysis does not include multipoles $\ell <100$). Instead, we focus on improvements in the simulations, optimality of the lensing reconstruction, foreground masking, and new results such as the polarization-only reconstruction, joint analysis with the CIB, and delensing.
\begin{figure*}[]
\centering
\includegraphics[width = \textwidth]{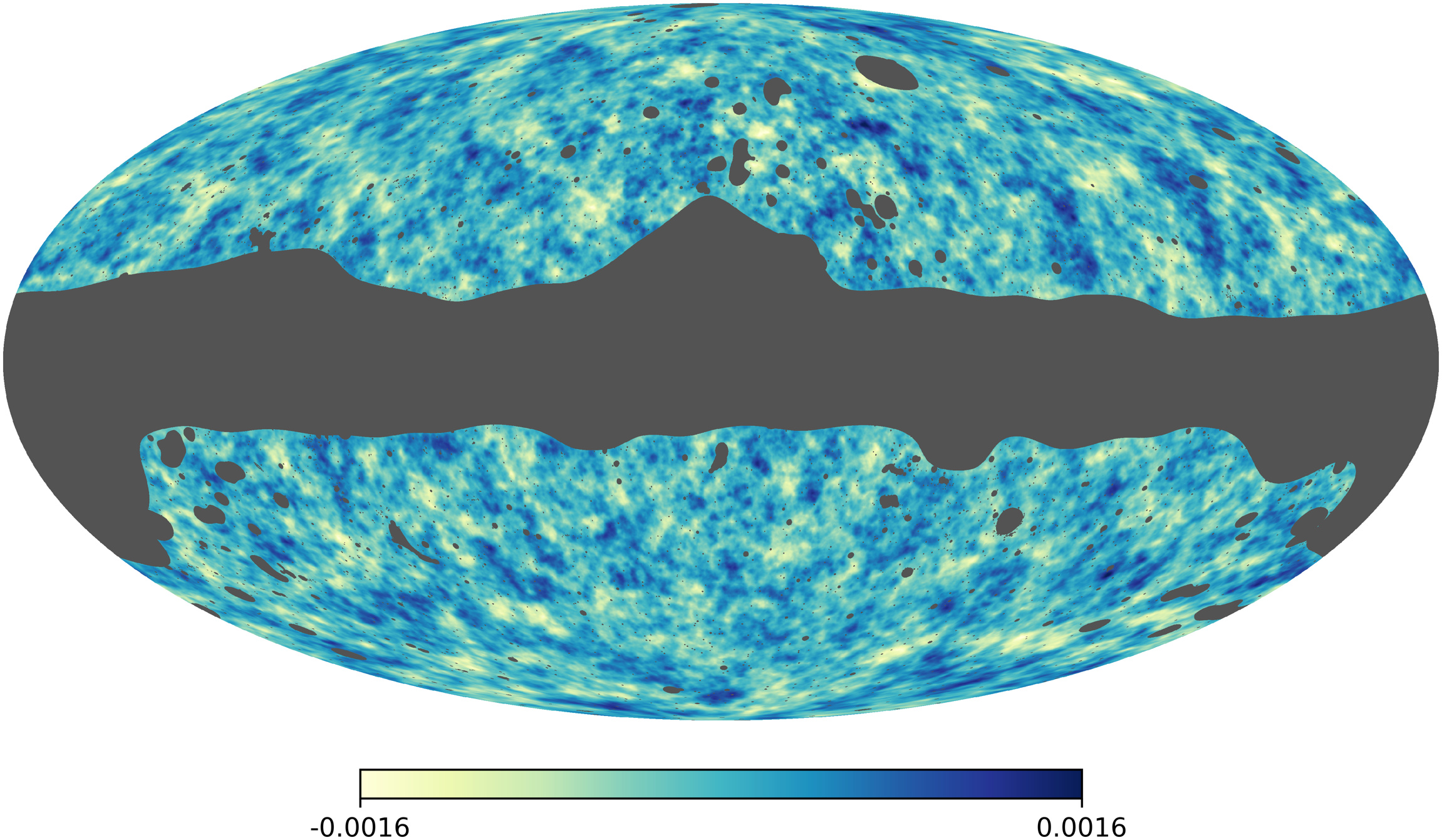}
\caption{Mollweide projection in Galactic coordinates of the lensing-deflection reconstruction map from our baseline minimum-variance (MV) analysis. We show the Wiener-filtered displacement-like scalar field with multipoles $\hat\alpha^{\rm MV}_{LM} = \sqrt{L (L + 1) }\hat\phi^{\rm MV}_{LM}$, corresponding to the gradient mode (or $E$ mode) of the lensing deflection angle. Modes with $L<8$ have been filtered out.
\label{fig:lensmap}
}
\end{figure*}
\noindent We highlight the following main results.
\begin{unindentedlist}
\item The most significant measurement of the CMB lensing power spectrum to date, $9\,\sigma$ from polarization alone, and $40\,\sigma$ using the minimum-variance estimate combining temperature and polarization data on 67\,\% of the sky, over the (conservative) multipole range $8 \leq L \leq 400$.
\item A new best estimate of the lensing potential over 58\,\% of the sky by combining information from the \planck\ CMB lensing reconstruction and high-frequency maps as a probe of the CIB. The CIB is expected to be highly correlated with the CMB lensing potential, and although the CIB does not provide robust independent information on the lensing power spectrum, the map can provide an improved estimate of the actual realization of lensing modes down to small scales. The joint estimate gives the best picture we currently have of the lensing potential.
\item Using the lensing-reconstruction maps, we demonstrate that the CMB acoustic peaks can be delensed, detecting peak sharpening at $11\,\sigma$ from the \planck\ reconstruction alone and $15\,\sigma$ on further combining with the CIB (corresponding to removing about 40\,\% of the lensing effect). We also detect at $9\,\sigma$ a decrease in power of the $B$-mode polarization after delensing. 
\item Using the \planck\ lensing likelihood alone we place a 3.5\,\% constraint on the parameter combination $\sigma_8 \Omega_{\rm m}^{0.25}$. This has comparable statistical power to current constraints from galaxy lensing, but the high-redshift CMB source plane gives a different degeneracy direction compared to the $\sigma_8 \Omega_{\rm m}^{0.5}$ combination from galaxy lensing at lower redshift. Combining a baryon density prior with measurements of baryon acoustic oscillations (BAOs) in the galaxy distribution gives a competitive measurement of $\sigma_8$, $\Omega_{\rm m}$, and $H_0$. We can also break the degeneracy by combining our likelihood with the first-year lensing results from the Dark Energy Survey~\citep[DES; ][]{Troxel:2017xyo}, giving the tightest lensing-only constraints on these parameters.
\end{unindentedlist}

Our baseline lensing reconstruction map is shown in Fig.~\ref{fig:lensmap}.
In Sect.~\ref{sec:pipe} we explain how this was obtained, and the changes compared to our analysis in  \PlanckLensTwo. We also describe the new optimal filtering approach used for our best polarization analysis.
In Sect.~\ref{sec:results} we present our main results, including power-spectrum estimates, cosmological parameter constraints, and a joint estimation of the lensing potential using the CIB. We end the section by using the estimates of the lensing map to delens the CMB, reducing the $B$-mode polarization power and sharpening the acoustic peaks. In Sect.~\ref{sec:tests} we describe in detail a number of null and consistency tests, explaining the motivation for our data cuts and the limits of our understanding of the data. We also discuss possible contaminating signals, and assess whether they are potentially important for our results. In Sect.~\ref{sec:products} we briefly describe the various data products that are made available to the community, and we end with conclusions in Sect.~\ref{sec:conclude}. A series of appendices describe some technical details of the calculation of various biases that are subtracted, and derive the error model for the Monte Carlo estimates.

\section{Data and methodology}
\label{sec:pipe}
\newcommand{\transfmultipolestopix}{\mathcal{T}}
\newcommand{\transfpixtopix}{\mathcal{B}}
\renewcommand{\Res}{\bar X}

This final \planck\ lensing analysis is based on the 2018 \planck\ HFI maps as described in detail in~\cite{planck2016-l03}. Our baseline analysis uses the \smica\ foreground-cleaned CMB map described in
\cite{planck2016-l04}, and includes both temperature and polarization information.
We use the \planck\ Full Focal Plane (\FFP) simulations, described in detail in \cite{planck2016-l03}, to remove a number of bias terms and correctly normalize the lensing power-spectrum estimates.
Our analysis methodology is based on the previous \Planck\ analyses, as described in \PlanckLensOne\ and \PlanckLensTwo.
After a summary of the methodology, Sect.~\ref{subsec:reconstruction} also lists the changes and improvements with respect to \PlanckLensTwo. Some details of the covariance matrix are discussed in Sect.~\ref{subsec:cov}, and details of the filtering in Sect.~\ref{subsec:inhogfilt}. The main set of codes applying the quadratic estimators will be made public as part of the CMB lensing toolbox \textsc{LensIt}.\footnote{\url{https://github.com/carronj/LensIt}}

\subsection{Lensing reconstruction}\label{subsec:reconstruction}
The five main steps of the lensing reconstruction are as follows.
\begin{enumerate}
	\item {\it Filtering of the CMB maps}. The observed sky maps are cut by a Galactic mask and have noise, so filtering is applied to remove the mask and approximately optimally weight for the noise.
The lensing quadratic estimators use as input optimal Wiener-filtered $X = T$, $E$, and $B$\, CMB multipoles, as well as inverse-variance-weighted CMB maps. The latter maps can be obtained easily from the Wiener-filtered multipoles by dividing by the fiducial CMB power spectra $C_\ell^{\rm fid}$ before projecting onto maps. We write the observed temperature $T$ and polarization (written as the spin $\pm 2$ combination of Stokes parameters ${}_{\pm 2} P \equiv Q \pm i U$) pixelized data as
\begin{equation}
		\begin{pmatrix}
		T^{\rm dat} \\ {}_2P^{\rm dat} \\ {}_{-2}P^{\rm dat}	
		\end{pmatrix} = {\transfpixtopix} \mathcal{Y} \begin{pmatrix} T \\ E \\ B \end{pmatrix} + \text{noise},
\end{equation}
where $T$, $E$, and $B$ on the right-hand side are the multipole coefficients of the true temperature and $E$- and $B$-mode polarization. The matrix $\mathcal{Y}$ contains the appropriate (spin-weighted) spherical harmonic functions to map from multipoles to the sky, and the matrix $\transfpixtopix$ accounts for the real-space operations of beam and pixel convolution. We further use the notation $\transfmultipolestopix \equiv \transfpixtopix \mathcal{Y}$ for the complete transfer function from multipoles to the pixelized sky. The Wiener-filtered multipoles are obtained from the pixelized data as

	\begin{equation}\label{eq:filt}
		\begin{pmatrix}
		T^{\rm WF} \\ E^{\rm WF} \\ B^{\rm WF}	
		\end{pmatrix} \equiv C^{\rm fid} \transfmultipolestopix^\dagger {\Cov}^{-1} 		\begin{pmatrix} T^{\rm dat} \\ _2P^{\rm dat} \\ _{-2}P^{\rm dat} \end{pmatrix},
	\end{equation}
	where the pixel-space covariance is $\Cov =\transfmultipolestopix C^{\rm fid} \transfmultipolestopix^\dagger + N$. Here, $C^{\rm fid}$ is a fiducial set of CMB spectra and $N$ is the pixel-space noise covariance matrix, which we approximate as diagonal.
As in previous releases, our baseline results use independently-filtered temperature and polarization maps (i.e., we always neglect $C^{TE}_\ell$ in $\Cov^{-1}$ in Eq.~\ref{eq:filt}) at the cost of a 3\,\% increase in reconstruction noise on our conservative multipole range ($L\leq 400$).
The large matrix inversion is performed with a multigrid-preconditioned conjugate-gradient search~\citep{Smith:2007rg}. The temperature monopole and dipole are projected out, being assigned formally infinite noise. As in \PlanckLensTwo, we use only CMB multipoles $100 \le \ell \le 2048$ from these filtered maps. Our baseline analysis approximates the noise as isotropic in the filtering, which has the advantage of making the lensing estimator normalization roughly isotropic across the sky at the expense of some loss of optimality. In this case we also slightly rescale the filtered multipoles so that the effective full-sky transfer function matches the one seen empirically on the filtered simulations, with a minimal impact on the band powers. We also present new more optimally-filtered results, as discussed in Sect.~\ref{subsec:inhogfilt}.

	\item {\it Construction of the quadratic lensing estimator.}  We determine
 $\hat \phi$ from pairs of filtered maps, and our implementation now follows~\citet{Carron:2017mqf}.  This differs slightly from \PlanckLensTwo, allowing us to produce minimum-variance (MV) estimators from filtered maps much faster, which is useful given the variety of tests performed for this release.  We calculate a spin-1 real-space (unnormalized) lensing displacement estimate

	\begin{equation}\label{eq:lensing2pt}
	    \begin{split}
		_1\hat d(\hn) &= - \sum_{s = 0,\pm 2} {_{-s}}\Res(\hn) \left[\eth_s X^{\rm WF} \right](\hn), \\
       \end{split}
	\end{equation}
    where $\eth$ is the spin-raising operator, and the pre-subscript $s$ on a field denotes the spin.
	 The quadratic estimator involves products of the real-space inverse-variance filtered maps
	\beq
	\label{eq:res}
	\Res(\hn) \equiv \left[ \transfpixtopix^\dagger \Cov^{-1} X^{\rm dat} \right](\hn) ,
	\enq
	and the gradients of the Wiener-filtered maps
	\beq
	\label{eq:gmap}
	\begin{split}
		\left[\eth {}_0X^{\rm WF}\right](\hn) &\equiv \sum_{\ell m}\sqrt{\ell (\ell + 1)} T^{\rm WF}_{\ell m} \left._1Y_{\ell m}(\hn)\right. ,\\	
	\left[\eth {}_{-2}X^{\rm WF}\right](\hn) &\equiv -\sum_{\ell m} \sqrt{(\ell+2)(\ell - 1)} \left[ E^{\rm WF}_{\ell m}- i B^{\rm WF}_{\ell m}\right]\left._{-1}Y_{\ell m}(\hn)\right. ,\\
	\left[\eth {}_2X^{\rm WF}\right](\hn) &\equiv-\sum_{\ell m} \sqrt{(\ell-2)(\ell+ 3)} \left[ E^{\rm WF}_{\ell m}+ i B^{\rm WF}_{\ell m}\right] \left._3Y_{\ell m}(\hn)\right. .\\
	\end{split}
	\enq
	The deflection estimate $_1\hat d(\hn)$ is decomposed directly into gradient $(g)$ and curl $(c)$ components by using a spin-1 harmonic transform, where the gradient piece contains the information on the lensing potential and the curl component is expected to be zero:\footnote{We follow the standard convention and use $L,M$ rather than $\ell,m$ for lensing multipoles.}
	\begin{equation}\label{eq:gcLM}
_{\pm 1}\hat d(\hn) \equiv \mp \sum_{L M} \left( \frac{\hat g_{L M} \pm i \hat c_{L M}}{\sqrt{L (L + 1)}}\right)\: _{\pm 1}Y_{L M}(\hn).
	\end{equation}

	By default we produce three estimators, namely temperature-only ($s = 0$), polarization-only ($s = \pm 2 $), and MV ($s = 0,\pm 2 $), rather than the traditional full set $TT,TE,TB,EE$, and $EB$ estimators of \cite{Okamoto:2003zw}. The temperature-polarization coupling $C_\ell^{TE}$ is neglected in the $C^{\rm fid}$ factor that appears in the Wiener-filter of Eq.~\eqref{eq:filt} for temperature- and polarization-only estimators, but is included in the MV reconstruction. We use lensed CMB spectra in Eq.~\eqref{eq:filt}
to make the estimator nearly unbiased to non-perturbative order~\citep{Hanson:2010rp,Lewis:2011fk}.
	When producing the full set of individual quadratic estimators, we simply use the same equations after setting to zero the appropriate set of filtered maps entering Eq.~\eqref{eq:lensing2pt}.


The estimators described above only differ from the implementation described in \PlanckLensTwo\ by the presence of the filtered $B$ modes, $B^{\rm WF}$, in Eq.~\eqref{eq:gmap}. This affects the $TB$ and $EB$ estimators, and introduces a $BB$ component in the polarization and MV estimators, which yields no lensing information to leading order in a cosmology with only lensing $B$ modes. However, these modifications have a negligible impact on the reconstruction band powers and their covariance (a maximal fractional change of 0.4\,\% for polarization only, 0.05\,\% for MV) and we make no attempt at further optimization here.

\item {\it Mean-field subtraction and normalization.}  This involves modification of the lensing deflection estimators in Eq.~\eqref{eq:lensing2pt}. Masking and other anisotropies bias the reconstruction and complicate the estimator's response to the lensing potential, which is diagonal in the harmonic domain only under idealized conditions.
   The mean field is the map-level signal expected from mask, noise, and other anisotropic features of the map in the absence of lensing; we subtract this mean-field bias after estimating it using the quadratic estimator mean over our most faithful set of simulations.
   As in \PlanckLensTwo, we apply an approximate isotropic normalization at the map level, calculated analytically for the full sky following~\cite{Okamoto:2003zw}, using isotropic effective beams and noise levels in the filters.
   Our lensing map estimate becomes
\begin{equation}\label{MF}
	\hat \phi_{LM} \equiv \frac{1}{\mathcal R^\phi_L} \Big(\hat g_{LM} - \left\langle \hat g_{LM} \right\rangle_{\rm MC}\Big),
\end{equation}
and similarly for the lensing curl. With the notational conventions adopted above, the responses are identical to those defined in \PlanckLensTwo.
 The isotropic normalization is fairly accurate on average: cross-spectra between reconstructions from masked simulations and the true input lensing realizations match expectations to sub-percent levels on all but the largest scales.
 For the released lensing maps, the subtracted mean field is calculated across the entire available set of 300 simulations (see below), and is also provided. For power-spectrum estimation, we use cross-spectrum estimators of maps with independent Monte Carlo noise on the mean-field subtraction, obtained using $30$ independent simulations for the mean field subtracted from each map (60 simulations in total).
 This number is motivated by a nearly optimal trade-off between uncertainties in the Monte Carlo estimates of the mean field and biases that affect the reconstruction band-power covariance matrix (see Sect.~\ref{subsec:cov}).

 \item {\it Calculation of the power spectrum of the lensing map and subtraction of additional biases.}  More specifically we need to perform subtraction of the so-called $N^{(0)}$ and $N^{(1)}$ lensing biases, as well as point-source contamination. For a pair of lensing map estimates $\hat \phi_1$ and $\hat \phi_2$, we use the same simple cross-spectrum estimator as in \PlanckLensTwo,
\begin{equation}
     \label{eq:naive_phispec}
	\hat C^{\hat{\phi}_1\hat{\phi}_2}_L \equiv\frac{1}{(2L + 1) f_{\rm sky}} \sum_{M = -L}^L\hat\phi_{1,LM}\hat\phi^{*}_{2,LM},\end{equation}
from which biases are subtracted:
\begin{equation}\label{BPdata}
\hat C_L^{\phi\phi} \equiv	\hat C^{\hat \phi_1 \hat \phi_2}_L - \left.\Delta C_L^{\hat \phi_1 \hat \phi_2}\right|_{\rm RDN0} - \left.\Delta C_L^{\hat \phi_1 \hat \phi_2}\right|_{\rm N1} - \left.\Delta C_L^{\hat \phi_1 \hat \phi_2}\right|_{\rm{PS}}.
\end{equation}
Lensing power-spectrum estimation is designed to probe the connected 4-point function of the data that is induced by lensing. The combination of the mean-field subtraction and the first bias term $\Delta C_L^{\hat \phi_1 \hat \phi_2}|_{\rm RDN0}$ ($N^{(0)}$) in Eq.~\eqref{BPdata} subtracts the disconnected signal expected from Gaussian fluctuations even in the absence of lensing, and is calculated using the same realization-dependent $N^{(0)}$ $(\RDNzero)$ estimator described in \PlanckLensTwo\ (and summarized in Appendix~\ref{app:biases} for the baseline cases). The $\Delta C_L^{\hat \phi_1 \hat \phi_2}|_{\rm N1}$ term subtracts
an $\clo(C_L^{\phi\phi})$ signal term ($N^{(1)}$) coming from non-primary couplings of the connected 4-point function~\citep{Kesden:2003cc},
and in our baseline analysis this is calculated using a full-sky analytic approximation in the fiducial model as described in Appendix~\ref{app:biases}. The signal-dependence of this term is handled consistently in the likelihood as described in Sect.~\ref{sec:params}. We tested an alternative simulation-based $N^{(1)}$ calculation and a deconvolution technique as described in Sect.~\ref{subsec:N1tests}. The point-source (PS) bias term
$\Delta C_L^{\hat \phi_1 \hat \phi_2}|_{\rm{PS}}$ subtracts the contribution from the connected 4-point function of unclustered point sources. The amplitude of this correction is estimated from the data, as described in \PlanckLensTwo.

The MV estimator empirically has very slightly larger reconstruction noise $N^{(0)}$ than the $TT$ estimator for lensing multipoles $L\,{\simeq}\,1000$ and beyond, as was also the case in \PlanckLensTwo: the simple combination of the various quadratic estimators in Eq.~\eqref{eq:lensing2pt} is slightly suboptimal if the fiducial covariance matrix does not exactly match that of the data. This is at most a $2\,\%$ effect at the highest multipoles of the reconstruction, and is sourced by our choice of independently filtering the temperature and polarization data (i.e., the neglect of $C_\ell^{TE}$ in $\Cov$ in Eq.~\ref{eq:filt}). Therefore, we have not attempted further optimization in Eq.~\eqref{eq:lensing2pt}.

\item {\it Binning, and application of a multiplicative correction.}  This final
correction is obtained through Monte Carlo simulations to
account for various approximations made in the previous steps, including correcting the approximate isotropic normalization assumed.
After converting the lensing potential spectra to convergence ($\kappa$) spectra ($C_L^{\kappa\kappa} = L^2 (L + 1)^2C_L^{\phi\phi} / 4$), we define our band-power estimates
\begin{equation}\label{Binning}
\hat C^{\kappa\kappa}_{L_b} \equiv \left( \sum_L\mathcal B_{b}^L \hat C_L^{\kappa \kappa} \right)\left(\frac{\sum_L \mathcal B_b^L \clkkfid}{\sum_L \mathcal B_b^L \left\langle\hat C^{\kappa\kappa}_L \right\rangle_{\rm MC}}\right).
\end{equation}
The binning functions $\mathcal B_b^L$ use an approximate inverse-variance weighting $V^{-1}_L \propto \left(2 L + 1 \right)f_{\rm sky}{\mathcal R_L^2}/[2L^4 (L + 1)^4]$ to produce roughly optimal signal amplitudes:
\begin{equation}\label{binfunc}
	\mathcal B_b^L = C^{\kappa\kappa,\rm fid}_{L_b}\frac{\clkkfid V^{-1}_{L}}{\sum_{L'}\left(C_{L'}^{\kappa\kappa,\rm fid}\right)^2 V^{-1}_{L'} }, \quad L_{\rm min}^b \leq L \leq L_{\rm max}^b.
\end{equation}
Equation~(\ref{binfunc}) rescales the amplitude measurements by the fiducial convergence spectrum interpolated to the bin multipole $L = L_b$, where the bin multipoles are the weighted means,
\begin{equation}
L_b \equiv \frac{ \sum_{L} L \:\mathcal B_b^L}{ \sum_{L'} \mathcal B_b^{L'}}.
\label{eq:Lb}
\end{equation}
This choice ensures that the binned fiducial spectrum goes exactly through the fiducial model at $L=L_b$, so plotting band-power bin values at $L=L_b$ against the unbinned fiducial model gives a fair visual comparison of whether the observed band power is higher or lower than the fiducial one (assuming that the true spectrum shape is close to the fiducial shape). For a flat convergence spectrum, Eq.~\eqref{eq:Lb} gives the centre of mass $L$ of the bin.

In a change to the earlier analyses, we no longer subtract a Monte Carlo correction from the estimated lensing power spectrum, instead making a multiplicative correction.
The ratio on the right-hand side of Eq.~\eqref{Binning} is our multiplicative Monte Carlo correction, which corrects for the various isotropic and simplifying approximations we make in constructing the unbinned power-spectrum estimator.
The simulation-averaged band powers $\left\langle\hat C_{L}^{\kappa\kappa} \right\rangle_{\rm MC}$ in Eq.~\eqref{Binning} are built from simulations as from the data according to Eq.~\eqref{BPdata}, but with a cheaper Monte Carlo $N^{(0)}$ $(\MCNzero)$ estimation described in Appendix~\ref{app:biases}, and no point-source correction (since the simulations are free of point sources).
The (reciprocal of the) Monte Carlo correction for our baseline MV band powers is illustrated later in Sect.~\ref{subsec:inhogfilt} (see Fig.~\ref{fig:MCcorr} there).
\end{enumerate}

Other differences to the analysis of \PlanckLensTwo\ include the following points.
\begin{unindentedlist}
\item An improved mask, with reduced point-source contamination for the same sky fraction. The amplitude of the point-source correction decreased by a factor of 1.9, and the detection of this point-source contamination is now marginal at $1.7\,\sigma$. The 2013 and 2015 lensing analyses used essentially the same mask, constructed as described in \PlanckLensOne.
    This is now updated using a combination of unapodized masks: a \smica-based confidence mask;\footnote{The \smica\ mask was a preliminary mask constructed for the \smica\ 2018 analysis; it differs from the final 2018 component-separation mask described in~\cite{planck2016-l04}, since this was finalized later. We make the mask used for the lensing analysis available with the other lensing products, and show results using the final component-separation mask in Table~\ref{table:paramvars} for comparison.} the 2015 70\,\% Galactic mask; and the point-source masks at 143\,GHz and 217\,GHz.
    We also consider a mask targeted at the resolved Sunyaev-Zeldovich (SZ) clusters with ${\rm S/N}>5$ listed in the 2015 SZ catalogue.\footnote{\url{https://wiki.cosmos.esa.int/planckpla2015/index.php/Catalogues}}
    This has little impact on the results, but is included in the baseline analysis, leaving a total unmasked sky fraction $f_{\rm sky}= 0.671$. A reconstruction map
    without the SZ mask is also made available for use in SZ studies.

\item We continue to use the foreground-cleaned \smica\ maps for our baseline analysis; however, the details of the \smica\ processing have changed, as described in~\cite{planck2016-l04}. Specifically, the \smica\ weights at high $\ell$ relevant for lensing are now optimized over a region of the sky away from the Galaxy (but larger than the area included in the lensing mask), significantly changing the relative weighting of the frequency channels on small scales. This changes the noise and residual foreground realization in the \smica\ maps compared to the 2015 analysis, and hence the lensing reconstruction data points scatter with respect to 2015 by more than would be expected from individual frequencies. The 2018 \smica\ maps also correct an LFI map calibration issue in the 2015 maps that affected the amplitude around the first peak; however, this had little impact on the 2015 lensing analysis, since the great majority of the signal comes from smaller scales.

\item Monte Carlo evaluation of the mean field and bias terms are
now based on \planck\ \FFP\ simulations, described in detail in \cite{planck2016-l03}.
In addition to many processing changes, the simulations fix an error in the FFP8 simulation pipeline used for \PlanckLensTwo, which led to aberration (due to the motion of the Solar System relative to the CMB rest frame) not being simulated; the new simulations include the expected level of aberration and the associated modulation~\citep{planck2013-pipaberration}, although this has minimal impact on the lensing analysis. There are only 300 noise \FFP\ simulations, so we now include various sources of Monte Carlo error from the finite number of simulations as additional contributions to the covariance matrix. The noise simulations were generated using a single fiducial foreground and CMB realization, which is subtracted before adding to the signal simulations. Small nonlinearities in the processing cause a weakly correlated residual between simulations.
	This residual can be detected both in the temperature and polarization simulation mean fields at very high lensing multipoles (see Appendix~\ref{subsec:subpix}), where it can be seen that the impact on our band powers is completely negligible compared to the error bars.

 There is a roughly 3\,\% mismatch in power at multipoles $\ell\simeq 2000$ between the data and the FFP10 temperature simulations (which have no variance from residual foregrounds). We account for this by adding isotropic Gaussian noise to the simulations, with a spectrum given by the power difference. The size of this component is roughly 5\,\muKa\ \JCrev{ with a weak scale dependence}. In polarization, as discussed in \cite{planck2016-l04}, the simulation power can be slightly larger than the data power. In this case, for consistency, we add a small additional noise component to the data maps.
\label{subsec:FFP}

\item The lensing maps that we release are provided to higher $L_{\rm max} = 4096$ than in 2015. Multipoles at $L\gg 60$ become increasingly noise dominated, but some residual signal is present at $L>2048$, so we increase the
range, following requests related to cross-correlation and cluster analyses~\citep{Geach:2017crt,Singh:2016xey}. The reconstruction with the full multipole range is made publicly available, but in this paper we only show results for the power spectrum at multipoles up to $L_{\rm max}=2048$; we have not studied the reliability of reconstructions at higher multipoles, so we recommend they be used with caution.
\item The treatment of the Monte Carlo (MC) correction differs, being now multiplicative instead of additive. After subtraction of the lensing biases and formation of the band powers, we calculate the MC correction by taking the ratio to the appropriately binned fiducial $C_L^{\rm {fid},\phi\phi}$. The choice of a multiplicative correction is more appropriate for mode-mixing effects, where corrections are expected to scale with the signal, and for calibration of the quadratic estimator responses when using inhomogeneous filtering. Our baseline reconstruction MC correction is most important on large scales (where it is around 10\,\%), but only has a small impact on the band-power errors.
\item The lensing likelihood is constructed as before, following appendix~C of \PlanckLensTwo. We now include $L \le 4096$ in the calculation of the fiducial $\None$ bias that we subtract, and include $L \le 2500$ in the linear correction to account for the model-dependence of $\None$ relative to the fiducial lensing power. The 2015 MV likelihood contained an almost inconsequential error in the calculation of the response of $\None$ to the polarization power that has now been corrected.
    The construction of the covariance matrix also differs slightly, with additional small terms to take into account uncertainties in several factors that are calibrated in simulations (see Sect.~\ref{subsec:cov}).
    For ``lensing-only'' parameter results we now adopt slightly tighter priors, and marginalize out the
    dependence on the CMB spectra given the observed \planck\ data, as described in Sect.~\ref{sec:paramslensing}.
\item Our fiducial model, the same as used to generate the FFP10 simulations, is now a spatially-flat \lcdm\ cosmology with: baryon density $\omega_{\rm b} \equiv \Omega_{\rm b}h^2 = 0.02216$;
cold dark matter density
$\omega_{\rm c} \equiv\Omega_{\rm c}h^2 = 0.1203$; two massless neutrinos and one massive with mass $0.06\,{\rm eV}$;
Hubble constant $H_0 = 100 h \,\mathrm{km}\,\mathrm{s}^{-1}\,\mathrm{Mpc}^{-1}$ with $h=0.670$;
spectral index of the power spectrum of the primordial curvature perturbation
$n_{\rm s} = 0.964$; amplitude of the primordial power spectrum (at $k=0.05\,\mathrm{Mpc}^{-1}$) $A_{\rm s} = 2.119\times 10^{-9}$; and Thomson optical depth through reionization $\tau=0.060$.
\end{unindentedlist}

\subsection{Covariance matrix}\label{subsec:cov}

Our band-power covariance matrix is obtained from the FFP10 simulation suite. Out of the 300 simulations, 60 are used for the mean-field subtraction ($30$ for each of the quadratic reconstructions that are correlated to form the power spectrum), and 240 for estimation of the lensing biases, Monte Carlo correction, and band-power covariance matrix.  It is impractical to perform the same, expensive, realization-dependent $N^{(0)}$ ($\RDNzero$) subtraction on all these simulations for evaluation of the covariance matrix, and therefore, as in previous releases, we use a cheaper semi-analytic calculation, as detailed in \PlanckLensTwo, which only requires empirical spectra of the CMB data. This semi-analytic calculation is only accurate to 1--2\,\%, which is not enough for debiasing where sub-percent accuracy is required to recover the lensing signal at high lensing multipoles; however, it is sufficient for the covariance matrix calculation.  
	
New to this release are two corrections to the covariance matrix that slightly increase the error bars. First, we take into account Monte Carlo uncertainties in the mean-field, RDN0, and MC corrections. As detailed in Appendix~\ref{app:cov},
for \planck\ noise levels the additional variance $\sigma^2_{\rm MC}$ caused by the finite number of simulations can be written to a good approximation in terms of the band-power statistical\footnote{That is, with perfect knowledge of the mean field, biases and MC correction.}
errors $\sigma^2_{\rm BP}$ as
\begin{equation}
\label{eq:cov_corr}
\sigma^2_{\rm MC} \simeq \left(\frac{2}{N_{\rm MF}} + \frac 9 {N_{\rm Bias}} \right) \sigma^2_{\rm BP}.
\end{equation}
Here, $N_{\rm MF}$ is the number of simulations entering the mean-field subtraction, and $N_{\rm Bias}$ the number used for the noise biases and MC correction. Our choice ($N_{\rm MF} = 60$ and $N_{\rm Bias} = 240 $) is close to optimal, given the 300 simulations at our disposal and our choice of \Nzero estimator. To account for the finite number of simulations, we have simply rescaled the entire covariance matrix by this factor, a $7\,\%$ increase in covariance, irrespective of binning.
Second, we also rescale our inverse covariance matrix by the factor \citep{Hartlap:2006kj}
\begin{equation}
\alpha_{\rm cov} = \frac{ N_{\rm var} - N_{\rm bins}- 2} {N_{\rm var} - 1} ,
\end{equation}
where $N_{\rm var}$ (which equals $N_{\rm bias}$ in our analysis) is the number of simulations used to estimate the covariance matrix, to correct for the bias that would otherwise be present in the inverse covariance matrix that is used in the likelihood.
We construct two likelihoods: one based on the \emph{conservative} multipole range $8 \leq L \leq 400$, for which the number of band-power bins $N_{\rm bins} = 9$ and $1/\alpha_{\rm cov} = 1.035$ (i.e., effectively a 3.5\,\% increase in band-power covariance); and one on the \emph{aggressive} multipole range $8 \leq L \leq 2048$, for which $N_{\rm bins} = 16$ and $1/\alpha_{\rm cov} = 1.06$. \JCrev{After our semi-analytical realization-dependent debiasing, the covariance matrix shows no obvious structure on either multipole range. We find all individual cross-correlation coefficients to be smaller than 10\,\% and consistent with zero, a constraint limited by the number of simulations available. We choose to include the off-diagonal elements in the likelihood.}

Following \PlanckLensTwo~we subtract a point-source template correction from our band powers, with an internally measured amplitude (the point-source shot-noise trispectrum $\hat S_4$).
We neglect the contribution to the error from point-source subtraction uncertainty, since for
for this release the estimated error on $\hat S_4$ would formally inflate band-power errors by at most $0.4\,\%$ at $L \simeq 300$, where the correction is strongest, and much less elsewhere.
\subsection{Inhomogeneous filtering}\label{subsec:inhogfilt}
\begin{figure}[ht!]
\includegraphics[width = \columnwidth]{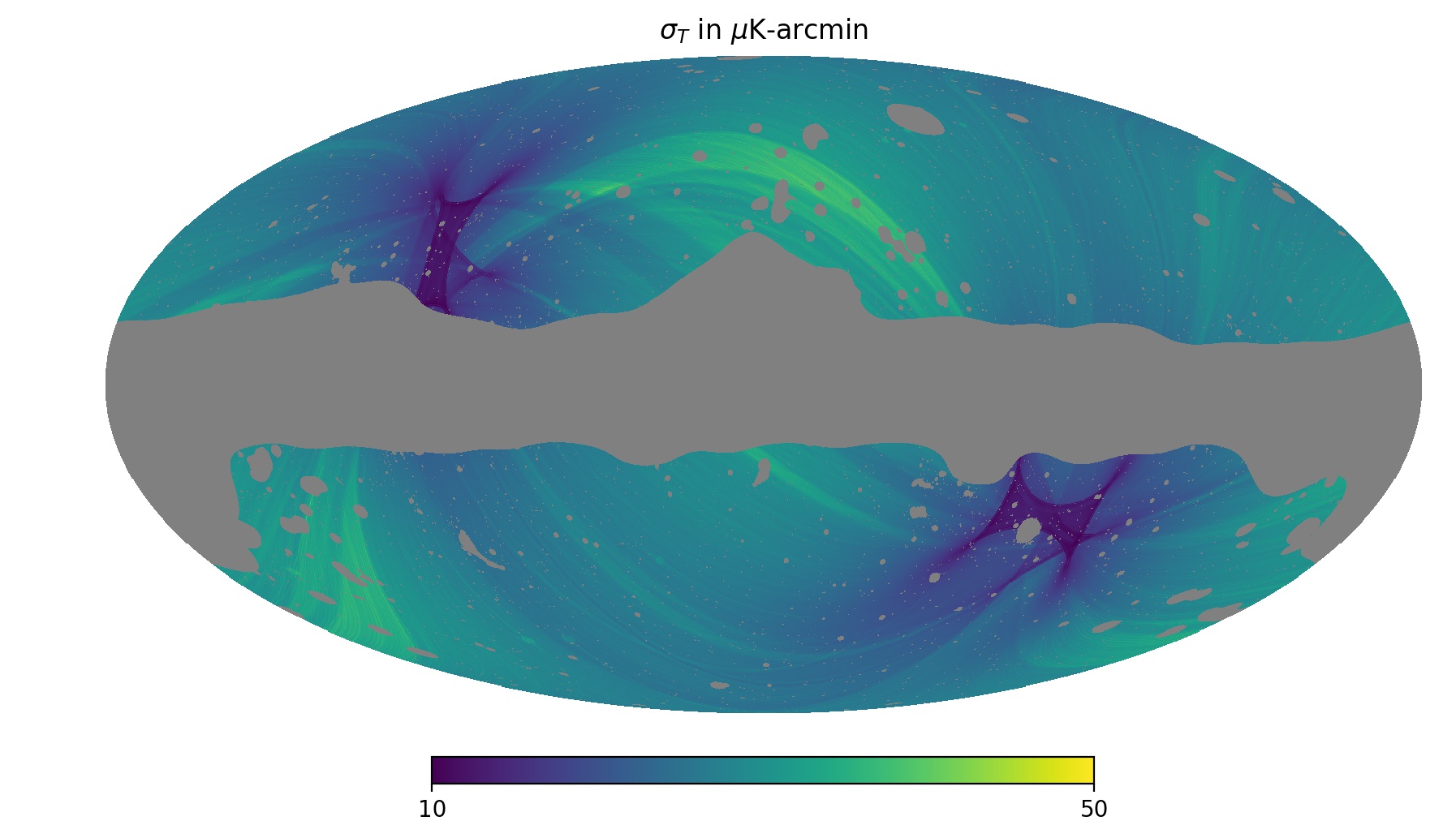}
\includegraphics[width = \columnwidth]{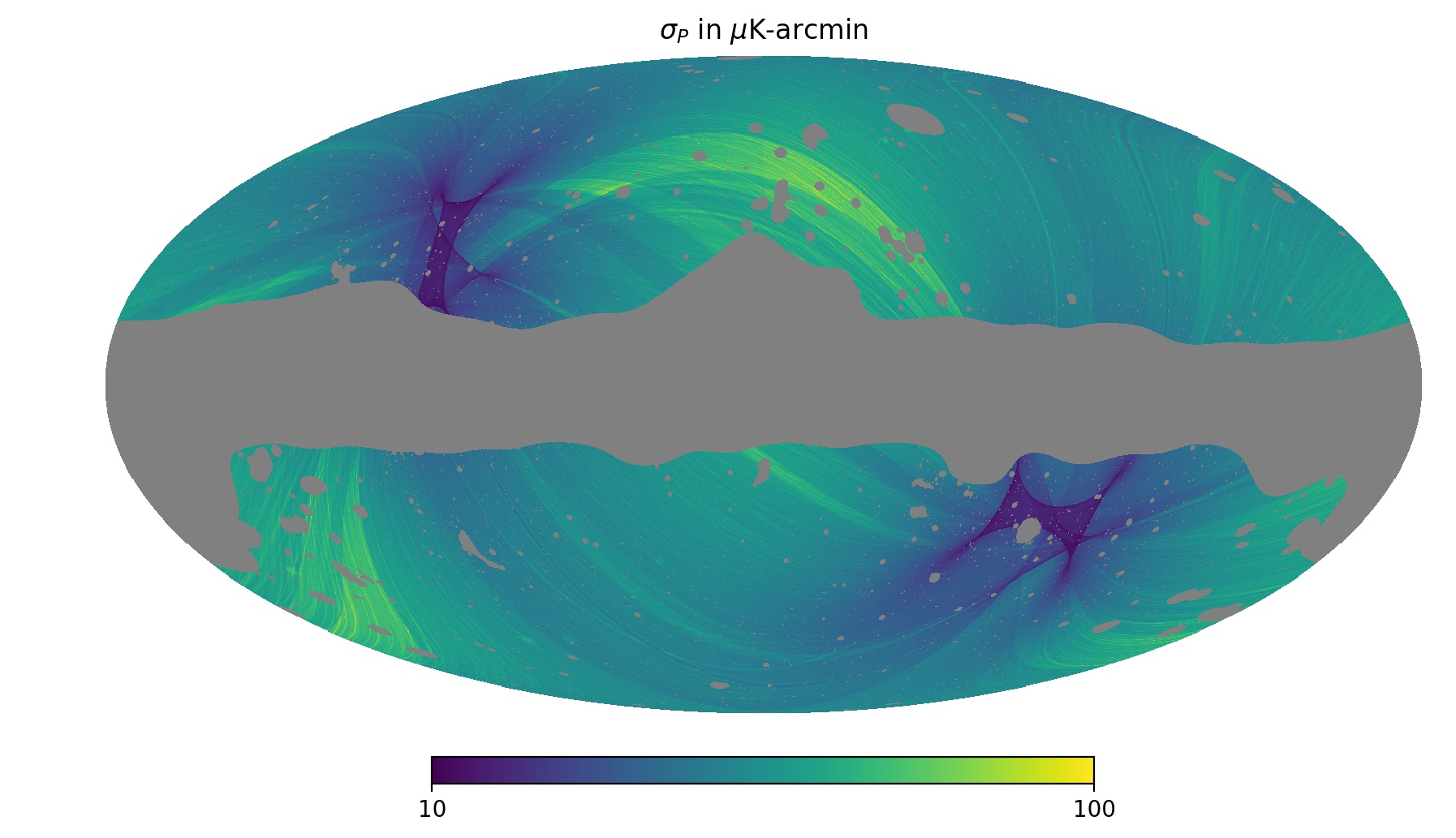}
\caption{Noise-variance maps (shown as noise rms in $\muKa$) that we use to filter the \smica\ CMB maps that are fed into the quadratic estimators when performing inhomogeneous filtering. The upper panel shows the temperature noise map, with median over our unmasked sky area of $27\,\muKa$. We use a common noise map for $Q$ and $U$ polarization, also neglecting $Q$ and $U$ noise correlations, shown in the lower panel, which spans an entire order of magnitude, with median $52\,\muKa$ (larger than $\sqrt{2}$ times the temperature noise because not all the \planck\ detectors are polarized). In temperature, the variance map has a homogeneous (approximately $5\,\muKa$) contribution from the isotropic additional Gaussian power that we add to the simulations to account for residual foreground contamination.}
\label{fig:vmapsmica}
\end{figure}
Approximating the noise as isotropic for filtering is suboptimal because the \planck\ scanning results in significant noise anisotropy with a dynamic range of $10$ for polarization and $5$ for temperature, after allowing for residual foregrounds (see Fig.~\ref{fig:vmapsmica}).
In this section we describe a new polarization-only reconstruction using inhomogeneous filtering, demonstrating a large improvement over polarization results that use homogeneous filtering. However, inhomogeneous filtering
is not used for our main cosmology results including temperature, where it makes little difference but would complicate the interpretation.

The \smica\ CMB map is constructed from \planck\ frequency maps using isotropic weights $w^{X,f}_\ell$ per frequency channel $f$. To construct the \ACrev{noise variance map used in the inhomogeneous filtering}, we first combine the variance maps from the individual frequency maps with the \ACrev{\smica\ weights} to obtain the total noise variance in each pixel of the \smica\ map. More specifically, in polarization, neglecting $Q,U$ noise correlations, defining \JCrev{the pixel noise variance} $\sigma^2_P \equiv \left(\sigma^2_Q + \sigma^2_U\right)/2$, \ACrev{and neglecting differences between $\sigma^2_Q$ and $\sigma^2_U$}, we have
\begin{equation}
	\sigma^2_P(\hn) = \sum_{\text{freq.}\:f} \int_{S_2} d \hn' \sigma^2_{P,f}(\hn') \sum_{s = \pm 2} \frac 14 \left[\xi^{E,f}_{2,s}(\hn \cdot \hn') \pm \xi^{B,f}_{2,s}(\hn \cdot \hn')\right]^2,
\end{equation}
where
\begin{equation}
\xi^{X,f}_{s,s'}(\mu) \equiv \sum_{\ell}\left(\frac{2 \ell + 1}{4\pi}\right) w^{X,f}_\ell d^{\ell}_{s,s'}(\mu),
\end{equation}
and $d_{s,s'}^\ell$ are reduced Wigner $d$-matrices. \ACrev{This equation follows from transforming the noise maps at each frequency $f$, which have pixel variance $\sigma^2_{P, f}(\hn)$ and are further assumed uncorrelated across frequencies,
into their $E$- and $B$-modes, applying the \smica\ weights $w^{X,E}_\ell$ and $w^{B,f}_\ell$, and transforming back to $Q$ and $U$ in pixel space. Averaging the pixel variances of the resulting $Q$ and $U$ noise maps yields $\sigma^2_P$. Physically, the variances of the frequency maps are combined non-locally with kernels that derive from the convolutions implied by the \smica\ weights.}
In temperature,
\begin{equation}
	\sigma^2_T(\hn) = \sum_{\text{freq.}\:f} \int_{S_2} d \hn' \sigma^2_{T,f}(\hn') \left[\xi^{T,f}_{0,0}(\hn \cdot \hn')\right]^2.
\end{equation}

The pixel noise is expected to be correlated to some degree, both due to the mapmaking process and the \smica\ weighting. We have seen no evidence that this is relevant to the reconstruction and for simplicity neglect it for the filter. \smica\ uses a hybrid method with two different set of weights in temperature. Only the second set, used for high multipoles and well away from the Galactic plane, is relevant for the sky area and multipoles used by the lensing analysis.
 The small mismatch in the power in the FFP10 simulations and the data is corrected as described in Sect.~\ref{subsec:FFP}.
 The noise variance maps are shown in Fig.~\ref{fig:vmapsmica}.

 Using the variance maps in our filtering slightly slows down convergence to the solution, especially in temperature.
 Empirically, we found that using smoothed variance maps produces reconstructions with the same signal-to-noise ratio (S/N), as long as the mostly quadrupolar structure of the map is resolved. We use the variance maps of Fig.~\ref{fig:vmapsmica} smoothed with a Gaussian width of $10\deg$ for our quoted results; the impact on the execution time compared to homogeneous filtering is negligible, with the outputs virtually identical to filtering with the high-resolution variance maps.
 Using the noise anisotropy in the filter downweights modes in more noisy regions of the map, and hence improves the optimality of the estimator, especially in polarization, where the \planck\ data are noise dominated on most scales. The expected cross-correlation coefficient of the lensing potential estimate to the true signal improves by 20\,\% over the conservative multipole range $8 \leq L \leq 400$, and the S/N of the band powers increases by 30\,\%. The filter has less impact on the temperature, which is signal dominated, and the reconstruction is of essentially the same quality with or without inhomogeneous filtering.

 One disadvantage of the anisotropic filter is that it complicates the estimator's response: the correct normalization of the estimator becomes position dependent. We have not attempted to perform a full map-level normalization, instead simply making the additional correction as part of the Monte Carlo correction we apply to our band powers. At \Planck's noise levels we expect this procedure to be very close to optimal for polarization, and for most (but not all) scales for temperature.
Approximating the sky as a collection of independent patches with roughly constant noise within a patch, a full-sky optimal lensing spectrum estimation is obtained by inverse-variance weighting correctly-normalized spectra in each patch. Since the estimators are optimal in each patch, the estimator normalization in each patch is identical to the reconstruction noise level $N^{(0)}$. Therefore, whenever the reconstruction noise dominates the lens cosmic variance in the band-power errors, inverse-variance weighting of the patch spectra is equivalent to uniform weighting of the unnormalized estimators' spectra.

\begin{figure}[]
\includegraphics[width = \columnwidth]{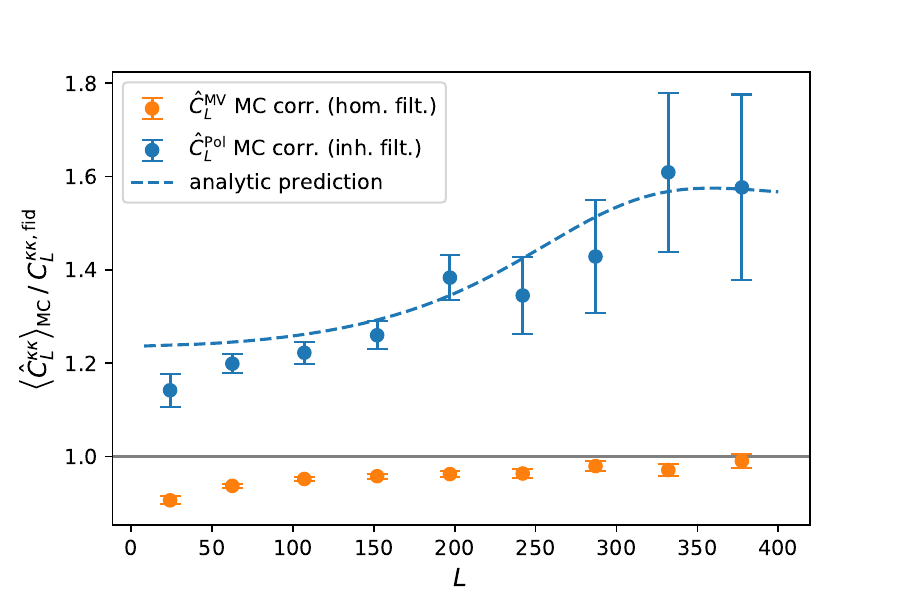}
\caption{Monte Carlo-derived multiplicative normalization corrections for the polarization reconstruction, using inhomogeneous filtering (blue points). Band powers are divided by these numbers to provide our final estimates. This correction is not small, and is sourced by the large spatial variation of the estimator response; however, it is very well reproduced by the approximate analytic model of
Eq.~\eqref{eq:MCcorr}, shown as the dashed blue line. The correction for our baseline MV band powers using homogeneous filtering is shown as the orange points.}
\label{fig:MCcorr}
\end{figure}

 We can predict the lensing spectrum's Monte Carlo correction fairly accurately using the simple independent-patch approximation. Let $\mathcal R_L$ denote the response of the quadratic estimator to the lensing signal, such that under idealized conditions the properly normalized lensing map estimate is (as in Eq.~\ref{MF})
\begin{equation}
\hat \phi_{LM} \equiv \frac{\hat g_{LM}} {\mathcal R_{L}}.
\end{equation}
Applying a single fiducial response $\mathcal R_L^{\rm fid}$ on the full-sky estimate, the local estimate in a patch centred on $\hn$ is biased by a factor $\mathcal R_L(\hn)/ \mathcal R_L^{\rm fid}$, where $\mathcal R_L(\hn)$ is the true response according to the local temperature and polarization filtering noise levels. We may then write the multipoles of the full-sky lensing map as a sum of multipoles extracted over the patches,
\begin{equation}
\hat \phi_{LM} \simeq \sum_{\textrm{patches p}} \frac{\mathcal R_L(\hn^p)}{\mathcal R_L^{\rm fid}} \hat \phi^{p}_{LM},
\end{equation}
where each unbiased component $\hat \phi_{LM}^p$ is obtained from the patch $p$.
Using a large number of patches, neglecting correlations between patches, and turning the sum into an integral gives the following useful approximate result for the correlation of the estimator with the input
\begin{equation}\label{eq:MCcorr_map}
\frac{\av{\hat C_L^{\hat \phi \phi_{\rm in}}}}{C_L^{\phi\phi,\rm fid}} \simeq \int \frac{d \hn}{4\pi} \left( \frac{\mathcal R_L(\hn)}{\mathcal R_L^{\rm fid}} \right),
\end{equation}
and equivalently for the estimator power spectrum,
\begin{equation}\label{eq:MCcorr}
\frac{\av{\hat C_L^{ \phi\phi}}}{C_L^{\phi\phi,\rm fid}} \simeq \int \frac{d \hn}{4\pi} \left( \frac{\mathcal R_L(\hn)}{\mathcal R_L^{\rm fid}} \right)^2.
\end{equation}

The spectrum-level correction of Eq.~\eqref{eq:MCcorr} is only close to the squared map-level correction of Eq.~\eqref{eq:MCcorr_map} (which can be made close to unity using a refined choice of fiducial response) if the true responses do not vary strongly across the sky. 
This is the case for the signal-dominated temperature map; however, the responses vary by almost an order of magnitude in the polarization map. The blue points in Fig.~\ref{fig:MCcorr} show the empirical Monte Carlo correction we apply to our inhomogeneously-filtered polarization band powers, together with the prediction from Eq.~\eqref{eq:MCcorr}. The agreement is visually very good, with a residual at low-$L$ that originates from masking, also found on our baseline, homogeneously-filtered MV band powers (orange points).  This large-scale MC correction has a significant dependence on the sky cut, but little dependence on other analysis choices; the lensing reconstruction is close to local in real space, but this breaks down near the mask boundaries. \JCrev{Finally, while our power-spectrum estimator in Eq.~\eqref{eq:naive_phispec} does not attempt to remove any mode-mixing
effect of masking on the lensing estimate, we note that the large-scale MC correction
is not simply just a $\phi$ mode-mixing effect: using a pseudo-$C_\ell$ inversion to construct the band powers from the masked $\phi$ map makes almost no difference to the MC correction.}

As part of the \Planck\ 2018 release, we make available products containing both versions of lensing maps. Optimally-weighted maps can be used if S/N is critical, while the isotropically-weighted maps can be used to simplify cross-correlation analyses if required. Our baseline results and likelihoods are still derived from the simpler isotropic filtering, but there is a substantial improvement in the polarization-only reconstruction when using the more optimal weighting. In all cases the reconstruction noise properties of the maps are best assessed using the corresponding set of released simulations.

\section{Results}
\label{sec:results}

\subsection{Lensing-reconstruction map and power spectrum}
\label{sec:lensmap}

\begin{figure*}
\includegraphics[width = \columnwidth]{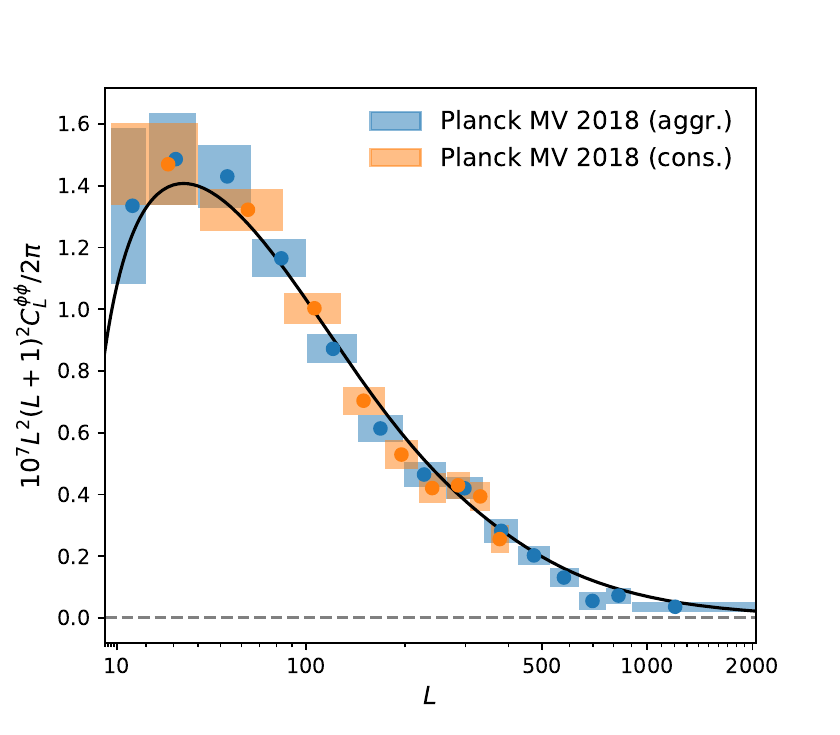}
\includegraphics[width = \columnwidth]{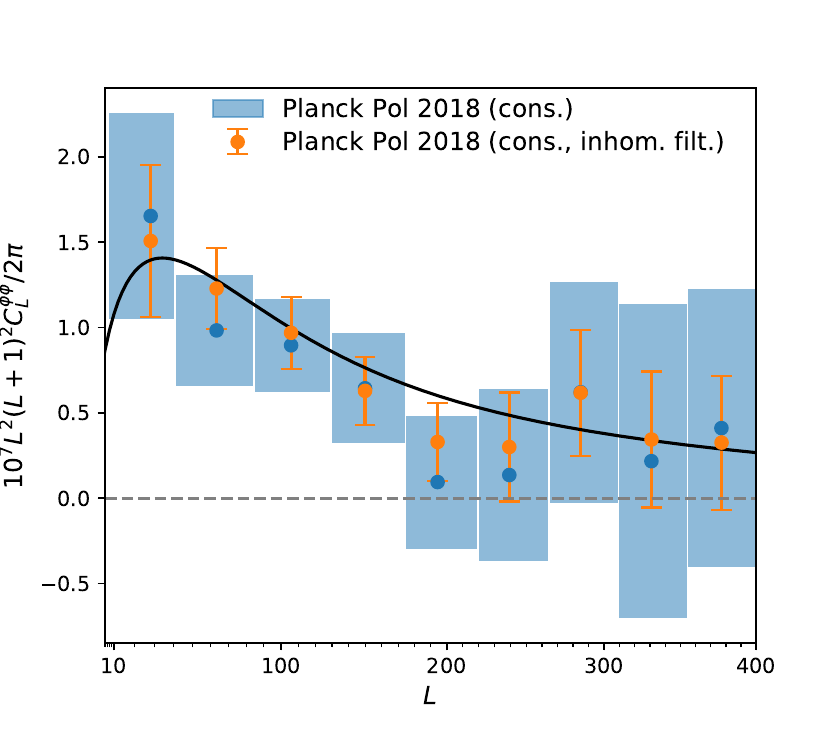}
\caption{\planck\ 2018 lensing reconstruction band powers (values and multipole ranges are listed in Table~\ref{table:BPMV}).
\emph{Left:} The minimum-variance (MV) lensing band powers, shown here using the aggressive (blue, $8 \le L \le 2048$) and conservative (orange, $8 \le L \le 400$) multipole ranges. The dots show the weighted bin centres and the fiducial lensing power spectrum is shown as the black line. \label{fig:MV2018BP}
\newline\emph{Right:} Comparison of \emph{polarization-only} band powers using homogeneous map filtering (blue boxes, with dots showing the weighted bin centres) and the more optimal inhomogeneous filtering (orange error bars). The inhomogeneous filtering gives a scale-dependent increase in S/N, amounting to a reduction of 30\,\% in the error on the amplitude of the power spectrum over the conservative multipole range shown. The black line is the fiducial lensing power spectrum.}
\label{fig:BPpol}
\end{figure*}

In Fig.~\ref{fig:lensmap} we show our baseline Wiener-filtered minimum-variance lensing deflection estimate
from the \planck\ temperature and polarization \smica\ CMB maps.  This is shown as a map of
	\be
	\hat{\alpha}^{\rm WF}_{LM} = \sqrt{L(L+1)}\frac{\clppfid}{\clppfid+ N_L^{\phi\phi}} \hat{\phi}^{\rm MV}_{LM},
	\label{eqn:alphawf}
	\ee
	where $\clppfid$ is the lensing potential power spectrum in our fiducial model
	and $N_L^{\phi\phi}$ is the noise power spectrum of the reconstruction.
    The quantity $\alpha_{LM}^{\rm WF}=\sqrt{L(L+1)}\phi^{\rm WF}_{LM}$ is equivalent to the Wiener-filtered gradient mode (or $E$ mode) of the lensing deflection angle. For power-spectrum estimates we plot
    $[L(L+1)]^2C_L^{\phi\phi}/2\pi = L(L+1)C^{\alpha\alpha}_L/2\pi$, so that a map of $\alpha$ has the same relation
    to the plotted power spectrum as the CMB temperature map does to $\ell(\ell+1)C^{TT}_\ell/2\pi$. As in 2015 we exclude $L<8$ due to the high sensitivity to the mean-field subtraction there.
    The characteristic scale of the lensing modes visible in the reconstruction is $L\simeq 60$, corresponding to the peak of the deflection power spectrum, where the S/N is of order 1.
The left panel in Fig.~\ref{fig:MV2018BP} shows our corresponding baseline MV reconstruction power spectra over the conservative and aggressive multipole ranges.

%
%
%

In addition to the MV reconstruction, we also provide temperature-only results, as well as two variants of the polarization-only reconstruction: the first polarization reconstruction uses the same homogeneous filtering as the temperature and MV results; the second uses the more optimal filter, based on the variance maps shown in Fig.~\ref{fig:vmapsmica}.
The inhomogeneous filtering gives a large improvement in the precision of the polarization-only reconstruction, as shown in the right panel in Fig.~\ref{fig:BPpol}. No significant improvement with inhomogeneous filtering is expected (or found) for the temperature and MV reconstructions, so we do not give results for them.

Table~\ref{table:BPMV} lists our band-power measurements.
For each spectrum, we provide the amplitude relative to the fiducial band powers in the first column, and the fiducial band powers in the second. The binning function was given in Eq.~\eqref{binfunc} and uses an approximate analytic inverse-variance weighting of the unbinned spectra. The fiducial band powers can therefore show slight variations because the noise varies between the different reconstructions.

\begin{table*}[]
\caption{Lensing-reconstruction power-spectrum band-power amplitudes and errors for the temperature-only $(\rm TT)$, combined temperature-and-polarization minimum-variance $(\rm MV)$, and polarization estimators $(\rm PP)$. The last two columns show the polarization reconstruction with inhomogeneous filtering, using the variance maps displayed in Fig.~\ref{fig:vmapsmica}. Amplitudes $\hat{A}$ are quoted in units of the FFP10 fiducial cosmology band powers of $ 10^7\:L^2(L + 1)^2 C_L^{\phi\phi} /2\pi \: $, displayed in the adjacent column. These fiducial band powers are obtained from a suitably-defined bin centre multipole, as described in Sect.~\ref{sec:pipe}, and can differ slightly for the different reconstructions as the binning functions are constructed using approximate inverse-variance weighting of the unbinned spectra. The two polarization-only reconstructions differ by the use of homogeneous or inhomogeneous noise filtering. They do share the same fiducial band powers given in the last column.}
\label{table:BPMV}
\begingroup
\newdimen\tblskip \tblskip=5pt
\setbox\tablebox=\vbox{
 \newdimen\digitwidth
 \setbox0=\hbox{\rm 0}
 \digitwidth=\wd0
 \catcode`*=\active
 \def*{\kern\digitwidth}
 \newdimen\signwidth
 \setbox0=\hbox{+}
 \signwidth=\wd0
 \catcode`!=\active
 \def!{\kern\signwidth}
\halign{\hbox to 1.25in{#\leaderfil}\tabskip 1em&
\hfil#\hfil& \hfil#\hfil& \hfil#\hfil& \hfil#\hfil&
\hfil#\hfil& \hfil#\hfil\tabskip=0pt& \hfil#\hfil& \hfil#\hfil\tabskip=0pt\cr
\noalign{\doubleline}
\omit $L_{\rm min}$--$L_{\rm max}$\hfil& $\fidprefix\hat A^{\phi,\rm TT}$ & $\rm TT$-fid.& $\fidprefix\hat A^{\phi,{\rm MV}}$& MV-fid.& $\fidprefix\hat A^{\phi,\rm{PP}}$& $\fidprefix\hat A^{\phi,{\rm PP}}$(inhom. filt.) &$\rm PP$-fid.\cr
\noalign{\vskip 3pt\hrule\vskip 5pt}
\omit&\multispan8\hfil Conservative multipole range $(8 \leq L \leq 400)$\hfil\cr
\noalign{\vskip -5pt}
\omit&\multispan8\hrulefill\cr
\noalign{\vskip 3pt}
 **8--*40& $1.10\pm0.12$& 1.40& $1.05\pm0.09$& 1.40& $1.18\pm0.43$& $1.08\pm0.32$& 1.40\cr
 *41--*84& $1.12\pm0.07$& 1.28& $1.04\pm0.05$& 1.28& $0.77\pm0.25$& $0.96\pm0.19$& 1.28\cr
 *85--129& $1.02\pm0.07$& $9.90\times10^{-1}$& $1.01\pm0.05$& $9.92\times10^{-1}$& $0.90\pm0.28$& $0.97\pm0.21$& $9.95\times10^{-1}$\cr
 130--174& $0.91\pm0.08$& $7.59\times10^{-1}$& $0.92\pm0.06$& $7.61\times10^{-1}$& $0.84\pm0.42$& $0.82\pm0.26$& $7.65\times10^{-1}$\cr
 175--219& $0.84\pm0.09$& $5.97\times10^{-1}$& $0.88\pm0.08$& $5.98\times10^{-1}$& $0.16\pm0.65$& $0.55\pm0.38$& $6.01\times10^{-1}$\cr
 220--264& $0.93\pm0.12$& $4.83\times10^{-1}$& $0.87\pm0.10$& $4.84\times10^{-1}$& $0.28\pm1.03$& $0.62\pm0.66$& $4.86\times10^{-1}$\cr
 265--309& $1.15\pm0.13$& $4.00\times10^{-1}$& $1.07\pm0.11$& $4.01\times10^{-1}$& $1.54\pm1.61$& $1.54\pm0.92$& $4.02\times10^{-1}$\cr
 310--354& $1.10\pm0.15$& $3.38\times10^{-1}$& $1.17\pm0.14$& $3.38\times10^{-1}$& $0.64\pm2.71$& $1.02\pm1.18$& $3.38\times10^{-1}$\cr
 355--400& $0.74\pm0.16$& $2.88\times10^{-1}$& $0.89\pm0.16$& $2.88\times10^{-1}$& $1.42\pm2.83$& $1.13\pm1.36$& $2.89\times10^{-1}$\cr
\noalign{\vskip 3pt\hrule\vskip 5pt}
\omit&\multispan8\hfil Aggressive multipole range $(8 \leq L \leq 2048)$\hfil\cr
\noalign{\vskip -5pt}
\omit&\multispan8\hrulefill\cr
\noalign{\vskip 3pt}
**8--**20& $1.05\pm0.27$& 1.24 & 1.07 $\pm$ 0.20 & 1.24& & & & \cr
*21--**39& $1.13\pm0.13$& 1.40 & 1.06 $\pm$ 0.11 & 1.40& & & & \cr
*40--**65& $1.23\pm0.09$& 1.34 & 1.07 $\pm$ 0.08 & 1.34& & & & \cr
*66--*100& $1.02\pm0.07$& 1.14 & 1.02 $\pm$ 0.05 & 1.14& & & & \cr
101--*144& $0.98\pm0.07$& $9.02\times10^{-1}$& $0.96\pm0.05$& $9.04\times10^{-1}$& & & & \cr
145--*198& $0.83\pm0.08$& $6.83\times10^{-1}$& $0.89\pm0.06$& $6.86\times10^{-1}$& & & & \cr
199--*263& $0.91\pm0.09$& $5.10\times10^{-1}$& $0.91\pm0.08$& $5.13\times10^{-1}$& & & & \cr
264--*338& $1.14\pm0.11$& $3.80\times10^{-1}$& $1.10\pm0.10$& $3.82\times10^{-1}$& & & & \cr
339--*425& $0.92\pm0.14$& $2.85\times10^{-1}$& $0.99\pm0.13$& $2.85\times10^{-1}$& & & & \cr
426--*525& $0.89\pm0.16$& $2.13\times10^{-1}$& $0.95\pm0.14$& $2.13\times10^{-1}$& & & & \cr
526--*637& $0.77\pm0.20$& $1.60\times10^{-1}$& $0.82\pm0.19$& $1.60\times10^{-1}$& & & & \cr
638--*762& $0.29\pm0.24$& $1.21\times10^{-1}$& $0.45\pm0.23$& $1.21\times10^{-1}$& & & & \cr
763--*901& $0.53\pm0.28$& $9.34\times10^{-2}$& $0.77\pm0.28$& $9.34\times10^{-2}$& & & & \cr
902--2048& $0.66\pm0.32$& $5.18\times10^{-2}$& $0.70\pm0.30$& $5.18\times10^{-2}$& & & & \cr
\noalign{\vskip 5pt\hrule\vskip 3pt}}}
\endPlancktablewide
\endgroup
\end{table*}

Section~\ref{sec:paramslensing} introduces our new lensing-only likelihood, marginalizing over the CMB spectra. Using this likelihood to obtain lensing amplitude summary statistics $\hat A$, (with $\hat{A}=1$ for $\hat{C}_{L}^{\phi\phi}$ equal to the best-fit \lcdm\ model to the \planck\ temperature and polarization power spectra and the reconstructed lensing power\footnote{This likelihood combination corresponds to that denoted $\planckalllensing$ in~\citet{planck2016-l06}; this is the baseline combination advocated there for parameter constraints.}), we obtain
\begin{align}
\bfprefix\hat{A}^{\phi,{\rm MV}}_{8 \rightarrow 400} &= 1.011 \pm 0.028 \textrm{ (CMB marginalized)}
\end{align}
over the conservative multipole range $L=8$--$400$.
This corresponds to a slightly higher value of the lensing spectrum than \PlanckLensTwo\ (for which $\bfprefix\hat{A}^{\phi,\rm MV}_{40 \rightarrow 400} = 0.995 \pm 0.026 $) with a similar significance. The shift is mostly driven by the temperature reconstruction, whose amplitude is higher than 2015 by $0.8\,\sigma$. This shift is consistent with that expected from the change in methodology and data: the bin $8 \leq L \leq 40$ was not included in the 2015 lensing likelihood, and is
around $1\,\sigma$ high in temperature (causing a $0.3\,\sigma$ amplitude shift), and the mask and \smica\ weights have changed. We have evaluated the expected deviation for these two changes by comparison to reconstructions on the new \smica\ maps with the 2015 mask, and on the new mask, using the 2015 \smica\ weights. Using the observed shifts in the simulated reconstructions after the indicated changes, we find expected amplitude differences of $0.18\,\sigma$ and $0.33\,\sigma$, respectively. Discarding any additional changes in the data processing, adding these in quadrature results in the observed total amplitude shift of $1.3\,\sigma$. Over the aggressive multipole range, the measured amplitude is
\begin{equation}
\bfprefix\hat{A}^{\phi,{\rm MV}}_{8 \rightarrow 2048}= 0.995 \pm 0.026 \textrm{ (CMB marginalized)}.
\end{equation}
As discussed in detail in Sect.~\ref{sec:tests}, the high-$L$ range fails a pair of consistency tests and we advise against using the full range for parameter constraints.

The temperature reconstruction still largely dominates our MV estimate, with amplitudes
\begin{align}
\bfprefix\hat{A}^{\phi,\rm TT}_{8 \rightarrow 400} &= 1.026 \pm 0.035 \textrm{ (CMB marginalized)}, \\
\bfprefix\hat{A}^{\phi,\rm TT}_{8 \rightarrow 2048}&= 1.004 \pm 0.033 \textrm{ (CMB marginalized)}.
\end{align}
Amplitude statistics for the polarization-only reconstructions are as follows (neglecting the CMB marginalization and other very small likelihood linear corrections):
\begin{align}
\bfprefix\hat{A}^{\phi,\rm PP}_{8 \rightarrow 400} &= 0.85 \pm 0.16 \textrm{ (homogeneous filtering)}; \\
\bfprefix\hat{A}^{\phi,\rm PP}_{8 \rightarrow 400} &= 0.95 \pm 0.11 \textrm{ (inhomogeneous filtering)}.
\end{align}
This is formally a $5\,\sigma$ measurement for our baseline filtering, and roughly $9\,\sigma$ with the optimized filtering. As can be seen in Fig.~\ref{fig:BPpol}, the improvement of the polarization reconstruction is scale-dependent, with most gain achieved on small scales. This behaviour is consistent with analytic expectations, calculated using the independent-patch approximation introduced in Sect.~\ref{subsec:inhogfilt}.
\AC{There, a 30\,\% reduction in band-power errors was noted, but no discussion of the dependence on scale was given (although the latter is shown for the MC correction).}\JC{Edited, though not detailing the calculation nor introducing a figure.}
\begin{figure*}
\centering
\includegraphics[width = \textwidth]{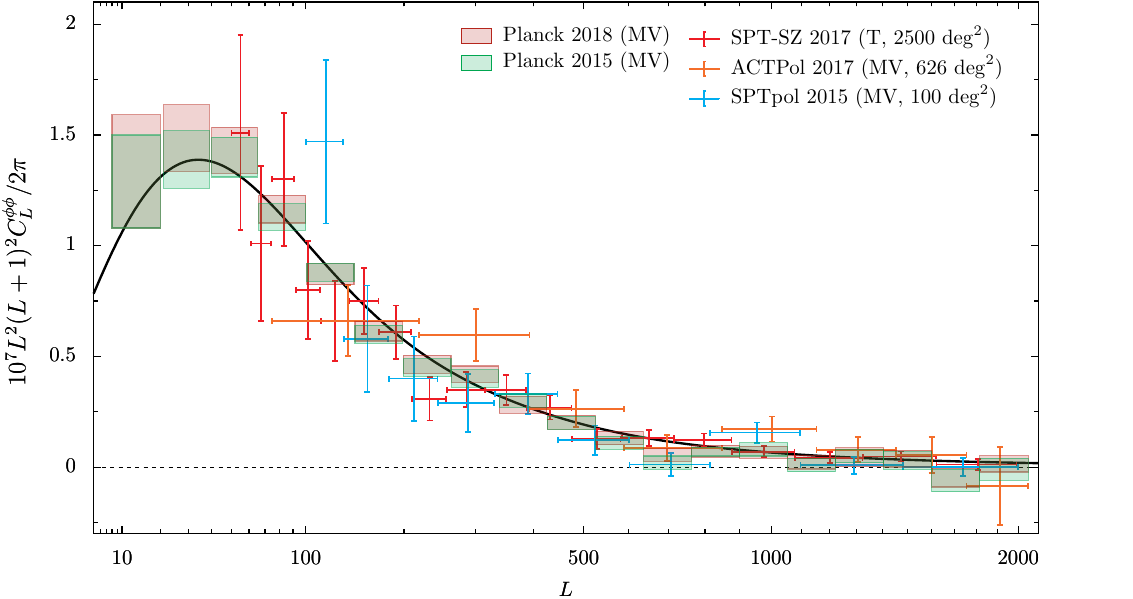}
\caption{\planck\ 2018 lensing power-spectrum band powers (pink boxes) over the aggressive multipole range.
The 2015 analysis band powers (green) were calculated assuming a slightly different fiducial model and have not been (linearly) corrected to the 2018 model. Also shown are recent measurements by the ACTPol \citep{Sherwin:2016tyf}, SPTpol \citep{Story:2014hni}, and SPT-SZ \citep{Simard:2017xtw} collaborations. The SPT-SZ measurement is not completely independent, since the SPT-SZ reconstruction also uses temperature data from \Planck, but with subdominant weight over the smaller sky area used. The black line shows the lensing potential power spectrum for the \LCDM\ best-fit parameters to the \planck\ 2018 likelihoods (\planckall, which excludes the lensing reconstruction).
\AC{Not a big deal, but figure uses different fonts (and tick-mark locations) to other plots in the paper.}\JC{tried for a while, did not manage to get the right fonts on the labels. Anyone familiar with Pyx? (other plots done with matplotlib} \AC{Don't worry -- hadn't realised it wasn't matplotlib.}
}
\label{fig:BPcompil}
\end{figure*}

All reconstruction band powers are consistent with a \lcdm\ cosmology fit to the \planck\ CMB power spectra. Figure~\ref{fig:BPcompil} presents a summary plot of our new MV band powers together with a compilation of other recent measurements, and the previous results from \PlanckLensTwo.

\subsection{Likelihood and parameter constraints}
\label{sec:params}

We construct a lensing likelihood from the power-spectrum reconstruction following the same method as \PlanckLensTwo.
We now expand the default conservative multipole range to $8\le L\le 400$, but exclude the higher multipoles to be conservative, given marginal evidence for null-test failures in the lensing curl and frequency consistency at $L>400$.
Multipoles $L<8$ are very sensitive to the fidelity of the simulations due to the large mean field there, and we continue to exclude them for robustness (though only $L=2$ looks clearly anomalous).
The likelihood for the band powers over $8\le L \le 400$
is approximated as Gaussian, with a fixed covariance estimated from simulations, but the power-spectrum band powers are corrected perturbatively for changes in normalization and \None\ due to parameter-dependent deviations from the fiducial model. We neglect a possible dependence on cosmology of the small Monte Carlo correction.

The final likelihood is of the form\footnote{\JCrev{\cite{planck2013-p12} lists in Appendix C several arguments and tests (performed with more simulations than we are using in this paper) that justifies our use of a Gaussian likelihood. These tests performed on the updated simulations do not show any qualitative difference.}}
	\be
		-2 \log {\cal L}_{\phi} = \bin_i^{\elp}
		( \hat{C}_{\elp}^{\phi\phi} - C_{\elp}^{\phi\phi, {\rm th}} )
		\left[ \Sigma^{-1} \right]^{ij} \bin_j^{\elp'}
		( \hat{C}_{\elp'}^{\phi\phi} - C_{\elp'}^{\phi\phi, {\rm th}} ),
\label{eq:binlike}
	\ee
where $\Sigma$ is the covariance matrix and the binning functions $\bin_i^{\elp}$ are defined in Eq.~\eqref{binfunc}.
The binned ``theory'' power spectrum for cosmological parameters $\theta$ is given in the linear approximation by
\be
 \bin_i^{\elp} C_{\elp}^{\phi\phi, {\rm th}}
\simeq \bin_i^{\elp} \left.C_{\elp}^{\phi\phi}\right|_\vtheta + M_i^{a,\ell'}\left(
 \left.C^a_{\ell'}\right|_\vtheta - \left.C^a_{\ell'}\right|_{\rm fid}\right),
 \label{eq:fastlike}
\ee
where $a$ sums over both the $C_{L}^{\phi\phi}$ and CMB power-spectra terms, and the linear correction matrix $M_i^{a,\ell'}$ can be pre-computed in the fiducial model. The linear correction accounts for the $\None$ dependence on $C_{L}^{\phi\phi}$, and the dependence of the lensing response \ALrev{and $\None$} on the CMB power spectra; explicitly,
\ALrev{
\begin{eqnarray}
M_i^{\phi,\elp'} &=&  \bin_i^{\elp}\frac{\partial}{\partial C_{\elp'}^{\phi}} \left.\Delta C_L^{\hat \phi_1 \hat \phi_2}\right|_{\rm N1} \\
M_i^{X,\ell'} & =& \bin_i^{\elp}\frac{\partial}{\partial C_{\ell'}^{X}} \left( \left.\Delta C_L^{\hat \phi_1 \hat \phi_2}\right|_{\rm N1} +
\ln\left([\mathcal R^\phi_L]^2\right) \left.C_L^\phi\right|_{\rm fid}\right),
\end{eqnarray}
where $C_L^\phi$ derivatives are understood not to act on the lensing contribution to lensed power spectra, $X$ is one of the CMB power spectra, and $C_\ell^X$ derivatives do not act on the fiducial power spectra in the estimator weights.
}

The 2015 CMB likelihoods were based on an LFI polarization likelihood at low multipoles, but the new 2018 low-$\ell$ likelihood now uses the HFI data for the low-$\ell$ polarization and gives a constraint on the optical depth with a considerably smaller uncertainty~\citep{planck2016-l05}. Since uncertainty in the optical depth is the main limitation to inferring the perturbation power-spectrum amplitude from CMB power-spectrum measurements, this means that the 2018 CMB likelihoods can constrain amplitude-related parameters significantly better than in 2015, reducing the relative impact of the information coming from the lensing likelihood. However, it is still important to check consistency using the lensing power spectrum. The lensing spectrum also contains some shape information and can probe extensions to \LCDM\ in some directions of parameter space that cannot be constrained with primary CMB power-spectrum measurements alone (e.g., due to the geometric degeneracy). The new $8\le L < 40$ bin only has a small impact on \lcdm\ parameter constraints, but is valuable for some extended models. For example, some modified gravity models, or the presence of compensated isocurvature modes, can give substantial changes to the lensing spectrum at low multipoles as discussed in \paramsIII\ and~\cite{planck2016-l10}.

\subsubsection{Constraints from lensing alone and comparison with CMB}
\label{sec:paramslensing}

\begin{figure}
\centering
\includegraphics[width = \columnwidth]{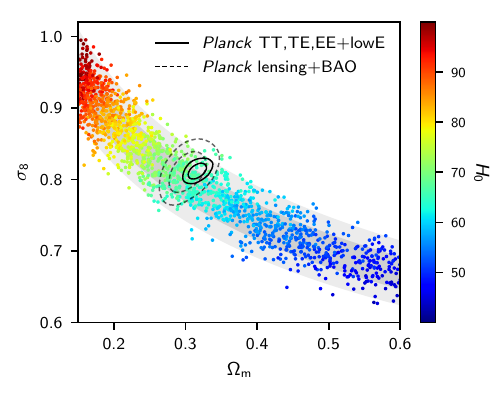}
\caption{Constraints on $\Omm$ and $\sigma_8$ in the base-\LCDM\ model from CMB lensing alone (posterior sample points coloured by the value of the Hubble constant in units of $\Hunit$) using the priors described in the text.
Grey bands give corresponding $1\,\sigma$ and $2\,\sigma$ lensing-only constraints using the approximate fit $\sigma_8 \Omm^{0.25} = 0.589\pm 0.020$.
The joint 68\,\% and 95\,\% constraints from CMB lensing with the addition of BAO data \citep{Beutler:2011hx, Ross:2014qpa,Alam:2016hwk} are shown as the dashed contours, and the constraint from the \planck\ CMB power-spectrum data is shown for comparison as the solid contours.
\label{fig:LCDMlensonly}
}
\end{figure}
To do a ``lensing-only'' analysis, in \PlanckLensTwo\ we fixed the theoretical CMB power spectra, which are required for the linear \ACrev{correction} in Eq.~\eqref{eq:fastlike}, to a \lcdm\ fit to the CMB power-spectrum data. We now remove any dependence on the theoretical model of the CMB power spectra by marginalizing out the theoretical $C_\ell^{\rm CMB}$ by approximating their distribution as Gaussian, where up to a constant
\be
-2\ln P(C^{\rm CMB}|\hat{C}^{\rm CMB}) \simeq \left(C^{\rm CMB}-\hat{C}^{\rm CMB}\right) \cov_{\rm CMB}^{-1}
\left(C^{\rm CMB}-\hat{C}^{\rm CMB}\right).
\ee
Here, $C^{\rm CMB}$ and $\hat{C}^{\rm CMB}$ are vectors of CMB $TT$, $TE$, and $EE$ power-spectrum values at each multipole, with
$\hat{C}^{\rm CMB}$ a data estimate of the CMB power spectra without foregrounds (or noise), which could be measured in various ways. The covariance matrix of the CMB power spectra is $\cov_{\rm CMB}$.
Integrating out $C^{\rm CMB}$, the likelihood then takes the form of Eq.~\eqref{eq:binlike} with covariance increased to account for the uncertainty in the CMB power spectra,
\be
\bar\Sigma_{ij} = \Sigma_{ij} + M_i^{X,\ell}\cov_{\rm CMB}^{X\,\ell;Y\,\ell'} M_j^{Y,\ell'},
\label{eq:sigmamarge}
\ee
and the theory spectrum shifted by the linear correction for the observed CMB power,
\begin{multline}
 \bin_i^{\elp} \bar{C}_{\elp}^{\phi\phi, {\rm th}}
\simeq \left(\bin_i^{\elp}+M_i^{\phi,\elp}\right)\left.C_{\elp}^{\phi\phi}\right|_\vtheta \\
 - M_i^{\phi,\elp}\left.C^{\phi\phi}_{\elp}\right|_{\rm fid}
   + M_i^{X,\ell}\left(
 \hat{C}^X_{\ell} - \left.C^X_{\ell}\right|_{\rm fid}\right),
 \label{eq:like}
\end{multline}
where $X,Y$ are summed only over CMB spectra.
The combined term on the second line is now a constant, so the likelihood only depends on cosmological parameters via $C^{\phi\phi}_{L}|_\vtheta$.
We evaluate the CMB power correction using the \pliklite\ band powers, which are calculated from the full {\tt plik} high-$\ell$ likelihood by marginalizing over the foreground model without any further assumptions about cosmology~\citep{planck2014-a13,planck2016-l05}. To relate \pliklite\ bins to the $M_i^{X,\ell'}$ bins, we assume that the underlying CMB power spectra are represented only by modes that are smooth over $\Delta \ell =50$.
The \pliklite\ \ALrev{bandpower covariance $\cov_{\rm CMB}$} is similarly used to calculate Eq.~\eqref{eq:sigmamarge}. The increase in the diagonal of the covariance is about 6\,\% at its largest, and the linear correction shifts lensing amplitude estimates slightly compared to using a \lcdm\ best fit. The shift is largely explained because, over the $\ell$ range that the lensing reconstruction is sensitive to, the CMB $TT$ data are somewhat less sharply peaked than the \lcdm\ model (which also shows up in a preference for the phenomenological lensing amplitude parameter
$\Alens>1$ when fitting just CMB power-spectrum data, as discussed in~\citealt{planck2016-l06});
smaller $dC_\ell/d\ell$ between the acoustic peaks leads to a smaller lensing signal response, so the theory model value $\bar{C}_{\elp}^{\phi\phi, {\rm th}}$ is decreased (by approximately 1.5\,\% compared to the \lcdm\ best fit).

We follow~\PlanckLensTwo\ in adopting some weak priors for constraining parameters from the lensing likelihood without using the \planck\ CMB power-spectrum data. Specifically, we
fix the optical depth to reionization to be $\tau = 0.055$, put a prior on the spectral index of $\ns = 0.96\pm 0.02$,
and limit the range of the reduced Hubble constant to $0.4 < h < 1$.
We also place a prior on the baryon density of $\Omb h^2 = 0.0222\pm 0.0005$, motivated by D/H measurements in quasar absorption-line systems combined with the predictions of big-bang nucleosynthesis (BBN).\footnote{From a set of seven quasar absorption-line observations, \cite{Cooke:2017cwo} estimate a primordial deuterium ratio $10^5 {\rm D/H} = 2.527\pm 0.030$.
Assuming that standard BBN can be solved exactly, the D/H measurement can be converted into an $\Omb h^2$ measurement with notional $1\,\sigma$ statistical error of $1.6 \times 10^{-4}$.
However, as discussed in \paramsIII, the central value depends on various nuclear rate parameters that are uncertain at this level of accuracy.
For example, adopting the theoretical rate of~\cite{Marcucci:2015yla}, rather than the defaults in the {\tt PArthENoPE} code~\citep{Pisanti:2007hk}, results in a central value shifted to $\Omb h^2=0.02198$ compared to $\Omb h^2= 0.02270$, while \cite{Cooke:2017cwo} quote a central value of $\Omb h^2=0.02166$ using \cite{Marcucci:2015yla} but a different BBN code.
Our conservative BBN prior is centred at the mid-point of these two differences, with error bar increased so that the different results (and other rate uncertainties) lie within approximately $1\,\sigma$ of each other.
} The exact choice of $\Omb h^2$ prior has very little effect on lensing-only constraints, but the prior is useful to constrain the sound horizon (since this has a weak but important dependence on $\Omb h^2$) for joint combination with baryon oscillation (BAO) data.
 We adopt the same methodology and other priors as \citet[][hereafter PCP18]{planck2016-l06}, using \CAMB~\citep{Lewis:1999bs} to calculate theoretical predictions with {\tt HMcode} to correct for nonlinear growth~\citep{Mead:2016zqy}.
 Our \COSMOMC~\citep{Lewis:2013hha} parameter chains are available on the Planck Legacy Archive,\footnote{Chains at \url{https://pla.esac.esa.int}, description and parameter tables in~\cite{planck2016-ES}.} where for comparison we also provide alternative results with a different set of cosmological priors consistent with those used by the DES collaboration~\citep{Abbott:2017wau}.
 Parameter limits, confidence contours and marginalized constraints are calculated from the chains using the {\tt GetDist} package,\footnote{\url{https://getdist.readthedocs.io/}} following the same conventions as \citet{planck2014-a15}.


Figure~\ref{fig:LCDMlensonly} shows the lensing-only \LCDM\ constraint on $\sigma_8$ and $\Omm$. As discussed in detail in \PlanckLensTwo, the lensing data constrain a narrow band in the 3-dimensional $\sigma_8$--$\Omm$--$H_0$ parameter space, corresponding to
\begin{equation}
  \frac{\sigma_8}{0.8}\left(\frac{h}{0.67}\right)^{-1}\left(\frac{\Omm}{0.3}\right)^{-0.27} = 0.999\pm 0.026 \quad (\text{68\,\%, lensing only}),
\label{threePCA}
\end{equation}
or the tighter and slightly less prior-dependent $2\,\%$ constraint
\begin{equation}
  \frac{\sigma_8}{0.8}\left(\frac{\Omm}{0.3}\right)^{0.23}\left(\frac{\Omm h^2}{0.13}\right)^{-0.32} = 0.986\pm 0.020 \quad (\text{68\,\%, lensing only}).
\label{threePCAtwo}
\end{equation}
The allowed region projects into a band in the $\Omm$--$\sigma_8$ plane with
\begin{equation}
  \sigma_8 \Omm^{0.25} = 0.589\pm 0.020 \quad (\text{68\,\%, lensing only}).
\label{eq:lensonlyband}
\end{equation}
The corresponding result using a fixed CMB fit for the CMB power spectrum is
$ \sigma_8 \Omm^{0.25} = 0.586\pm 0.020 $, which is consistent with the similar constraint, $\sigma_8 \Omm^{0.25} = 0.591\pm 0.021$, found in \PlanckLensTwo.
  The roughly $0.25\,\sigma$ shift down in this parameter is consistent with the slight increase in $\hat{A}^\phi$ because of the anti-correlation of $\sigma_8 \Omm^{0.25}$ with the lensing deflection power, as discussed in \PlanckLensTwo.
While the tight three-parameter constraints of Eqs.~\eqref{threePCA} and \eqref{threePCAtwo} depend on our priors (for example
weakening by a factor of 2--4 if the baryon density prior and other priors are weakened substantially) the 2-dimensional projection of Eq.~\eqref{eq:lensonlyband} is much more stable (see Table~\ref{table:paramvars} for examples of prior sensitivity).
The \planck\ 2018 power-spectrum constraints give slightly lower values of $\sigma_8$ compared to the 2015 analysis due to the lower optical depth, which increases the overlap between the lensing-only and CMB power-spectrum contours, making them very consistent within the \LCDM\ model.

Combining CMB lensing with BAO data \citep{Beutler:2011hx, Ross:2014qpa,Alam:2016hwk}, and recalling that we are placing a prior on $\Omb h^2$ so that the sound horizon is fairly well constrained, we can break the main degeneracy and constrain individual parameters, giving the \lcdm\ constraints
\threeonesig{H_0 &= 67.9^{+1.2}_{-1.3}\,\Hunit}{\sigma_8 &= 0.811\pm 0.019}{\Omm &= 0.303^{+0.016}_{-0.018}}
 {\text{lensing+BAO}. \label{lensingBAOconstraint}}
The value of the Hubble constant inferred here assuming \LCDM\ is in good agreement with other inverse distance-ladder measurements~\citep{Aubourg:2014yra,Abbott:2017smn}, and the \LCDM\ result from \planck\ power spectra in \paramsIII, but is somewhat in tension with (i.e., lower than) more model-independent values obtained using distance-ladder measurements~\citep{Riess18}.

Massive neutrinos suppress the growth of structure on scales smaller than the neutrino free-streaming scale. The combination of CMB lensing and BAO data is expected to be a particularly clean way to measure the absolute neutrino mass scale via this effect.
Allowing for a varying neutrino mass, the constraints from lensing with BAO are very broad and peak away from the base $\sumnu = 0.06\eV$ we assumed for \lcdm, though not at a significant level (see Table~\ref{table:paramvars}).
Remaining degeneracies can be broken by using the acoustic-scale measurement from the \planck\ CMB power spectra. The acoustic scale parameter $\theta_*$, the ratio of the sound horizon at recombination to the angular diameter distance, is very robustly measured almost independently of the cosmological model (since many acoustic peaks are measured by \planck\ at high precision). Using $\theta_*$ is equivalent to using an additional high-precision BAO measurement at the recombination redshift. For convenience we use the $\thetaMC$ parameter, which is an accurate approximation to $\theta_*$, and conservatively take $100\thetaMC = 1.0409 \pm 0.0006$ (consistent with the \planck\ data in a wide range of non-\LCDM\ models).
Using this, we have a neutrino mass constraint based only on lensing and geometric measurements combined with our priors:
\begin{equation}
  \sumnu< 0.60\eV \quad (\text{95\,\%, lensing+BAO+$\thetaMC$}).
\end{equation}
The other parameters determining the background geometry are very tightly constrained by the inverse distance ladder, and the amplitude parameter is still well measured, though with lower mean value, due to the effect of neutrinos suppressing structure growth:
\threeonesig{H_0 &= (67.4\pm 0.8)\,\Hunit}{\sigma_8 &= 0.786^{+0.028}_{-0.023}}{\Omm &= 0.306\pm 0.009}
{\text{lensing+BAO+$\thetaMC$}. \label{lensingBAOthetaconstraint}}

\subsubsection{Joint \planck\ parameter constraints}

\begin{figure}
\centering
\includegraphics[width = 0.45\textwidth]{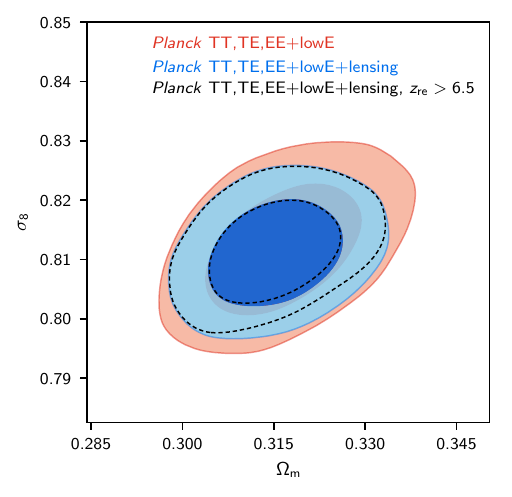}
\caption{Constraints on $\Omm$ and $\sigma_8$ in the base-\LCDM\ model from \planck\ temperature and polarization power spectra (red), and the tighter combined constraint with CMB lensing (blue).
The dashed line shows the joint result when the reionization redshift is restricted to $\zre > 6.5$ to be consistent with observations of high-redshift quasars~\citep{Fan:2006dp}.
Contours contain 68\,\% and 95\,\% of the probability.
\label{fig:sigma8joint}
}
\end{figure}

The CMB lensing power spectrum is consistent with expectations from the \planck\ CMB power-spectrum results, and combining the two can tighten constraints on the amplitude parameters and geometrical parameters that are limited by geometrical degeneracies when measured from the anisotropy power spectrum alone. Figure~\ref{fig:sigma8joint} shows the combined constraint in the $\sigma_8$--$\Omm$ plane, giving the tight (sub-percent) amplitude \LCDM\ result
\begin{equation}
  \sigma_8 = 0.811\pm 0.006 \quad (\text{68\,\%, \planckall$+$lensing}).
\end{equation}
This result uses the temperature and $E$-mode polarization CMB power spectrum likelihoods, including both at low-$\ell$, which we denote by $\planckall$ following \paramsIII.
Constraints on some other parameters, also in combination with BAO, are shown in Table~\ref{table:paramvars}; for many more results and further discussion see \paramsIII.
As in the previous analyses, the \planck\ high-$\ell$ CMB power spectra continue to prefer larger fluctuation amplitudes (related to the continuing preference for high $\Alens$ at a level nearing $3\,\sigma$; see \paramsIII), with the low-$\ell$ optical depth constraint from $E$-mode polarization pulling the amplitude back to values that are more consistent with the lensing analyses. Even with the low-$\ell$ polarization, the power-spectrum data prefer around $1\,\sigma$ higher $\sigma_8$ than the lensing data alone, with the joint constraint lying in between. Values of $\sigma_8$ inferred from the CMB power-spectrum data in a given theoretical model cannot be arbitrarily low because the reionization optical depth cannot be arbitrarily small; observations of high-redshift quasars~\citep{Fan:2006dp} indicate that reionization was largely complete by redshift $z\simeq 6.5$, and this additional constraint is shown in the dotted line in Fig.~\ref{fig:sigma8joint}.

\subsubsection{Joint CMB-lensing and galaxy-lensing constraints}
\label{sec:paramsDES}

\begin{figure}
\centering
\includegraphics[width = \columnwidth]{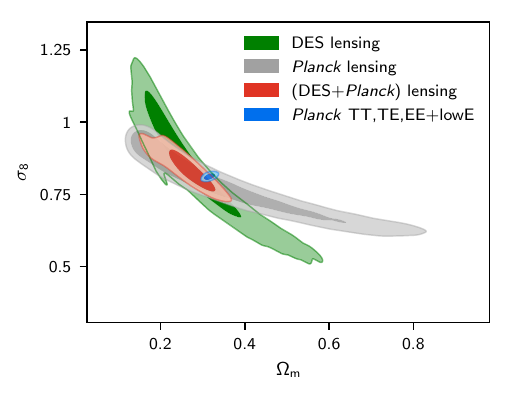}
\caption{Constraints on $\Omm$ and $\sigma_8$ in the base \lcdm\ model from DES galaxy lensing (green),
\planck\ CMB lensing (grey), and the joint constraint (red). The \planck\ power-spectrum constraint
is shown in blue. Here, we adopt cosmological parameter priors consistent with the CMB lensing-only analysis, which differ from the priors assumed by the DES collaboration~\citep{Troxel:2017xyo}. The odd shape of the DES lensing and joint contours is due to a non-trivial degeneracy with the intrinsic alignment parameters, giving a region of parameter space with large negative intrinsic alignment amplitudes that cannot be excluded by current lensing data alone
(but is reduced by different choices of cosmological parameter priors; see \paramsIII).
Contours contain 68\,\% and 95\,\% of the probability.
\label{fig:DESJoint}
\AL{This figure looks worse than the old one, but actually has more samples and (notionally) better converged. The covariance did increase a bit, and bug in the redshift parameter may also have shifting things a bit}
}
\end{figure}

Cosmic shear of galaxies can be used to measure the lensing potential with lower-redshift sources than the CMB. Since the source galaxies and lines of sight to the CMB partly overlap, in general the signals are correlated due to both correlated lensing shear and the intrinsic alignment of the source galaxies in the tidal shear field probed by CMB lensing. Since our CMB lensing reconstruction covers approximately $70\,\%$ of the sky, the area will overlap with most surveys. The cross-correlation signal has been detected with a variety of lensing data~\citep{Hand:2013xua,Liu:2015xfa,Kirk:2015dpw,Harnois-Deraps:2017kfd,Singh:2016xey} and may ultimately be a useful way to improve parameter constraints and constrain galaxy-lensing systematic effects~\citep{Vallinotto:2012,Das:2013aia,Larsen:2015aoa,Schaan:2016ois}.

Here we do not study the correlation directly, but simply consider constraints from combining the CMB lensing likelihood with the cosmic shear likelihood from the Dark Energy Survey~\citep[DES; ][]{Troxel:2017xyo}. In principle the likelihoods are not independent because of the cross-correlation; however, since the fractional overlap of the full CMB lensing map with the DES survey area is relatively small, and since the \planck\ lensing reconstruction is noise dominated on most scales, the correlation should be a small correction in practice and we neglect it here. We use the DES lensing (cosmic shear) likelihood, data cuts, nuisance parameters, and nuisance parameter priors as described by~\citet{Troxel:2017xyo,Abbott:2017wau,Krause:2017ekm}. However we use cosmological parameter priors consistent with our own CMB lensing-only analysis described in Sect.~\ref{sec:paramslensing}.\footnote{In particular, our flat parameter prior $0.4 < h < 1$ on the Hubble constant $100h\Hunit$ is wider than the $0.55 < h < 0.91$ range assumed by~\citet{Troxel:2017xyo}, and our flat prior on $\Omc h^2$ shifts some of the probability weight with respect to the flat prior on $\Omm$ used by DES. We also approximate the contribution of massive neutrinos as a single mass eigenstate with $m_\nu = 0.06\eV$ in our base-\lcdm\ model, rather than marginalizing over the sum of the neutrino masses with a flat prior, and use {\tt HMcode}~\citep{Mead:2016zqy} for nonlinear corrections.}

Figure~\ref{fig:DESJoint} shows the \planck\ and DES lensing-only \lcdm\ constraints in the $\Omm$--$\sigma_8$ plane, together with the joint constraint, compared to the result from the \planck\ CMB power spectra.
The DES lensing constraint is of comparable statistical power to CMB lensing, but due to the significantly lower mean source redshift the degeneracy directions are different (with DES cosmic shear approximately constraining $\sigma_8 \Omm^{0.5}$ and CMB lensing constraining $\sigma_8 \Omm^{0.25}$).
The combination of the two lensing results therefore breaks a large part of the degeneracy, giving a substantially tighter constraint than either alone. The lensing results separately, and jointly, are both consistent with the main \planck\ power-spectrum results. The joint result in the $\Omm$--$\sigma_8$ plane constrains the combined direction
\begin{equation}\label{eq:DESlensband}
\sigma_8(\Omm/0.3)^{0.35} = 0.798^{+0.024}_{-0.019}\quad [\text{68\,\%, (DES+\planck) lensing}],
\end{equation}
although the posterior is not very Gaussian due to a non-trivial DES lensing-intrinsic-alignment parameter degeneracy.
If instead we adopt the cosmological parameter priors of~\citet{Troxel:2017xyo}, but fixing the neutrino mass,
then the lensing-only joint result is more Gaussian, with $\sigma_8(\Omm/0.3)^{0.4} = 0.797^{+0.022}_{-0.018}$; this is tighter than the constraint obtained by~\citet{Troxel:2017xyo} in combination with galaxy clustering data.\footnote{For fixed neutrino mass, the DES joint constraint using the DES priors described in the caption of Table~\ref{table:paramvars} is $\sigma_8(\Omm/0.3)^{0.5}=0.793\pm 0.024$. Marginalizing over neutrino mass, the joint DES-\planck\ lensing-only result using DES priors is $\sigma_8(\Omm/0.3)^{0.5}=0.786^{+0.023}_{-0.019}$. }

\begin{figure}
\centering
\includegraphics[width = 0.5\textwidth]{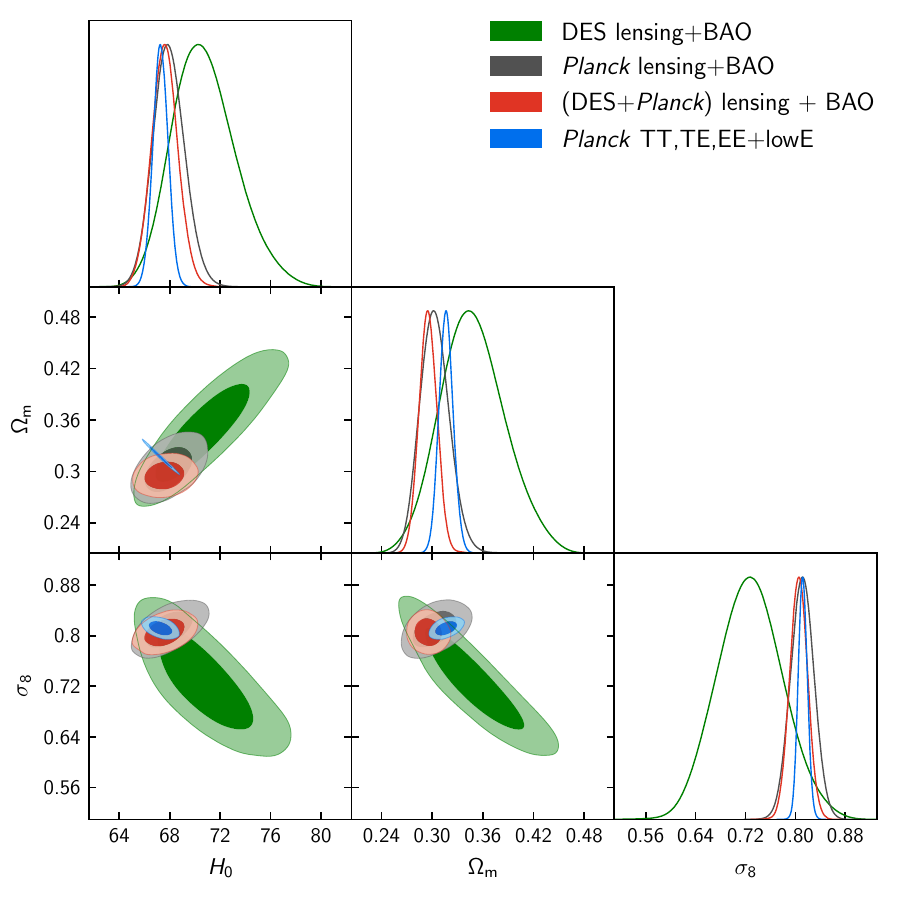}
\caption{\LCDM\ model constraints on $\Omm$, $\sigma_8$, and $H_0$ (in units of $\Hunit$) from DES galaxy lensing+BAO (green),
\planck\ CMB lensing+BAO (grey), and the joint constraint of DES lensing, CMB lensing, and galaxy BAO (red).
All these results use the standard lensing priors described in the text, including the baryon density prior $\Omb h^2 = 0.0222\pm 0.0005$. The \planck\ power-spectrum constraints are shown in blue. Contours contain 68\,\% and 95\,\% of the probability.
}
\label{fig:DESJointBAO}
\end{figure}

\begin{table*}[htbp!]
\caption{Parameter constraints for different lensing datasets and priors, with and without galaxy BAO data. The top block shows parameter constraints using the default conservative \planck\ lensing likelihood, alone and in combination with the DES lensing likelihood, the DES combined lensing and galaxy clustering likelihood,
the \planck\ CMB acoustic scale, as well as the full \planck\ CMB power spectra. The middle block shows results for lensing alone when changing the lensing multipole range, using only temperature reconstruction (``TT'') rather than the minimum-variance combination with polarization (``MV''), or changing the binning scheme. In this context, ``conservative'' and ``aggressive'' refer to the individual bin boundaries listed in the upper and lower parts, respectively, of Table~\ref{table:BPMV}, so that, e.g., ``MV aggressive $8 \le L \le 425$'' uses the first nine bins in the lower part of that table. The ``CompSep mask'' row shows results when
constructing the lensing mask using the final common mask from~\cite{planck2016-l04}, rather than the earlier \smica\ mask used by default throughout this paper.
The lower block gives results from the default conservative (or aggressive) lensing likelihood when varying assumptions. The best-fit $C_\ell^{\rm CMB}$ row shows the result of using a fixed \lcdm\ fit to \planckall\ for the CMB power spectra
(as in the 2015 analysis) rather than marginalizing out the theoretical CMB spectra.
The two ``DES prior'' rows \citep[following][]{Abbott:2017wau} use flat priors on $0.1<\Omm<0.9$, $0.03<\Omb<0.07$, $0.87<\ns<1.07$, $0.55<h<0.91$, $0.5<10^9\As<5$, and, when varying neutrino mass, $0.05\eV<\sum m_\nu < 1\eV$ (which is then unconstrained over this interval). All other results use flat priors on $0.001<\Omc h^2<0.99$, $0.5<\thetaMC<10$, $1.61<\log(10^{10}\As)<3.91$ and (except when combined with \planck\ CMB power spectra) our default lensing priors of $0.4<h<1$, $\ns=0.96\pm 0.02$, $\Omb h^2=0.0222\pm 0.0005$; when varying the neutrino mass, the flat prior is $\sumnu<5\eV$ with three degenerate neutrinos. Note these prior sensitivities have no impact on joint constraints with the CMB power spectra, where the lensing likelihood can be calculated self-consistently without additional priors.
The small sensitivity to nonlinear modelling is demonstrated by using the~\cite{Takahashi:2012em} variant of the \HALOFIT\ nonlinear model~\citep{Smith:2002dz} rather than the default {\tt HMcode}~\citep{Mead:2016zqy}, and by comparison to the linear theory result when the nonlinear corrections are entirely neglected.  All limits in this table are $68\,\%$ intervals, and $H_0$ is in units of ${\rm km}\,{\rm s}^{-1}\,{\rm Mpc}^{-1}$.
}
\label{table:paramvars}
\begingroup
\newdimen\tblskip \tblskip=5pt
\setbox\tablebox=\vbox{
 \newdimen\digitwidth
 \setbox0=\hbox{\rm 0}
 \digitwidth=\wd0
 \catcode`*=\active
 \def*{\kern\digitwidth}
 \newdimen\signwidth
 \setbox0=\hbox{+}
 \signwidth=\wd0
 \catcode`!=\active
 \def!{\kern\signwidth}
\halign{\hbox to 2.0in{#\leaderfil}\tabskip 0.5em&
\hfil#\hfil\tabskip 0.5em& \hfil#\hfil& \hfil#\hfil& \hfil#\hfil&
\hfil#\hfil& \hfil#\hfil& \hfil#\hfil\tabskip=0pt\cr
\noalign{\doubleline}
\omit& \multispan4\hfil\lcdm\hfil& \multispan3\hfil\lcdm+$\sum m_\nu$\hfil\cr
\noalign{\vskip -5pt}
\omit&\multispan4\hrulefill& \multispan3\hrulefill\cr
\noalign{\vskip 2pt}
\omit& Lensing& \multispan3\hfil Lensing$+$BAO\hfil& Lensing& \multispan2\hfil Lensing$+$BAO\hfil\cr
\noalign{\vskip -5pt}
\omit& \hrulefill& \multispan3\hrulefill& \hrulefill& \multispan2\hrulefill\cr
\noalign{\vskip 2pt}
\omit& $\sigma_8 \Omega_{\mathrm{m}}^{0.25}$& $\sigma_8$ & $H_0$ & $\Omega_{\mathrm{m}}$& $\sigma_8 \Omega_{\mathrm{m}}^{0.25}$& $\sigma_8$ & $\Sigma m_\nu\,[\mathrm{eV}]$\cr
\noalign{\vskip 3pt\hrule\vskip 3pt}
MV conservative $8\le L \le 400$ & $0.589\pm 0.020$ & $0.811\pm 0.019$ & $67.9^{+1.2}_{-1.3}$ & $0.303^{+0.016}_{-0.018}$ & $0.569\pm 0.023$ & $0.730\pm 0.041$ & $1.64^{+0.59}_{-1.2}$\cr
DES lensing joint & $0.599\pm 0.018$ & $0.805\pm 0.014$ & $67.6\pm 1.0$ & $0.295\pm 0.011$ & $0.586\pm 0.019$ & $0.755^{+0.037}_{-0.032}$ & $0.78^{+0.27}_{-0.66}$\cr
DES combined joint & $0.589\pm 0.014$ & $0.799\pm 0.013$ & $67.1\pm 1.0$ & $0.286\pm 0.009$ & $0.576\pm 0.015$ & $0.748\pm 0.028$ & $0.75^{+0.33}_{-0.46}$\cr
$100\thetaMC = 1.0409 \pm 0.0006$ joint & $0.592\pm 0.020$ & $0.812\pm 0.015$ & $68.0\pm 0.7$ & $0.304\pm 0.009$ & $0.570\pm 0.022$ & $0.786^{+0.028}_{-0.023}$ & $0.30^{+0.15}_{-0.21}$\cr
\planckTT\ joint & $0.609\pm 0.008$ & $0.809\pm 0.006$ & $67.5\pm 0.5$ & $0.311\pm 0.007$ & $0.603^{+0.01}_{-0.008}$ & $0.812^{+0.01}_{-0.007}$ & $< 0.063$\cr
\planckall\ joint & $0.608\pm 0.006$ & $0.810\pm 0.006$ & $67.7\pm 0.4$ & $0.311\pm 0.006$ & $0.606^{+0.009}_{-0.007}$ & $0.814^{+0.01}_{-0.007}$ & $< 0.058$\cr
\hline
MV conservative $40\le L \le 400$  & $0.588\pm 0.021$ & $0.813\pm 0.020$ & $68.1^{+1.3}_{-1.5}$ & $0.306^{+0.017}_{-0.021}$ & $0.569\pm 0.024$ & $0.729\pm 0.041$ & $1.62^{+0.59}_{-1.1}$\cr
MV aggressive $8\le L \le 425$ & $0.591\pm 0.019$ & $0.813\pm 0.019$ & $68.1^{+1.2}_{-1.3}$ & $0.305^{+0.016}_{-0.018}$ & $0.573\pm 0.023$ & $0.736\pm 0.039$ & $1.54^{+0.56}_{-1.1}$\cr
MV aggressive $8\le L \le 2048$ & $0.578\pm 0.016$ & $0.797\pm 0.016$ & $67.4\pm 1.1$ & $0.295^{+0.014}_{-0.016}$ & $0.559\pm 0.018$ & $0.715^{+0.031}_{-0.038}$ & $1.77^{+0.68}_{-1.1}$\cr
TT conservative $8\le L \le 400$ & $0.572\pm 0.022$ & $0.803\pm 0.021$ & $67.0^{+1.1}_{-1.3}$ & $0.288^{+0.015}_{-0.018}$ & $0.553\pm 0.023$ & $0.703^{+0.037}_{-0.045}$ & $1.95^{+0.68}_{-1.2}$\cr
TT aggressive $8\le L \le 2048$ & $0.557^{+0.019}_{-0.016}$ & $0.786\pm 0.017$ & $66.4\pm 1.1$ & $0.275\pm 0.014$ & $0.541\pm 0.018$ & $0.693^{+0.031}_{-0.036}$ & $1.93^{+0.68}_{-1.0}$\cr
CompSep mask $8\le L \le 400$ & $0.591\pm 0.020$ & $0.812\pm 0.019$ & $68.1^{+1.2}_{-1.3}$ & $0.306^{+0.016}_{-0.019}$ & $0.572\pm 0.023$ & $0.735\pm 0.040$ & $1.53^{+0.55}_{-1.1}$\cr
\hline
DES priors & $0.591\pm 0.020$ & $0.808^{+0.024}_{-0.028}$ & $69.5^{+4.9}_{-11}$ & $0.302^{+0.021}_{-0.026}$ & $0.586\pm 0.020$ & $0.775\pm 0.030$ & $\text{--}$\cr
$'$$'$ + ($\Omega_{\rm b}h^2=0.0222\pm 0.0005$) & $0.593\pm 0.020$ & $0.807^{+0.022}_{-0.026}$ & $68.0\pm 1.5$ & $0.306\pm 0.022$ & $0.586\pm 0.020$ & $0.773^{+0.026}_{-0.029}$ & $\text{--}$\cr
Best-fit $C^{\rm CMB}_\ell$ & $0.586\pm 0.020$ & $0.806\pm 0.019$ & $68.0^{+1.2}_{-1.3}$ & $0.305^{+0.016}_{-0.018}$ & $0.566\pm 0.023$ & $0.726\pm 0.041$ & $1.62^{+0.58}_{-1.2}$\cr
$'$$'$ (MV aggressive $8\le L \le 2048$) & $0.575\pm 0.016$ & $0.793\pm 0.016$ & $67.6\pm 1.1$ & $0.296^{+0.014}_{-0.016}$ & $0.557\pm 0.018$ & $0.712^{+0.032}_{-0.038}$ & $1.73^{+0.67}_{-1.1}$\cr
Takahashi {\HALOFIT} & $0.587\pm 0.020$ & $0.809\pm 0.020$ & $67.9^{+1.2}_{-1.3}$ & $0.302^{+0.016}_{-0.018}$ & $0.560\pm 0.025$ & $0.720\pm 0.044$ & $1.72^{+0.61}_{-1.2}$\cr
$'$$'$ (MV aggressive $8\le L \le 2048$) & $0.574\pm 0.017$ & $0.795\pm 0.017$ & $67.3\pm 1.1$ & $0.293^{+0.013}_{-0.016}$ & $0.548\pm 0.020$ & $0.703^{+0.036}_{-0.042}$ & $1.87^{+0.70}_{-1.2}$\cr
Linear theory & $0.597\pm 0.020$ & $0.820\pm 0.020$ & $68.3^{+1.2}_{-1.4}$ & $0.309^{+0.016}_{-0.020}$ & $0.578\pm 0.024$ & $0.742\pm 0.042$ & $1.56^{+0.57}_{-1.1}$\cr
\noalign{\vskip 3pt\hrule\vskip 3pt}}}
\endPlancktablewide
\endgroup
\end{table*}

Cosmological parameter degeneracies (and degeneracies with intrinsic-alignment and other nuisance parameters) limit the precision of the DES lensing-only results. Results can be tightened by using different priors \citep[as in the DES analysis][see also the analysis in \paramsIII\ using the DES priors]{Abbott:2017wau}. Adding additional data also substantially reduces the degeneracy. For example, adding BAO data gives the tighter contours shown in Fig.~\ref{fig:DESJointBAO}. DES lensing+ BAO gives the marginalized constraints on individual parameters
\threeonesig{H_0 &= 70.8^{+2.1}_{-2.8}\,\Hunit}{\sigma_8 &= 0.728\pm 0.052}{\Omm &= 0.348^{+0.034}_{-0.041}}
 {\text{DES lensing+BAO},}
also consistent with the main \planck\ power-spectrum parameter analysis.
However, as shown in Fig.~\ref{fig:DESJointBAO}, even after combining with BAO, the DES lensing results have a substantial remaining $\sigma_8$--$\Omm$--$H_0$ degeneracy that limits the individual constraints.
For CMB lensing the situation is much better, since the CMB lensing comes from a single well-defined source redshift plane that leads to a tight $\sigma_8$--$\Omm$--$H_0$ degeneracy that intersects with the BAO $\Omm$--$H_0$ degeneracy in a much narrower region of parameter space (giving the parameter constraints of Eq.~\ref{lensingBAOconstraint}). Adding CMB lensing data to the DES result therefore gives much tighter constraints on individual parameters:
\threeonesig{H_0 &= (67.6\pm 1.0) \,\Hunit}{\sigma_8 &= 0.805\pm 0.014}{\Omm &= 0.295\pm 0.011}
 {DES lensing\\+\planck\ lensing+BAO.}
The constraining power on individual parameters is dominated by CMB lensing+BAO, but the DES lensing data do
help partly to break the remaining CMB lensing degeneracy, giving tighter constraints than from CMB lensing+BAO alone.
The tight joint constraint on the Hubble constant here is consistent with the result of~\cite{Abbott:2017smn}, but without the use of galaxy clustering data (which are sensitive to details of bias modelling). Our combined result with the full DES joint likelihood, including galaxy clustering and galaxy-galaxy lensing~\citep{Abbott:2017wau}, is given in Table~\ref{table:paramvars}.

\subsubsection{Parameters from likelihood variations}

The baseline conservative likelihood, restricted to $8\le L\le 400$, was chosen to be robust; however, we cannot rule out the possibility that the moderate null-test failures at higher $L$ that motivate our choice $L \le 400$
are simply statistical fluctuations, in which case it would be perfectly legitimate to use the full multipole range for parameter constraints. Comparing different likelihood variations also allows us to assess how changes in the spectrum propagate into changes in parameters, and hence robustness of the parameter constraints.

Table~\ref{table:paramvars} compares parameter constraints from the baseline likelihood with a number of variations with different binning, multipole ranges, and with and without using polarization in the reconstruction. As a general trend the higher multipoles are below the best-fit to the conservative range, so including lensing multipoles up $L\le 2048$ tends to pull amplitude parameters to lower values, although for the MV reconstruction only to an extent that is compatible with expected shifts when including more data.
The shift between $TT$ and MV results is, however, more anomalous: over the full multipole range the mean value of $\sigma_8 \Omm^{0.25}$ shifts by more than $1\,\sigma$ despite the error bar only shrinking by a bit more than $10\,\%$ (assuming Gaussian statistics, the shift is greater than $2\,\sigma$ unusual). Over the conservative multipole range the shift is more acceptable, although still somewhat conspicuous.
 Even over the conservative range the band powers are slightly tilted with respect to a \lcdm\ best fit, which gives a mild (1--$2)\,\sigma$ preference for non-zero neutrino mass when combined with BAO.
 We discuss the consistency of the different data ranges in more detail at the data level in Sect.~\ref{sec:tests} below.

\subsection{Joint CIB-CMB lensing potential reconstruction}
\label{sec:CIBjoint}
\newcommand{\tra}{\mathcal{T}}
The \planck\ lensing reconstruction has S/N peaking at unity around the peak of the deflection-angle power spectrum at $L\simeq 60$, but is noise dominated on smaller scales. However, the cosmic infrared background (CIB) is a high-redshift tracer of the matter distribution and known to be correlated at the roughly 80\,\% level to the CMB lensing potential and hence, potentially, is a good proxy for it~\citep{Song:2002sg,planck2013-p13,Sherwin:2015baa}. The CIB is well measured by \planck's higher frequency channels~\citep{planck2011-6.6,planck2013-pip56,planck2016-XLVIII,Ying:2016eiz}, with high S/N to much smaller scales than probed directly by the lensing reconstruction. It has already been demonstrated that \planck's CIB measurement can be used to delens the CMB acoustic peaks with about the same efficiency as \planck's internal measurement~\citep{Larsen:2016wpa,Carron:2017vfg}, with the CIB carrying substantially more information on small scales.
For the purpose of delensing degree-scale $B$ modes, most of the lensing signal required is from lensing multipoles $L \simeq 500$, where the \planck\ MV lensing reconstruction map is fully noise dominated, making the CIB especially valuable until higher-sensitivity internal CMB-polarization-based reconstructions are available.
The main difficulty in using the CIB as a tracer is contamination by Galactic dust and modelling of the cross-correlation coefficient. The Galactic dust is more of a problem on large scales, just where the \planck\ lensing reconstruction S/N peaks. This suggests that a joint analysis could potentially give a substantial improvement to the lensing potential determination, and hence also improve the efficiency of delensing based on \planck\ data \citep{Sherwin:2015baa}.  A combined tracer map, including galaxy number counts from WISE in addition to the CIB and the 2015 \planck\ lensing map, has recently been presented over 43\,\% of the sky by~\citet{Yu:2017djs}. Here we focus on the \planck-based combination of the CIB and lensing reconstructions on 60\,\% of the sky, with similar maximum delensing efficiency, and use them to delens all four \planck\ CMB spectra.

\begin{figure}
\centering
\includegraphics[width = \columnwidth]{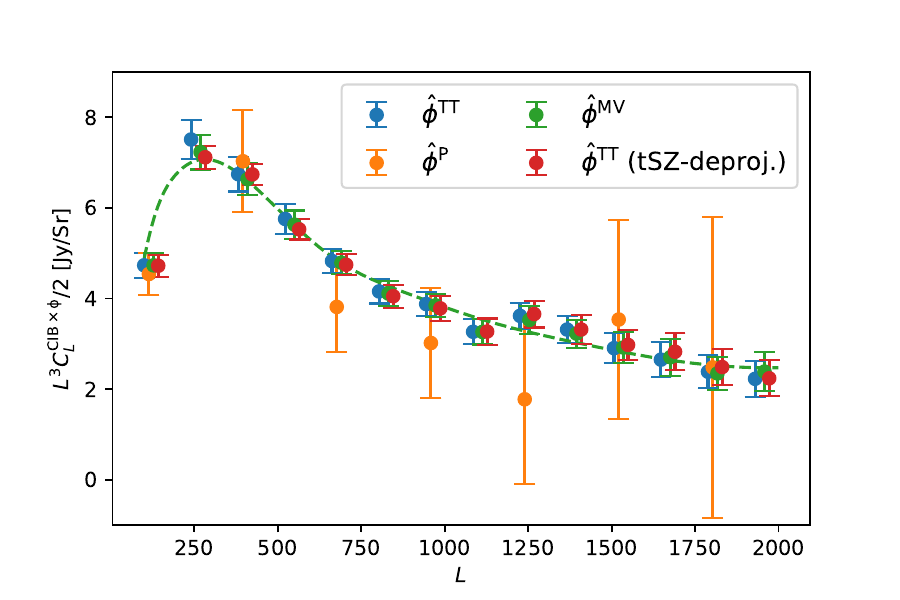}
\caption{Empirical cross-spectra between our lensing reconstructions (temperature-only in blue, polarization-only in orange, and MV in green) and the \GNILC\ 545-GHz CIB map. The red points use the temperature lensing reconstruction from the thermal-SZ-deprojected \smica\ CMB map. The dashed line shows the smooth spline fit to the green points, which we use to weight
the lensing tracers when forming our combined lensing potential estimate.}
\label{fig:qestxCIB}
\end{figure}

We combine the (internal) MV quadratic estimator reconstruction with the CIB map provided by the \GNILC\ (Generalized Needlet Internal Linear Combination) component-separation method at 353, 545, and 857\,GHz \citep[based on the 2015 \planck\ data release and available at the Planck Legacy Archive]{planck2016-XLVIII}. We use a mask given by the union of the lensing mask and the \GNILC\ mask, leaving 60\,\% of the sky unmasked (58\,\% after apodization). Given the strong coherence of the CIB across this range of frequencies, we see very little gain in using combinations of the three frequency maps, and our baseline results use the 545-GHz CIB map only. Figure~\ref{fig:qestxCIB} shows the cross-spectra of the lensing temperature-only, polarization-only and MV reconstructions with the \GNILC\ 545-GHz map. The bispectrum between residual foregrounds in the CMB temperature maps that enter the quadratic estimator and the CIB can potentially bias the CIB-lensing cross-spectra. Figure~\ref{fig:qestxCIB} shows two tests of such bias, neither of which shows any evidence for systematic contamination. The first performs lens reconstruction with a \smica\ CMB map constructed to deproject the thermal-SZ effect (i.e., the weighting across frequencies nulls any component with the frequency spectrum of the thermal SZ effect). The second uses only polarization in the lensing reconstruction, which is expected to be essentially free of extragalactic foregrounds, at the cost of significantly larger reconstruction noise for \planck. Contamination from residual CIB in the \smica\ temperature map is assessed below.

We build lensing estimates as follows. Consider a set of tracers $I^i$ of the lensing potential, with auto-power spectra $C_L^{I_i I_i}$, cross-correlation coefficient matrix $\boldsymbol{\rho}_L$, and cross-correlation coefficients to the true lensing potential $\rho^{i \phi}_L$. Assuming that we can treat the lensing and tracer fields as being approximately jointly Gaussian, it is straightforward to obtain a maximum a posteriori (MAP) estimate for the true lensing given the tracers, yielding an optimally-filtered potential map $\hat\phi^{\rm MAP}$. We can write this in terms of (isotropic) weights $w_L$ acting in harmonic space on renormalized maps with unit spectra, through
\begin{equation}
\frac{\hat\phi_{LM}^{\rm MAP}}{\sqrt{\clppfid}} \equiv \sum_i w_L^i \frac{I_{LM}^i}{\sqrt{C^{I_iI_i}_{L}}},
\label{Eq:normphi}	
\end{equation}
where the optimal weights are given at each lensing multipole by
\begin{equation}
\label{Eq:combw}
w_L^j = \sum_i\rho^{\phi i}_L (\boldsymbol{\rho}_L^{-1})_{ij}.
\end{equation}
The expected resulting squared cross-correlation coefficient $\rho^2_L$ is the weighted sum $\sum_i w^i_L \rho_L^{\phi i}$.
To compare the noise in the joint reconstruction to the quadratic-estimator reconstruction noise $N_L^{(0)}$, we can also define an effective noise level $N_L$ so that
\begin{equation}\label{eq:noiselevels}
\frac{\clppfid}{\clppfid + N_L } = \rho^2_L = \sum_i w^i_L \rho_L^{\phi i}.
\end{equation}

\begin{figure}
\includegraphics[width =\columnwidth]{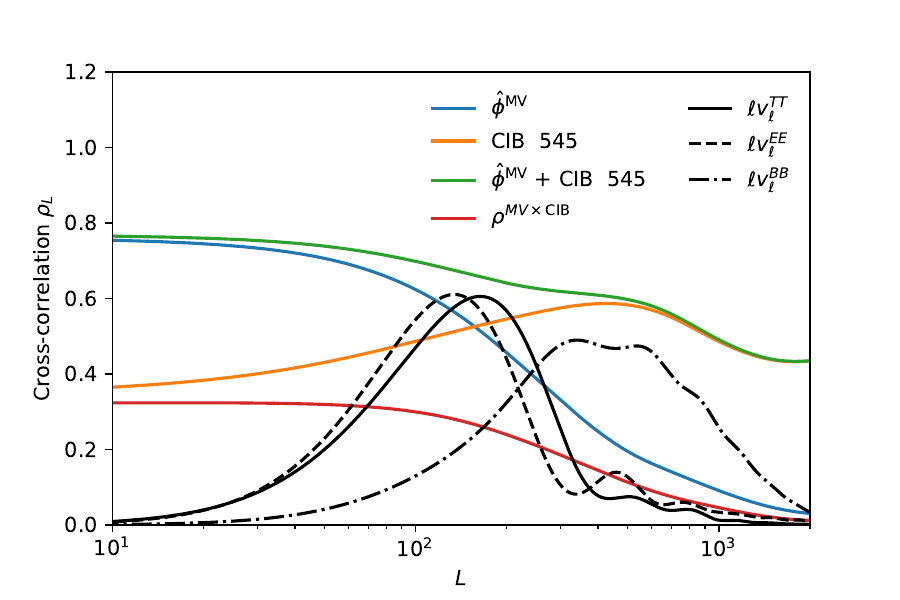}

\includegraphics[width =\columnwidth]{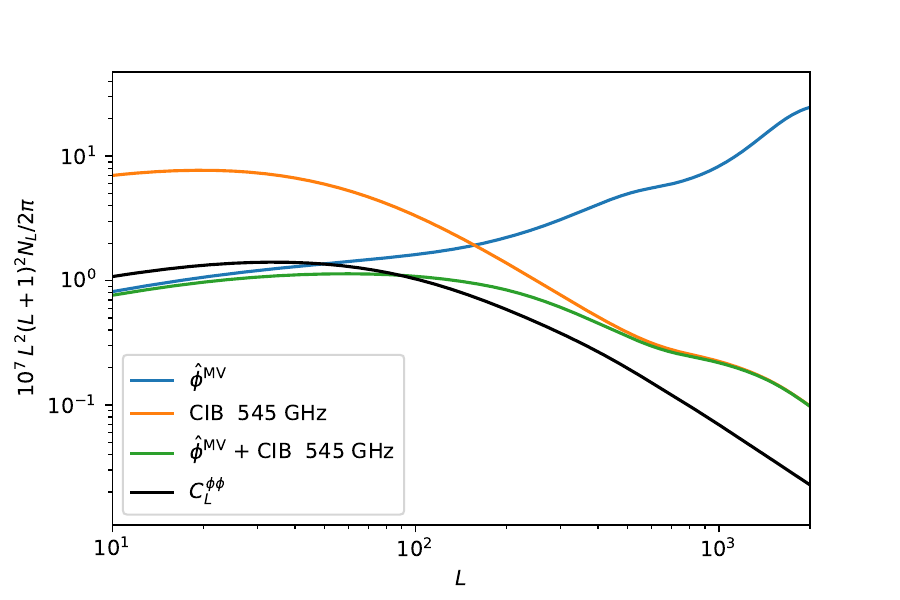}
\caption{
 {\it Top}: Expected cross-correlation coefficient of the true lensing potential to the minimum-variance (MV) estimator, the 545-GHz \GNILC\ CIB map (orange), and their combination (green). To build these curves, the cross-spectrum of the CIB map to the true lensing potential is approximated by its cross-spectrum to the MV (quadratic-estimator) lensing reconstruction, displayed in Fig.~\ref{fig:qestxCIB}, and \GNILC\ CIB curves are smooth spline fits. The red curve shows for comparison the cross-correlation coefficient of the CIB to the MV quadratic estimator. This cross-correlation is suppressed by instrument noise, foreground residuals, and shot noise in the CIB map and reconstruction noise in the lensing quadratic estimator.
The black curves show the lensing kernels
that contribute to the lensing of the temperature (solid), $E$-mode polarization (dashed), and $B$-mode polarization (dash-dotted) power spectra, as described in the text.
\JC{I removed this for the revision: Finally, the dashed lines show the forecast delensing efficiency $\epsilon_L = \rho_L^2$}. {\it Bottom}: Effective reconstruction noise levels $N_L$ for each of these tracers, as defined by Eq.~\eqref{eq:noiselevels}, and, for comparison, the theoretical lensing spectrum $C_L^{\phi\phi}$. The quadratic estimator reconstruction noise is slightly underestimated at low $L$ in this figure, since we have neglected Monte Carlo corrections when combining the tracers.
\label{fig:CIBrhos}
}
\end{figure}

We adopt a purely empirical approach to obtain the weights, together with the auto-spectra on the right-hand side of Eq.~\eqref{Eq:normphi}, and take $C_L^{\phi\phi,\rm fid}$ as our FFP10 fiducial lensing spectrum. For each pair of tracers, we perform a bicubic spline fit across 12 bins of the observed cross-correlation coefficient and auto-spectra, with errors for each bin determined from the observed scatter. The spectra and cross-spectra entering the cross-correlation coefficient are calculated by deconvolving the effects of the mask from pseudo-$C_\ell$ spectra \citep{Wandelt:2000av}, using a 12\arcm\ apodization window.
To estimate the cross-correlation of the tracers to the true lensing potential, we assume that the quadratic estimator is unbiased (which, according to the FFP10 simulations, is a very good approximation except at the very lowest multipoles); the cross-spectra of the CIB maps to the quadratic estimator are then used as proxies to the true $C^{\rm{CIB}\,\phi}$ at the corresponding frequency.

These CIB-lensing cross-spectra get a contribution from the CIB bispectrum from residual CIB present in the CMB maps. However, this is very small: Sect.~\ref{subsec:testforegroundscorr} discusses foreground contamination to our lensing estimates more generally using a dedicated simulation set. Using these simulations, we can cross-correlate the quadratic estimator applied to the expected residual CIB in the \smica\ maps to the CIB component at 545\,GHz. We find a bias of at most 1\,\% at $\ell \simeq 2000$, which we can safely neglect when building the tracers.

\begin{figure*}
\includegraphics[width = 0.95\textwidth]{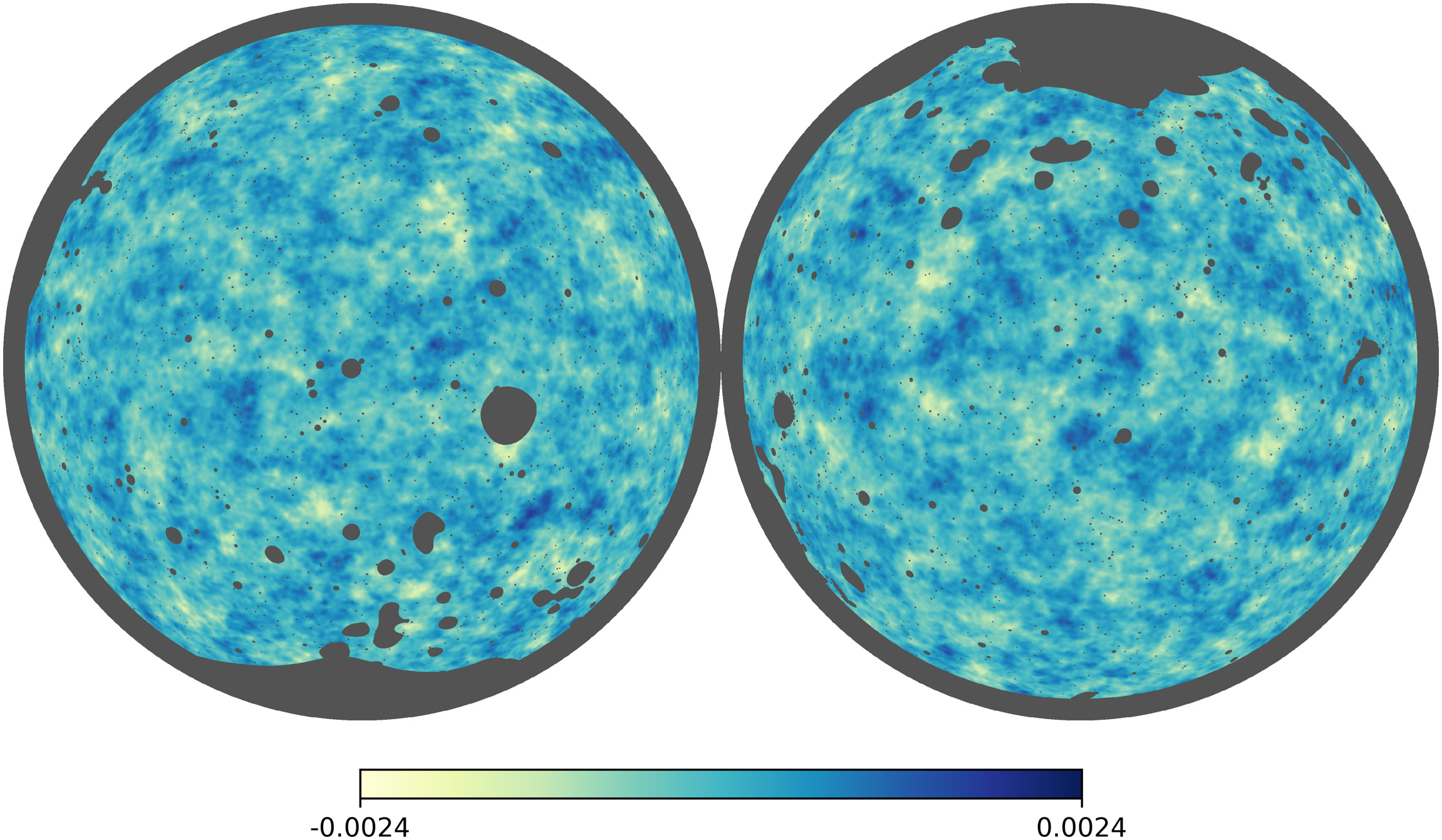}
\includegraphics[width = 0.95\textwidth]{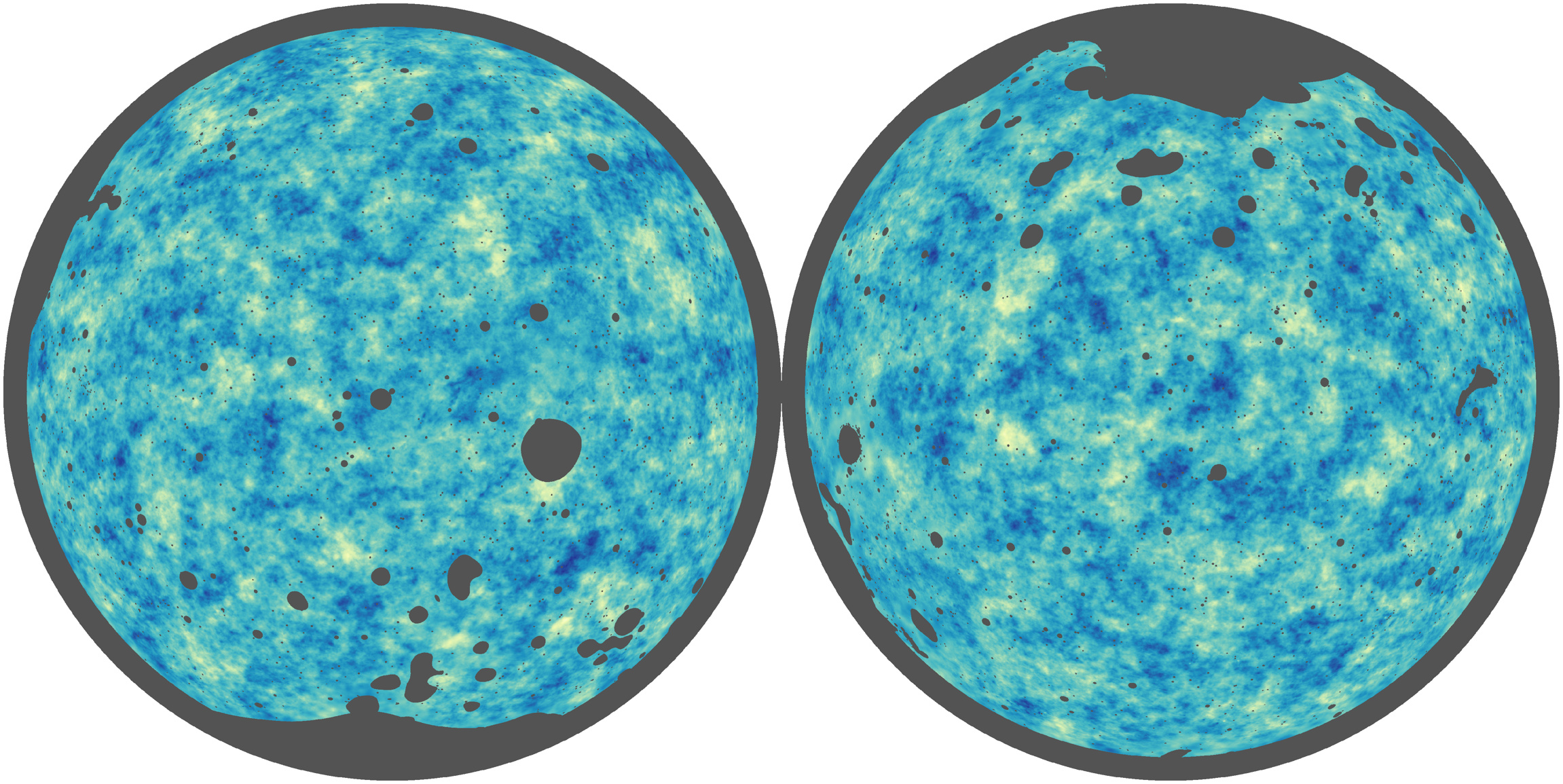}
\caption{Comparison of lensing maps constructed from the minimum-variance quadratic estimator alone (upper panel) and in combination with the CIB, as traced by the 545-GHz \GNILC\ frequency map (lower panel). The combination is performed on 60\,\% of the sky, defined as the union of the lensing mask and the \GNILC\ mask. Maps show the orthographic projection of the Wiener-filtered displacement $E$ mode, the scalar field with multipoles $\hat\alpha_{LM} = \sqrt{L(L+1)}\hat \phi_{LM}$, with $10\leq L \leq 2000$. The left and right panels are centred on the north and south Galactic poles, respectively. While the two reconstruction maps are clearly strongly correlated, the combined map has substantially more small-scale power, due to the higher S/N of the CIB on small scales.
\label{fig:jointmaps}
}
\end{figure*}

The cross-correlation coefficients $\rho_L^{\phi\, \rm GNILC}$ built in this way have significant noise, but can barely be distinguished from each other across frequencies. We choose to use the same fiducial $\rho^{\phi\, \rm CIB}$ at all frequencies, which we obtain by averaging the estimates at $353$ and $545$\,GHz. Figure~\ref{fig:CIBrhos} shows (smoothed) cross-correlation coefficients of the true lensing map to the MV estimator (blue), together with that of the 545-GHz \GNILC\ map (orange), and their combination (green). 
\JCrev{Delensing efficiencies, $\epsilon_L \equiv \rho_L^2$}, are a good measure of how much lensing can be removed at each lensing multipole if the maps are used for a delensing analysis (see Sect.~\ref{sec:delens}). Effective noise levels from Eq.~\eqref{eq:noiselevels} are shown in the lower plot for the MV reconstruction, the 545-GHz \GNILC\ lensing estimate, and their combination, showing how the CIB particularly improves the S/N at smaller scales.

We apply the weights given by Eq.~\eqref{Eq:combw} to the pseudo-$a_{L M}$s of the tracers, obtained using the apodized mask, keeping afterwards the multipole range $10 \leq L \leq 2000$. We do not perform any further low-$L$ cuts to the \GNILC\ maps, since, for our purposes at least, they do not display obvious signs of dust contamination, and they make only a small contribution to the combination with the internal lensing reconstruction on large scales. We note that our delensing analysis below is robust to the low-$L$ quadratic estimator mean field, as demonstrated by \citet{Carron:2017vfg}.

Our joint CIB and internal lensing reconstruction maps are shown in Fig.~\ref{fig:jointmaps}. As expected, adding the CIB tracer dramatically improves the resolution of the lensing reconstruction, effectively filling in the small-scale modes by using the phase information about the sky realization from the CIB, with amplitude determined from cross-correlation of the CIB with the lensing reconstruction. The joint map is available in the 2018 \Planck\ public release.
It could be used for delensing future low-$\ell$ polarization data, and is independent of the CMB at $\ell < 100$ (where we cut the input CMB maps), except for possibly a small contamination in the CIB map.
We have also constructed a version of the map that uses no input $B$ modes, to avoid delensing biases (see Sect.~\ref{sec:BBdelen} below for further discussion).

The joint CIB-lensing map is an excellent tracer of the realization of the lensing field; however, the lensing amplitude information that it contains all comes from the lensing reconstruction, since we do not have an accurate predictive model for the CIB signal. It cannot therefore be used directly to improve parameter constraints from its power spectrum.

\subsection{Delensing \planck\ power spectra}
\label{sec:delens}
CMB $B$-mode polarization from lensing is present on all scales, and for small tensor-to-scalar ratios could become an important source of confusion for the signal from primordial gravitational waves. However, using a lensing reconstruction it is possible to estimate and subtract most of the $B$-mode signal, a process known as delensing~\citep{Knox:2002pe,Kesden:2002ku,Hirata:2003ka}. This may ultimately be crucial for future observations to detect low levels of primordial gravitational waves.
Delensing can also be applied to the temperature and $E$-mode polarization maps, so that the corresponding peak-smoothing effect in the power spectra can be partly removed, sharpening the acoustic peaks and restoring more of the information to the power spectrum~\citep{Green:2016cjr}. Peak sharpening by delensing has been convincingly demonstrated by~\citet[][using CIB maps as a lensing tracer]{Larsen:2016wpa} and \citet[][using the internal \planck\ 2015 lensing reconstruction]{Carron:2017vfg}.
\citet{Carron:2017vfg}
and \citet{Manzotti:2017net} (the latter using the CIB as measured by \textit{Herschel} at 500\,$\mu\text{m}$ to delens SPT data) have also successfully demonstrated that delensing can reduce the power in the $B$-mode polarization, although currently instrumental noise rather than lensing is still limiting the detectable level of primordial $B$ modes.

The three black lines in the upper panel of Fig.~\ref{fig:CIBrhos} show the leading eigenvector of the matrix $\ell \partial C_\ell^{TT,EE,BB}/{\partial \ln C_L^{\phi\phi}}$, for $100 \le \ell \le 2048$, showing the scales over which lensing modes are relevant for lensing of the power spectra. Peak sharpening in $T,E$ requires good delensing efficiency at lensing multipoles $L \alt 250$, while removal of $BB$ lensing power requires smaller-scale lensing reconstruction. As discussed in Sect.~\ref{sec:CIBjoint}, a joint CIB/internal delensing analysis is expected to be significantly better for delensing all signals because of the complementarity of scales.

For characterization of our delensing analysis, we need to extend the FFP10 simulation suite to include CIB components. We perform this in the simplest manner, simulating the CIB as an isotropic Gaussian field, obeying our estimates of the various cross-correlations. To produce the CIB tracers $I_i$ at each frequency $\nu_i$, we first rescale the FFP10 input lensing potential harmonic coefficients according to $\rho^{\phi i}$, then add independent Gaussian noise $\epsilon_i$:
\begin{equation}
\label{eq:CIBnoise}
I^i_{LM} = \sqrt{\frac{\hat C_L^{I^iI^i}}{C^{\phi\phi,\rm fid}_L}} \rho_L^{\phi i} \phi^{\rm input}_{LM} + \sqrt{\hat C^{I^iI^i}_L}\epsilon^i_{LM},
\end{equation}
where the noise is generated according to the reduced covariance
\begin{equation}
\left\langle \epsilon^i_{LM} \epsilon^{j\ast}_{LM} \right\rangle =\rho_L^{ij} - \rho_L^{\phi i} \rho_L^{\phi j}.
\end{equation}
After convolving with the \GNILC\ 5\arcm\ FWHM Gaussian beam, the maps are masked and analysed in the same way as the data maps. We separately use a simulation suite that includes no lensing effects, but are generated with lensed CMB spectra. In this case the simulated CIB maps are just the noise components described by Eq.~\eqref{eq:CIBnoise}, with $\rho^{\phi i}_L=0$. The simulated harmonic coefficients are conservatively computed over a range of scales $\ell_{\rm max} \leq 2500$, slightly larger than the one used for the data combination to avoid edge effects due to masking.

We first delens the $BB$ power spectrum using a template subtraction method in Sect.~\ref{sec:BBdelen}. We then use a debiasing technique to isolate the delensing signature on the CMB signal for the $TT, TE, EE$ and $BB$ spectra, with the help of a direct remapping method in Sect.~\ref{subsec:peakdelensing}.

\subsubsection{$C_\ell^{BB}$-delensing}\label{sec:BBdelen}
Using a filtered version of the $E$-mode map, and a tracer of the lensing $\phi$ field, we can build a template for the lensed $B$ modes that we can subtract from the data to reduce the $B$-mode power.
To reduce the $B$-mode power as much as possible, the fields must be Wiener filtered \citep{Smith:2008an,Smith:2010gu,Sherwin:2015baa}.
The extent by which we can then reduce the $B$-mode power depends on the quality of the tracer and the noise in the $E$-mode map. The correlations of the lensing tracer maps to the true lensing field are shown in Fig.~\ref{fig:CIBrhos}. For the relevant lensing multipoles ($L \simeq 500$) about $0.6^2$, i.e., 35\,\% of the lensing can be removed using \GNILC, with a moderate improvement in combination with the \planck\ internal reconstruction. The \Planck\ small-scale polarization data are noisy, and at CMB $E$-mode scale $\ell \simeq 500$, from where the large-scale $B$-power takes significant contributions, the $E$-mode filter quality $C^{EE}_\ell / (C_\ell^{EE} + N_\ell^{EE} /b_\ell^2)$ (where $b_\ell$ is the combined beam and pixel window function) is no greater than 0.6, so we cannot expect to remove more than about 20\,\% of the lensing $B$-mode power.
This is in contrast to current ground-based experiments, for which the much lower polarization noise level makes the lensing map fidelity the main limiting factor, as demonstrated by \citet{Manzotti:2017net}.

We now describe our template in more detail.  We start with the Wiener-filtered $E$-mode map, the same as produced by our pipeline for input to the quadratic estimators ($E^{\rm WF}$ in Eq.~\ref{eq:filt}). We use the inhomogeneously-filtered version, since this visibly increases the local fidelity of the template and our delensing efficiencies.
 We build the polarization (Stokes parameters) $P^{(E)}\equiv Q^{(E)}+iU^{(E)}$ from this $E$-mode map, and simply remap the polarization according to the deflection angle determined by the lensing tracer, i.e.,
\begin{equation}
P^{\rm template}(\hn) = P^{(E)}(\hn + \vgrad \hat{\phi}_{\rm WF}).
\end{equation}
After projecting into the $B$-mode component, this forms our template for the lensed $B$ modes.
The simplified notation $\hn + \vgrad \hat{\phi}_{\rm WF}$ denotes the displacement of length $|\vgrad \hat \phi_{\rm WF}|$ on the sphere along the geodesic defined by $\vgrad \hat \phi_{\rm WF}$, including the small rotation of the spherical-polar polarization components induced by the parallel transport of the polarization tensor \citep{Challinor:2002cd,Lewis:2005tp}. We perform the interpolation using the {\tt Python} curved-sky lensing tools from \lensit,\footnote{\url{https://github.com/carronj/LensIt}} with a bicubic spline interpolation algorithm. The observed $E$-mode map we use to build the template is itself lensed, while ideally one would use an unlensed $E$-mode map. In principle, we could try and improve on that by obtaining an optimally-filtered $E$ map including our deflection tracer in the Wiener filtering, using the algorithm developed by \cite{Carron:2017mqf}; however, the difference is a second-order effect in the deflection, and we neglect it here.

\begin{figure}
\centering
\includegraphics[width = 0.5\textwidth]{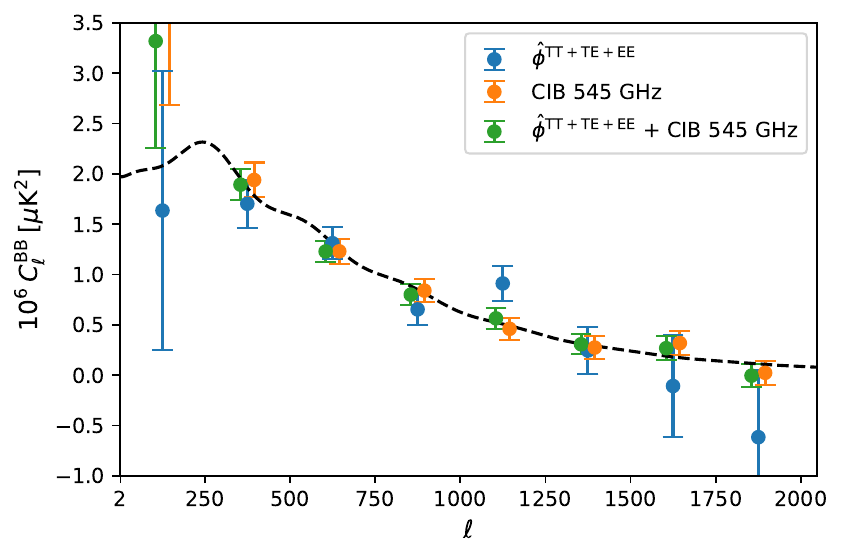}
\caption{\label{fig:phiExB} Estimates of the lensing $B$-mode power spectrum from cross-spectra between the $B$-mode polarization map and various $B$-mode templates constructed using different lensing tracers, as measured on $58\,\%$ of the sky: the $TT$+$TE$+$EE$ quadratic estimator (blue); the CIB map at 545 GHz (orange); and their combination (green). The raw cross-spectra are a filtered version of the $B$-mode spectrum and have been rescaled using the expected scaling computed from simulations assuming an isotropic filter. The dashed black line is the FFP10 fiducial lensing $B$-mode spectrum.}
\end{figure}

The cross-correlations of these templates to the $B$-mode CMB map provide strong indirect detections of the lensing $B$-mode power in the data. The cross-spectra are displayed in Fig.~\ref{fig:phiExB}, for templates constructed from the lensing quadratic estimator (blue), the CIB at 545 GHz (orange), and in combination (green). These templates are cross-correlated to our $B$-mode map, obtained from the Wiener-filtered $B^{\rm WF}_{\ell m}$ after dividing by $C_\ell^{\rm BB, fid} / \left( C_\ell^{\rm BB, fid} + N_\ell^{BB} / b_\ell^2 \right)$, where $b_\ell$ is the combined beam and pixel window function. This is an unbiased estimate of the $B$-mode map away from the mask boundaries. The black line shows the fiducial lensing $B$-power of roughly $(5\muKarcmin)^2$ at low multipoles. We use the quadratic $TT$+$TE$+$EE$ estimator, discarding the $EB$ and $TB$ parts, so that all tracers are statistically independent from the $B$-mode map. The raw cross-spectra actually provide only a filtered version of the $B$-mode power, since the Wiener-filtering of $E$ and $\phi$ maps reduces the power in the template. We follow \citet{Hanson:2013hsb}, assuming isotropy of the filtering, and simply rescale the cross-spectra by the ratio of the result calculated on the FFP10 simulation suite to the simulated $B$-mode power. In all cases, the data cross-correlation amplitudes are consistent with the fiducial model within $1\,\sigma$, with the lensing $B$-mode power detected at 13, 18, and $20\,\sigma$ for templates constructed from the lensing quadratic estimator, the CIB, and in combination, respectively.

\begin{figure}
\centering
\includegraphics[width = 0.5\textwidth]{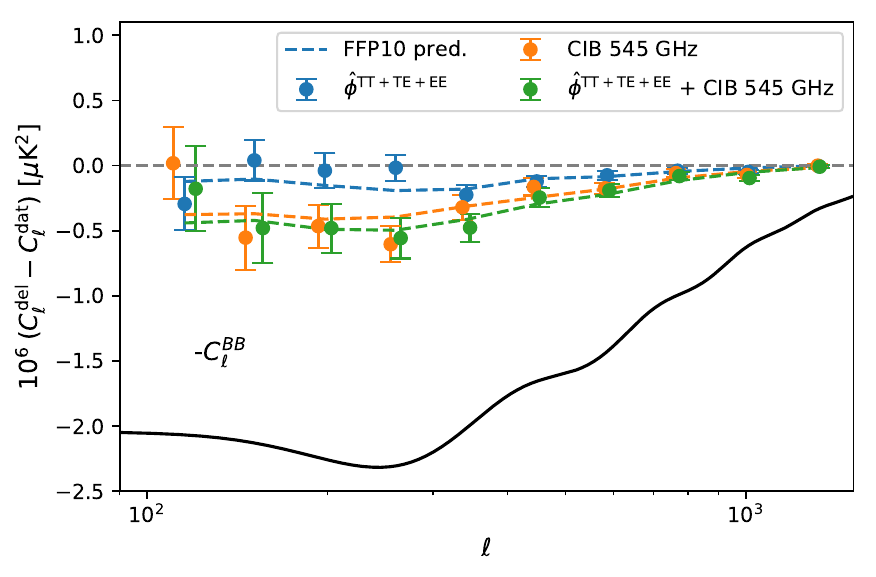}
\includegraphics[width = 0.5\textwidth]{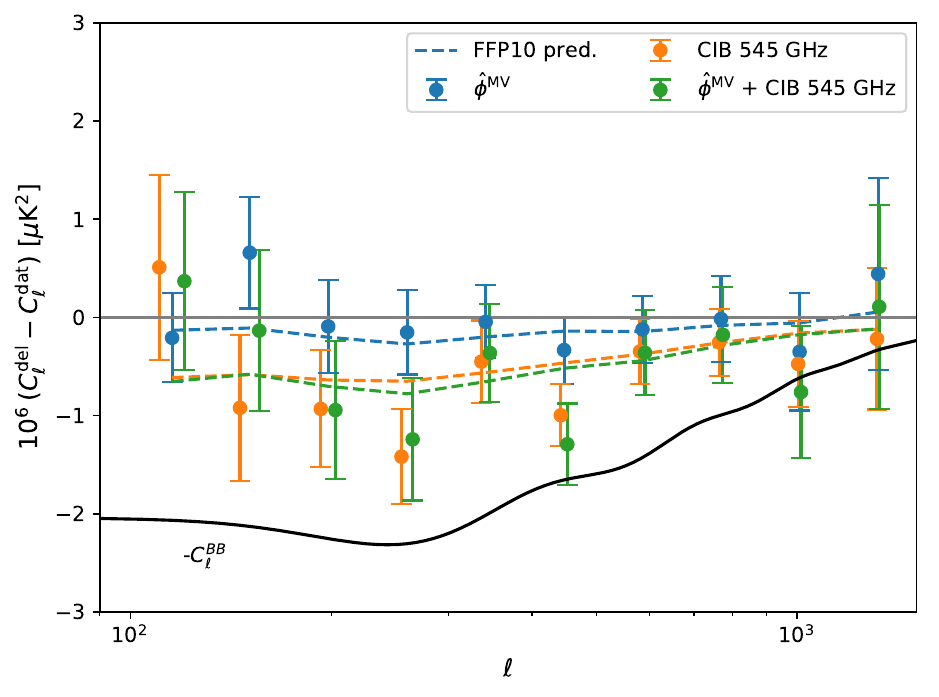}
\caption{Difference between the delensed and original $B$-mode power spectrum for the \smica\ CMB polarization maps. In the upper panel, the delensed $B$-mode map is obtained by template subtraction. The templates are constructed
using the labelled lensing tracers, and the Wiener-filtered $E$-mode map, as discussed in Sect.~\ref{sec:BBdelen}. The dashed lines show predictions obtained by repeating these operations on the FFP10 set of \planck\ simulations. The black curve, $-C_\ell^{BB}$ for our fiducial model, shows the difference expected for perfect delensing. Summary statistics are presented in Table~\ref{table:BBdelen}. In the lower panel, a remapping method is used, and the delensing signature is obtained after subtraction of biases, as described in Sect.~\ref{subsec:peakdelensing}, with summary statistics in Table~\ref{table:TTTEEEdelen}. In this case, the actual total $B$-mode power is not decreased after delensing.
\label{fig:delenBB}}
\end{figure}

To delens, the lensed $B$-mode template is simply subtracted from our $B$-mode map estimate, obtained as described above. Figure~\ref{fig:delenBB} shows the result of this procedure for different choices of tracer, after building the delensed difference spectrum that partly cancels cosmic variance and noise:
\begin{equation}
	\Delta \hat C_\ell \equiv \hat C_\ell^{\rm del} - \hat C_\ell^{\rm dat}.
\end{equation}
The spectra are obtained from the position-space $B$-mode maps before and after the $B$-mode template subtraction, using (scalar) deconvolution of the pseudo-$C_\ell$ spectra to account for the effect of the mask.
Table~\ref{table:BBdelen} lists the results of fitting a single delensing efficiency parameter $\Delta C_\ell^{BB} =-\epsilon C_\ell^{BB, \rm fid}$
across all the bins shown in Fig.~\ref{fig:delenBB}.
The efficiency defined in this way reaches 12\,\% for the combined tracers. Large-scale $B$ modes are better delensed, but with larger statistical errors. For instance, for $100 \leq \ell \leq 300$ we find a delensing efficiency of $0.217 \pm 0.047$ ($\hat{\phi}^{\rm{MV}}$+CIB at 545\,GHz), with a predicted value of $0.214$, in accordance with the naive expectations laid out at the beginning of this section based on the $E$ and $\phi$ map quality.

The presence of common $B$ modes in the quadratic estimator reconstruction noise
and the field being delensed
would create a spurious delensing signature coming from leading disconnected correlators \citep{Teng:2011xc,Carron:2017vfg}. As we did when estimating the lensing $B$-mode power spectrum above, we avoid this by using the quadratic estimator involving only $TT$+$TE$+$EE$, discarding the $EB$ and $TB$ parts. The lensing tracers used in Fig.~\ref{fig:delenBB} are then statistically independent from the $B$-mode map, at negligible cost in tracer fidelity.

\begin{table}[htbp!]
\caption{Reduction in the $BB$ power spectrum after lensing $B$-mode template subtraction, for combinations of different lensing tracers. The spectral differences $\Delta C^{BB}_\ell$ are calculated over 58\,\% of the sky, and a delensing efficiency $\epsilon$ is defined as a scaling of our fiducial lensing $B$-mode power, so that $\Delta C_\ell = -\epsilon C_\ell^{BB}$, computed over the $\ell$ range $100 \le \ell \le 2048$ (i.e., $\epsilon$ is the fraction by which the lensing $B$-mode power can be reduced). The right-hand column gives the expected efficiency, as estimated from FFP10 simulations. We neglect the $TB$ and $EB$ quadratic estimators in the default MV estimator when building the combined $TT$, $TE$, and $EE$ estimators used here; including them introduces large signals from disconnected correlators unrelated to actual delensing, for a negligible increase in delensing efficiency. The efficiency is scale-dependent, with a maximum close to 20\,\% on large scales. Also listed are reduced $\chi^2$ ($\chi^2_\nu$, for $\nu = 9$ degrees of freedom), for $\Delta \hat{C}_\ell$ compared to the best-fit model $-\epsilon C_\ell^{BB,{\rm fid}}$. This model does not capture very well the delensing results including the CIB. However, all results are fully consistent with the simulation predictions.}
\label{table:BBdelen}
\begingroup
\vskip -4mm
\newdimen\tblskip \tblskip=5pt
\setbox\tablebox=\vbox{
 \newdimen\digitwidth
 \setbox0=\hbox{\rm 0}
 \digitwidth=\wd0
 \catcode`*=\active
 \def*{\kern\digitwidth}
 \newdimen\signwidth
 \setbox0=\hbox{+}
 \signwidth=\wd0
 \catcode`!=\active
 \def!{\kern\signwidth}
\halign{\hbox to 1.35in{#\leaderfil}\tabskip 1em&
\hfil#\hfil& \hfil#\hfil\tabskip=0pt\cr
\noalign{\doubleline}
\omit \hfil $BB$-delensing template\hfil& Efficiency $\epsilon$, \Planck~data& FFP10 Pred.\cr
\noalign{\vskip 3pt\hrule\vskip 4pt}
$\hat{\phi}^{\rm{TT}}$& $0.048\pm0.008$ ($\chi_9^2 = 1.12$)& 0.036\cr
\noalign{\vskip 3pt}
$\hat{\phi}^{\rm{MV}}$& $0.051\pm0.009$ ($\chi_9^2 = 1.21$)& 0.044\cr
\noalign{\vskip 3pt}
CIB at 545& $0.098\pm0.013$ ($\chi_9^2 = 2.92$)& 0.106\cr
\noalign{\vskip 3pt}
$\hat{\phi}^{\rm{TT}}$ + CIB at 545& $0.122\pm0.014$ ($\chi_9^2 = 2.36$)& 0.120\cr
\quad $\Delta \epsilon$ CIB improv.& $0.024\pm0.007$& 0.014\cr
\noalign{\vskip 3pt}
$\hat{\phi}^{\rm{MV}}$ + CIB at 545& $0.122\pm0.014$ ($\chi_9^2 = 2.34$)& 0.125\cr
\quad $\Delta \epsilon$ CIB improv.& $0.024\pm0.008$ & 0.019\cr
\noalign{\vskip 3pt\hrule\vskip 3pt}}}
\endPlancktable
\endgroup
\end{table}

\subsubsection{Acoustic peak sharpening (de-smoothing)}\label{subsec:peakdelensing}
We now test the expected acoustic peak re-sharpening of the $T$ and $E$ power spectra after delensing. For the case of the quadratic estimator, internal delensing biases originate from the statistical dependence between the lensing reconstruction noise and the CMB maps. Here, the biases are more difficult to avoid than for $BB$-delensing in the previous section, and we refer the reader to \cite{Carron:2017vfg} for a detailed discussion of these biases. Proposed methods to correct for such biases include using a non-overlapping set of modes \citep{Sehgal:2016eag}, at the cost of some loss of S/N, or a much more elaborate realization-dependent higher-point function estimation \citep{Namikawa:2017iak}. We follow the procedure of \cite{Carron:2017vfg}, which we summarize below, to correct for these biases and evaluate our delensing efficiencies. This procedure has no impact on the covariance of the biased delensed spectra. This covariance has not been well studied, and this procedure will probably not be optimal for ambitious future experiments aiming to transfer information from the lensing higher-point statistics back to the CMB power spectrum, as advocated by \cite{Green:2016cjr}.

\begin{figure*}[t]
\vspace{0.5cm} 
\includegraphics[width = 1.0\textwidth]{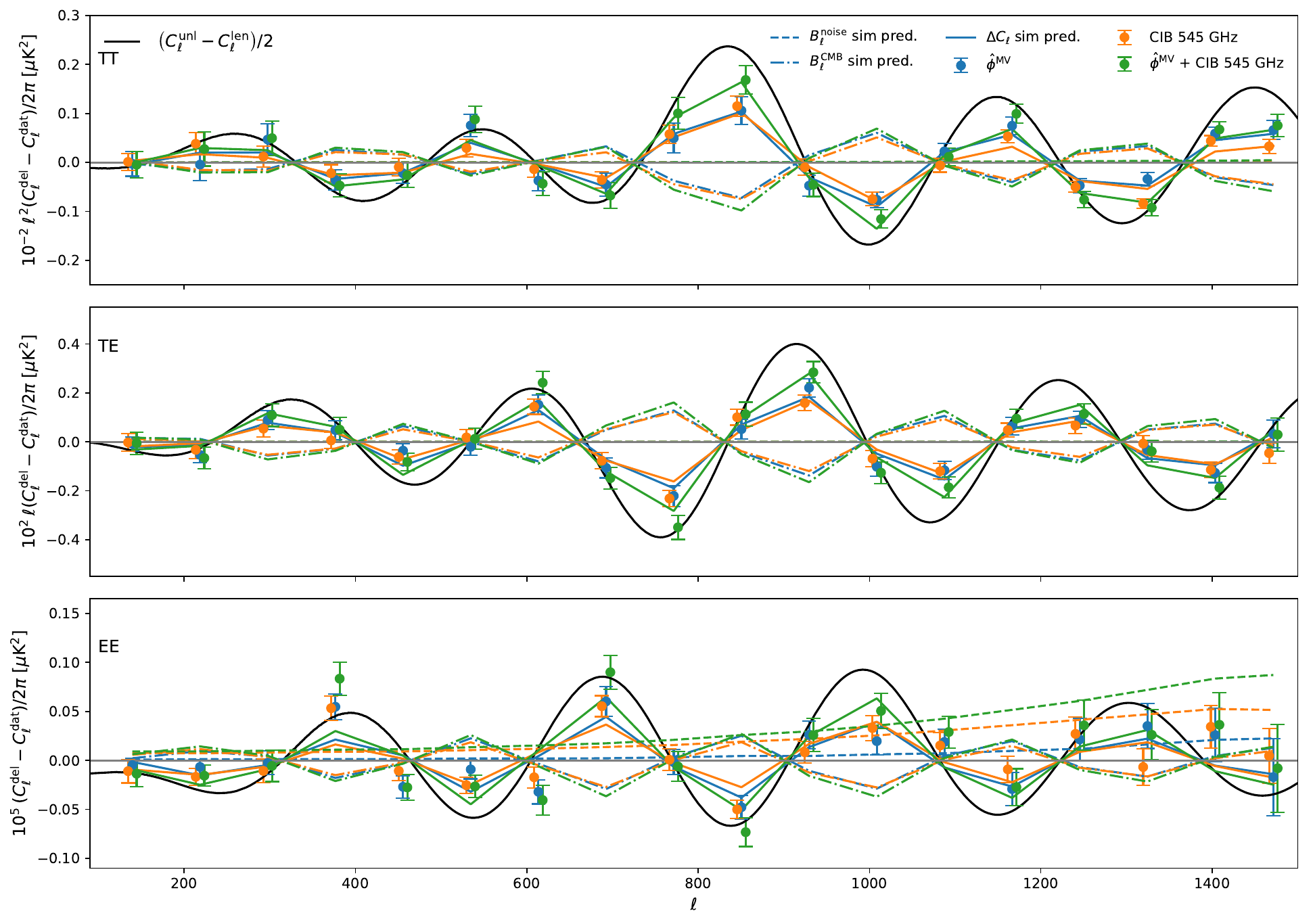}
\vspace{0.4cm}
\caption{\label{fig:delenTTTEEE}
Acoustic peak re-sharpening after delensing of the \smica~CMB maps. From top to bottom, the panels display the difference between the delensed and original $TT, TE$, and $EE$ spectra. Data points show delensing with our internal, minimum-variance, quadratic lensing estimate (blue), with the \GNILC~CIB map (orange), and with their optimal combination (green), as described in the text. For reference, the black line shows half the difference of the unlensed to lensed CMB spectra in the fiducial cosmology. In all cases, the solid coloured lines show the FFP10 simulation predictions, obtained after performing the same operations on each simulation as for the data. The dashed lines show the noise delensing bias defined by Eq.~\eqref{Bnoise} (only relevant here for $EE$ at high multipoles), while the dot-dashed lines show the CMB peak smoothing caused by the tracer reconstruction noise, Eq.~\eqref{BCMB}.}
\end{figure*}

The procedure is as follows: using our lensing tracers, we remap an estimate of the CMB temperature and polarization fields, filtered in the multipole range $100 \leq \ell \leq 2048$, according to the estimate of the inverse displacement $-\vgrad \hat \phi_{\rm WF}$. The CMB maps are built from the inverse-variance, homogeneously-filtered $T$, $E$, and $B$ CMB harmonic coefficients, rescaled by the isotropic limit of the filter (since we neglect the $TE$ correlation in the filtering, these limits are simply given by $C_\ell^{TT,EE,BB,\rm fid} + N_\ell^{TT,EE,BB}/b^2_\ell$); the resulting filtered maps are expected to be unbiased away from the mask boundaries. As before, we calculate differences of delensed and raw $T$, $E$, and $B$ spectra after deconvolution of their pseudo-$C_\ell$ power spectra for the effects of the mask.
	To assess the delensing effects and subtract the internal delensing biases, we subtract the same spectral differences obtained using the same delensing procedure, but on Gaussian simulations (generated using lensed CMB spectra) and uncorrelated CIB:
\begin{equation}
B_\ell^{\rm Gauss} \equiv \left\langle \hat C_\ell^{\rm del} - \hat C_\ell^{\rm dat} \right\rangle_{\rm Gauss}.
\end{equation}
   The combination $\hat{C}_\ell^{\rm del} - \hat{C}_\ell^{\rm dat} - B_\ell^{\rm Gauss}$ should vanish in the mean by construction in the absence of lensing, and we use these spectra to quantify the significance of the delensing that we detect. However, to assess how much delensing was really performed on the CMB, further corrections are required. First, $B_\ell^{\rm Gauss}$ subtracts not only the spurious delensing bias, but also the CMB peak \emph{smoothing} effect due to independent reconstruction noise of the tracer. Second, the spectral difference $\hat{C}_\ell^{\rm del} - \hat{C}_\ell^{\rm dat} - B_\ell^{\rm Gauss}$ still contains the difference of the noise spectra caused by the signal part of the tracer. To isolate the CMB peak-sharpening, we therefore build the combination
\begin{equation}
\Delta \hat C_{\ell, \rm{debias}} \equiv \hat C_\ell^{\rm del} - \hat C_\ell^{\rm dat} - B_\ell^{\rm Gauss} - B^{\rm Noise}_\ell + B^{\rm CMB}_\ell,
\label{eq:deltaClpeak}
\end{equation}
with
\begin{equation}
\label{Bnoise}
B^{\rm Noise}_\ell \equiv 	\left\langle \hat C_\ell^{\rm del} - \hat C_\ell^{\rm dat} \right\rangle_{\rm{signal} \:\hat \phi_{\rm WF}},
\end{equation}
and
\begin{equation}
\label{BCMB}
B^{\rm CMB}_\ell \equiv 	\left\langle \hat C_\ell^{\rm del} - \hat C_\ell^{\rm dat} \right\rangle_{\rm{independent} \:\hat \phi_{\rm WF}}.
\end{equation}
In Eq.~\eqref{Bnoise}, the lensing tracer only contains its signal part, and in Eq.~\eqref{BCMB} the lensing tracer is constructed using a Gaussian simulation independent of the one being delensed. Only the noise part of the simulation is considered in Eq.~\eqref{Bnoise}, and only the CMB part in Eq.~\eqref{BCMB}.

\begin{table*}[htbp!]
\caption{Delensing summary efficiencies using the remapping technique described in Sect.~\ref{subsec:peakdelensing}, calculated from $\chi^2$-minimization of Eq.~\eqref{eq:chi2del}.  The first three rows show results using as lensing tracer the temperature-only quadratic estimator ($\hat \phi^{\rm TT}$), the minimum-variance estimator combining temperature and polarization ($\hat \phi^{\rm MV}$), and the \GNILC~CIB map at 545\,GHz. We also show the expected efficiencies, as predicted from the FFP10 simulations, and the reduced $\chi^2$ ($\chi^2_\nu$, for $\nu = 17$ $(C_\ell^{TT}, C_\ell^{TE}, C_\ell^{EE})$ and $9$ $(C_\ell^{BB})$ degrees of freedom) for the best-fit efficiencies estimated from the delensed spectra. The next two rows show the results obtained combining these tracers, and the significance of the improvement found over the most efficient single tracer in the combination. This delensing of $C_\ell^{BB}$ (last column) does not actually reduce the total $B$-mode power (in contrast to the template delensing of Table~\ref{table:BBdelen}), but isolates the delensing signature using subtraction of the fiducial biases explained in the text.}
\label{table:TTTEEEdelen}
\begingroup
\vskip -4mm
\newdimen\tblskip \tblskip=5pt
\setbox\tablebox=\vbox{
 \newdimen\digitwidth
 \setbox0=\hbox{\rm 0}
 \digitwidth=\wd0
 \catcode`*=\active
 \def*{\kern\digitwidth}
 \newdimen\signwidth
 \setbox0=\hbox{+}
 \signwidth=\wd0
 \catcode`!=\active
 \def!{\kern\signwidth}
\halign{\hbox to 1.45in{#\leaderfil}\tabskip 0.5em&
\hfil#\hfil\tabskip 1em& \hfil#\hfil& \hfil#\hfil& \hfil#\hfil\tabskip=0pt\cr
\noalign{\doubleline}
\omit& $C_\ell^{TT}$& $C_\ell^{TE}$& $C_\ell^{EE}$& $C_\ell^{BB}$\cr
\noalign{\vskip 3pt\hrule\vskip 4pt}
$\hat{\phi}^{\rm{TT}}$&0.208 $\pm$ 0.022 ($\chi_{17}^2 = 0.90$)&0.212 $\pm$ 0.021 ($\chi_{17}^2 = 1.26$)&0.191 $\pm$ 0.032 ($\chi_{17}^2 = 2.17$)&0.048 $\pm$ 0.043 ($\chi_{9}^2 = 1.75$)\cr
\quad Prediction & 0.207& 0.222& 0.211& 0.080\cr
\noalign{\vskip 3pt}
$\hat{\phi}^{\rm{MV}}$&0.277 $\pm$ 0.025 ($\chi_{17}^2 = 0.98$)&0.276 $\pm$ 0.025 ($\chi_{17}^2 = 0.81$)&0.236 $\pm$ 0.038 ($\chi_{17}^2 = 2.32$)&0.042 $\pm$ 0.050 ($\chi_{9}^2 = 1.01$)\cr
\quad Prediction&0.267 & 0.271& 0.265& 0.091\cr
\noalign{\vskip 3pt}
\GNILC\ at 545 GHz&0.265 $\pm$ 0.019 ($\chi_{17}^2 = 1.74$)&0.247 $\pm$ 0.021 ($\chi_{17}^2 = 1.20$)&0.277 $\pm$ 0.033 ($\chi_{17}^2 = 1.89$)&0.385 $\pm$ 0.069 ($\chi_{9}^2 = 0.99$)\cr
\quad Prediction& 0.210& 0.228& 0.231& 0.275\cr
\noalign{\vskip 3pt}
$\hat{\phi}^{\rm{TT}}$ + \GNILC\ at 545 GHz &0.368 $\pm$ 0.026 ($\chi_{17}^2 = 1.05$)&0.371 $\pm$ 0.027 ($\chi_{17}^2 = 1.21$)&0.392 $\pm$ 0.042 ($\chi_{17}^2 = 2.52$)&0.392 $\pm$ 0.078 ($\chi_{9}^2 = 1.75$)\cr
\quad Prediction& 0.332& 0.372& 0.367& 0.313\cr
\quad Impr. $\Delta \epsilon$ (and pred.) & $0.103\pm0.019$ (0.121)& $0.124\pm0.019$ (0.144)& $0.115\pm0.027$ (0.136)& $0.008\pm0.032$ (0.038)\cr
\noalign{\vskip 3pt}
$\hat{\phi}^{\rm{MV}}$ + \GNILC\ at 545 GHz &0.411 $\pm$ 0.028 ($\chi_{17}^2 = 0.99$)&0.407 $\pm$ 0.030 ($\chi_{17}^2 = 1.04$)&0.402 $\pm$ 0.047 ($\chi_{17}^2 = 2.56$)&0.383 $\pm$ 0.080 ($\chi_{9}^2 = 1.26$)\cr
\quad Prediction& 0.375& 0.396& 0.398& 0.314\cr
\quad Impr. $\Delta \epsilon$ (and pred.)& $0.134\pm0.014$ (0.108)& $0.130\pm0.018$ (0.126)& $0.124\pm0.034$ (0.167)& $-0.002\pm0.038$ (0.039)\cr
\noalign{\vskip 3pt\hrule\vskip 3pt}}}
\endPlancktablewide
\endgroup
\end{table*}

Figure~\ref{fig:delenTTTEEE} shows the resulting spectral differences and Table~\ref{table:TTTEEEdelen} collects summary statistics. We list delensing efficiencies $\epsilon$, obtained as a straightforward $\chi^2$-minimization of the observed binned spectral differences
\begin{equation}\label{eq:chi2del}
	\chi^2(\epsilon) = \sum_{\rm{bins}\: b}\frac 1 {\sigma^2_{\ell_b}} \left[\Delta \hat C_{\ell_b, \rm{debias}} - \epsilon \left(C_{\ell_b}^{\rm unl, fid} -C_{\ell_b}^{\rm len, fid} \right)\right]^2,
\end{equation}
together with predictions obtained from our simulations and the best-fit reduced $\chi^2$.
Here, $C_{\ell_b}^{\rm unl, fid} -C_{\ell_b}^{\rm len, fid}$ is the binned difference of the unlensed and lensed power spectra in the fiducial model, and $\sigma_{\ell_b}^2$ is the variance of the binned $\Delta \hat C_{{\ell_b}, \rm{debias}}$ estimated from simulations.
We detect delensing at high significance in the difference between the delensed and raw spectra, which
removes most of the cosmic variance and noise common to both. When combining the tracers we also list the improvement achieved (defined with respect to the most powerful single tracer). The improvement is always significant except for the $B$-mode power, where the quadratic estimator provides only a little additional information on the required scales.

The last column of Table~\ref{table:TTTEEEdelen} and the lower panel of Fig.~\ref{fig:delenBB} show the delensing of the $B$-mode power with the remapping method. The procedure we follow in this section, using Eq.~\eqref{eq:deltaClpeak}, is designed isolate the change in the CMB \emph{signal} power spectrum due to the delensing remapping. This is not, by construction, the same as the actual change in the power spectrum of the maps: for example, for $B$-mode delensing, remapping by the reconstruction noise \emph{produces} $B$ modes, and remapping the instrument noise also changes the $B$-mode power. The delensing efficiency figures for $BB$ that are quoted in Table~\ref{table:TTTEEEdelen} are therefore not directly comparable to those reported in Table~\ref{table:BBdelen} for the template-delensing method, which are based on the actual change in $B$-mode power after delensing. The template method is also optimized to account for instrument noise (through the Wiener-filtering of the $E$-mode map in the construction of the template). The template-based efficiencies are, therefore, the relevant ones for assessing improvements in primordial $B$-mode limits from delensing. However, since the \planck\ $B$-mode measurement is noise dominated, there would be a negligible improvement in primordial $B$-mode limits due to the small decrease in the lensing component.


The peak sharpening in the other CMB spectra could in principle slightly improve parameter constraints~\citep{Green:2016cjr}; however, the likelihood model is complicated and the forecast improvement for \planck\ is very small, so we do not use the delensed spectra for cosmological parameter analyses. Our demonstration of peak sharpening is an important proof of principle, and also a useful consistency check on the lensing analysis.

\section{Null and consistency tests}\label{sec:tests}
\AL{At some point should comment on/test T to P leakage and polar efficiencies}

We now turn to various consistency tests of the lensing and lensing curl reconstruction band powers.
In Sect.~\ref{subsec:testBPpdf} we first discuss the empirical distribution of the band powers in simulations and the presence of features. In Sect.~\ref{subsec:matrices} we focus on the temperature reconstruction, and present detailed consistency tests based on comparisons of different reconstructions; Sect.~\ref{subsec:polestimators} then compares results from various polarization estimators with the temperature reconstruction. In Sect.~\ref{subsec:noisetests} we present cross-half-mission data splits and noise null tests. In Sect.~\ref{subsec:testforegroundscorr} we test for the impact of correlations between the foregrounds and lensing signal. Our entire analysis uses the FFP10 cosmology as the fiducial model; we test in Sect.~\ref{subsec:testfid} that this does not affect our results. Section~\ref{subsec:testnl} discusses the impact of the non-Gaussianity of the lensing field, and Sect.~\ref{subsec:N1tests} presents alternative calculations of the $N^{(1)}$ bias.
\subsection{Band-power distribution and features}\label{subsec:testBPpdf}
\begin{figure}[htp]
\includegraphics[width = \columnwidth]{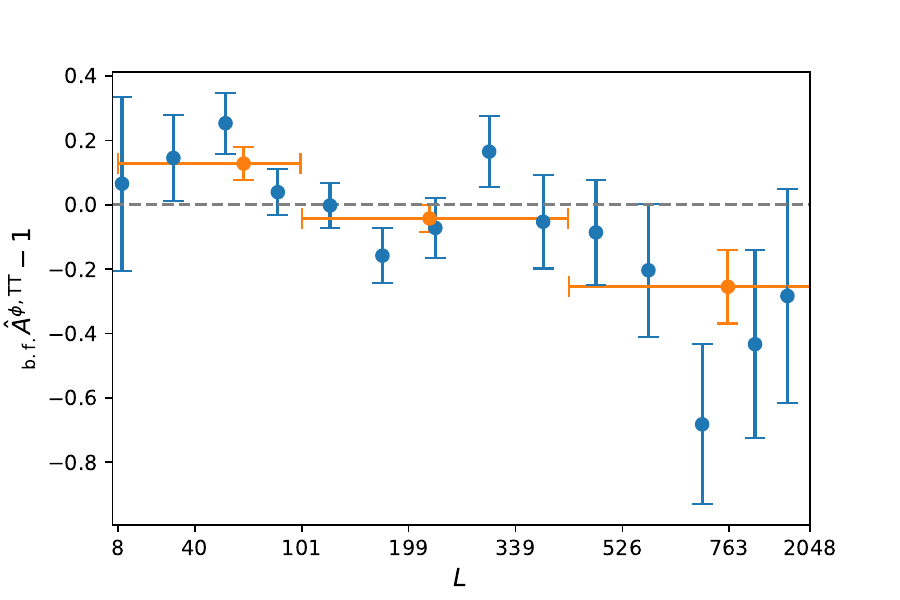}
\includegraphics[width = \columnwidth]{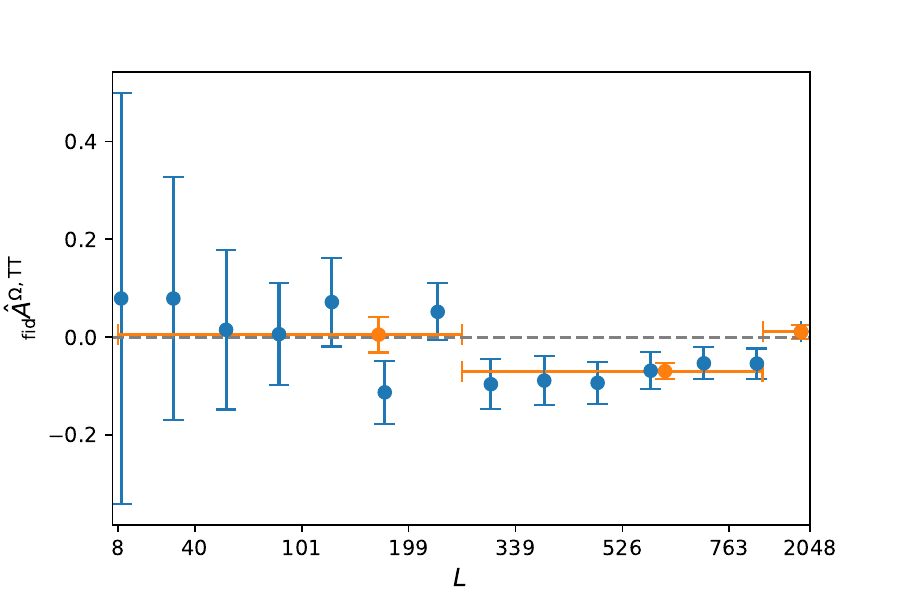}
\caption{\textit{Top}: Fractional differences between the lensing power spectrum band powers from the $\phi^{\rm TT}$ reconstruction over the aggressive multipole range and the \planck\ 2018 best-fit cosmology using \planckalllensing\ (blue points). \textit{Bottom}: Temperature-only curl reconstruction band-power amplitudes. In both cases, the orange points show the same data with a coarser binning, where the construction of the binning function and mid-points follows the same procedure as the blue points (see Sect.~\ref{subsec:reconstruction})\AL{explain why off centre to right?}. The curl bin at $264 \leq L \leq 901 $ deviates from zero with a formal significance of 4.3$\,\sigma$. This bin was chosen to maximize the deviation; after simple consideration of ``look-elsewhere'' effects as described in the text, we evaluate the PTE of that deviation to be $0.41\,\%$ (or roughly $2.9$ Gaussian $\sigma$), still suggestive of a problem in the curl reconstruction.
 \label{fig:features}
 }
\end{figure}

We first discuss the empirical statistical behaviour of our baseline data band powers compared to the result expected from FFP10 fiducial simulations. \PlanckLensTwo~already hinted at two possible features in the temperature-only lensing reconstruction (a sharp dip in the lensing spectrum at $L \simeq 700$, and a low curl spectrum over a broad range of scales). These two features are still present, as displayed in Fig.~\ref{fig:features}. The upper panel shows the relative deviation of $\hat{C}_L^{\phi\phi}$ to the \planckalllensing\ best-fit cosmology and the lower panel the curl spectrum, both on the aggressive multipole range.

The dip in the measured $\hat{C}_L^{\phi\phi}$ band powers
in the upper panel of Fig.~\ref{fig:features} is slightly less pronounced than in the previous release; the bin covering $638 \leq L \leq 762$ shows now a $2.8\,\sigma$ deviation from the best-fit spectrum, compared to $3.8\,\sigma$ in 2015 for the same best-fit spectrum. This significance is unchanged after marginalization of the CMB spectra uncertainties. Despite the large errors at high-$L$, the measurement can affect results if these multipoles are used for parameter inference. Can the dip be a statistical fluke? Empirically, we find that $5\,\%$ of our simulated temperature reconstructions show equal or stronger outliers (the most extreme $TT$ outlier across all aggressive bins and simulations is $-4\,\sigma$ at $L\simeq 50$).
The feature in the data is therefore consistent with a fluctuation that is only slightly unusual. A statistical fluctuation
is also consistent with Fig.~\ref{fig:qestxCIB}, which shows that the cross-spectrum with the \GNILC\ CIB tracer has no significant dip at these multipoles.

The lower panel of Fig.~\ref{fig:features} shows our lensing curl reconstruction, which is expected to be zero to within tiny post-Born corrections that can be safely neglected at \planck\ noise levels~\citep{Pratten:2016dsm, Fabbian:2017wfp}. Our curl amplitudes $\fidprefix\hat A^{\Omega,{\rm TT}}$ are built using the same binning procedure as for the lensing gradient in Eq.~\eqref{Binning}, using a reference flat spectrum
\begin{equation}
 \frac{L^2(L + 1)^2}{2\pi}\: C_L^{\Omega\Omega, \rm fid} \equiv 10^{-7}.
\label{eq:Clcurlfid}
  \end{equation}
Orange points use a coarser binning and emphasise the low curl found over a fairly wide range of scales at $300 \le L \le 900$, very similar to the result found in \PlanckLensTwo.  None of our original aggressive band powers (blue) are anomalous individually, but the resulting coarse central bin lies formally $4.2\,\sigma $ away from zero. This binning was custom-made in order to emphasize the feature, and any assessment of its true significance must take this into account. We have done so as follows: for each of our FFP10 simulations, we identified within the aggressive binning the longest sequence of adjacent positive or negative curl amplitudes, and inverse-variance weighted these points to produce the largest outlier possible.\footnote{Due to the lack of a sufficiently precise realization-dependent $\hat N^{(0)}$ de-biaser for all simulations, for this analysis we have used the covariance matrix built with a realization-independent Monte Carlo $N^{(0)}$ (defined similarly to the gradient-mode $\MCNzero$ in Appendix~\ref{app:biases}) consistently on simulations and data.} While there are plenty of longer sequences of consistently low or high amplitudes, only one of the 240 simulations shows a larger combined deviation. Hence we may assign an approximate revised PTE of $1/240 \simeq 0.4\,\%$ to the feature. This PTE roughly matches the analytic prediction from Gaussian statistics for our band powers (0.6\,\%). This is still very suggestive of the presence of an uncontrolled component affecting our curl band powers;
the following section presents additional curl consistency checks in more detail.

\subsection{Temperature lensing and lensing curl consistency and stability tests}\label{subsec:matrices}

\newcommand{\SMICA}[0]{\textbf{SMICA}}
\newcommand{\SMICAeighty}[0]{\textbf{f80}}
\newcommand{\SMICANOSZ}[0]{\textbf{SMICA noSZ}}
\newcommand{\SMICANOSZmask}[0]{\textbf{SZ-unm.}}
\newcommand{\SMICASZcons}[0]{\textbf{SZ-cons.}}
\newcommand{\NILC}[0]{\textbf{NILC}}
\newcommand{\SEVEM}[0]{\textbf{SEVEM}}
\newcommand{\COMMANDER}[0]{\textbf{COMMANDER}}
\newcommand{\ecleq}[0]{\textbf{ecl.eq.}}
\newcommand{\eclpol}[0]{\textbf{ecl.pol.}}
\newcommand{\SWNE}[0]{\textbf{SWNE}}
\newcommand{\SENW}[0]{\textbf{SENW}}
\newcommand{\onefourthree}[0]{\textbf{143 GHz}}
\newcommand{\twooneseven}[0]{\textbf{217 GHz}}
\newcommand{\onefourthreeg}[0]{\textbf{143 GHz du44}}
\newcommand{\twooneseveng}[0]{\textbf{217 GHz du44}}
\newcommand{\onefourthreeszcons}[0]{\textbf{143~GHz~SZ-cons.}}
\newcommand{\onefourthreeszunm}[0]{\textbf{143~GHz~SZ-unm.}}
\newcommand{\gfiveseven}[0]{\textbf{du57}}
\newcommand{\gfourfour}[0]{\textbf{du44}}
\newcommand{\ellmax}[1]{$\bf{\boldsymbol{\ell_{\rm max}} = #1}$}
\newcommand{\ellmin}[1]{$\bf{\boldsymbol{\ell_{\rm min}} = #1}$}
\newcommand{\MSC}{\textbf{PCL}}
\newcommand{\ftwofive}{\textbf{W2500}}
\newcommand{\SMICAbhs}{\textbf{S-hard.}}
\newcommand{\SMICAbhd}{\textbf{D-hard.}}
\newcommand{\SMICAbhf}{\textbf{M-hard.}}
\newcommand{\SMICAbhn}{\textbf{N-hard.}}
\newcommand{\twoonesevenbhd}{\textbf{217 GHz D-hard.}}

\begin{figure*}
\includegraphics[width=\textwidth]{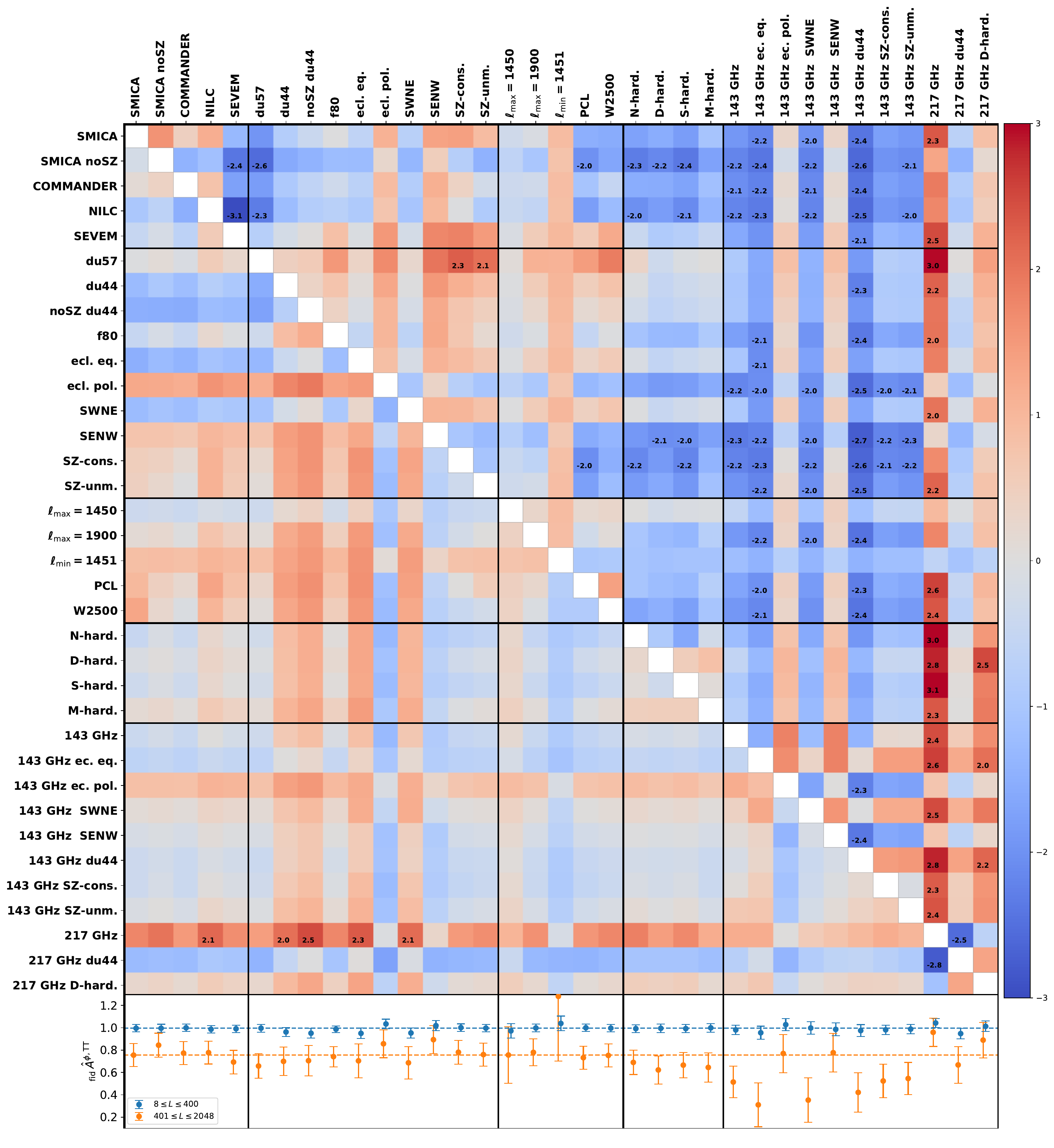}\\
\caption{\label{fig:diff_ptt_cons} Summary of lensing-amplitude-measurement consistency tests using temperature-only reconstructions. The upper panel shows the difference between the measured lensing amplitude from pairs of reconstructions in units of the standard deviation of the difference from simulations. This tests whether the observed shifts are compatible with the behaviour expected from the FFP10 simulations. The lower triangle shows the amplitude measured from the conservative range $8 \leq L \leq 400$, with a red colour indicating a larger amplitude of the reconstruction labelled on the left; the upper triangle uses the high-$L$ range ($401 \leq L \leq 2048$), with red indicating a larger amplitude of the reconstruction labelled at the top. Numerical significances are quoted for those tests that show differences greater than twice the expected standard deviation.
The first entry is our baseline temperature-only reconstruction; see the text for a description of all the other reconstructions. The lower panel shows the actual amplitudes (relative to our FFP10 fiducial model), with the result using the conservative range in blue and the high-$L$ range in orange. The dashed lines show for reference the values in our baseline reconstruction.
}
\end{figure*}

\begin{figure*}
\includegraphics[width = \textwidth]{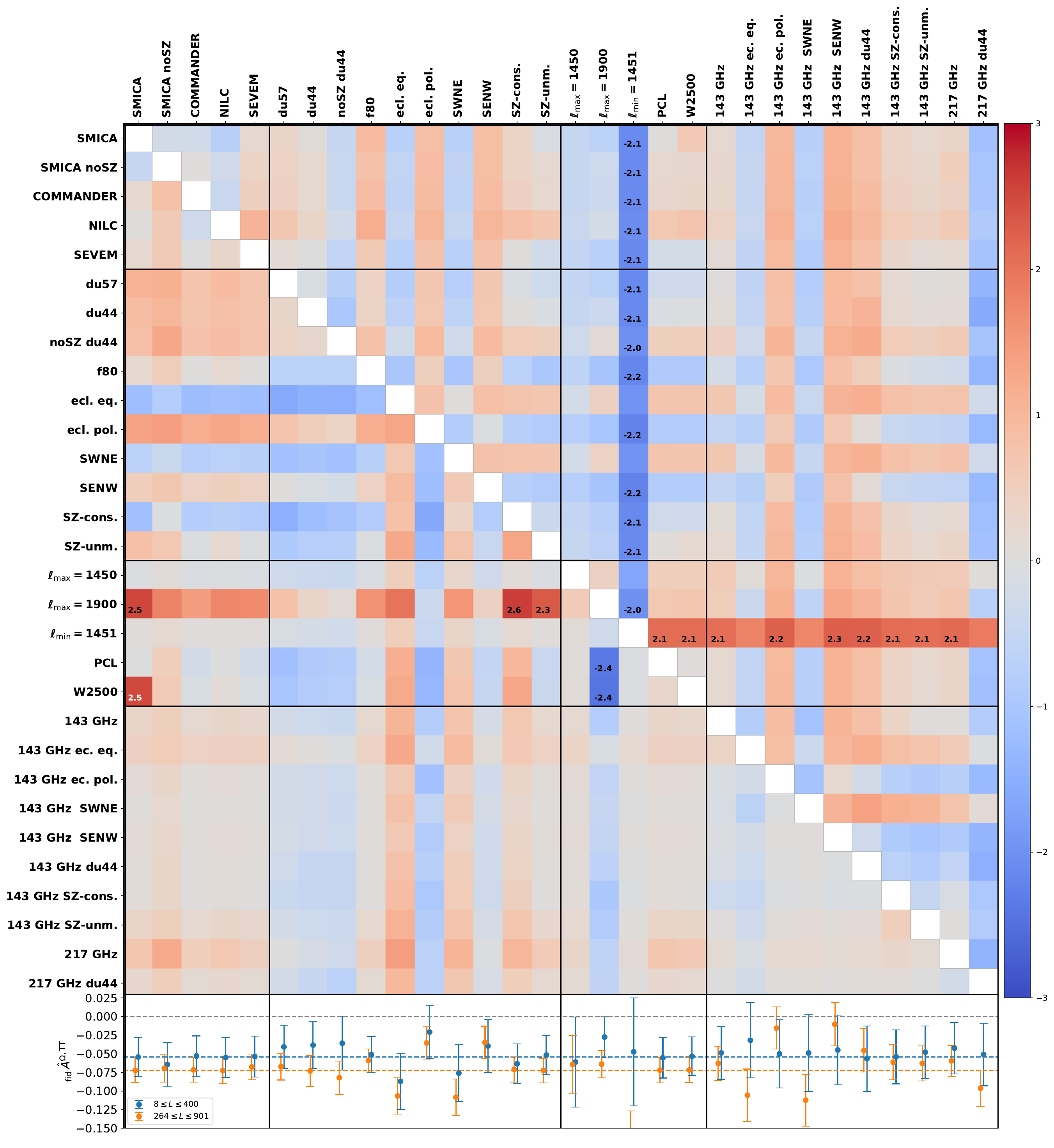}
\caption{\label{fig:diff_xtt_cons} Same as Fig.~\ref{fig:diff_ptt_cons}, but for the curl reconstruction.
Here the lower panel shows the curl amplitude over the conservative multipole range ($8 \le L \le 400$; blue) and over the range covering the negative curl feature shown in Fig.~\ref{fig:features} ($264 \leq L \leq 901$; orange). We use the same binning scheme as for the lensing gradient (Eq.~\ref{Binning}), with an amplitude of unity corresponding to a flat $L^2(L + 1)^2 /2\pi\: C_L^{\Omega\Omega, \rm fid} = 10^{-7}$ curl spectrum.}
\end{figure*}

We now turn to how the band powers differ between various reconstructions. We focus this subsection on temperature-only reconstructions, which dominate the signal and show stronger features than the MV estimator. \JCrev{For each one of these reconstructions, the methodology follows that described in Sect.~\ref{sec:pipe}, and uses the same set of 300 FFP10 simulations. All lensing biases, the point-source and Monte-Carlo corrections are recalculated consistently.}

 Figure~\ref{fig:diff_ptt_cons} presents a visual summary of our consistency tests based on gradient reconstructions, and Fig.~\ref{fig:diff_xtt_cons} for tests based on the curl. We now describe these in detail. For each pair of reconstructions, as listed below, we use the band powers to fit an amplitude relative to the fiducial model  spectrum (or the flat spectrum of Eq.~\ref{eq:Clcurlfid} for the curl),
and build the difference of the amplitudes between the two reconstruction, both on data and FFP10 simulations. In the simulations, this difference is very well described by a Gaussian centred on zero.
   The matrices of differences shown in Figs.~\ref{fig:diff_ptt_cons} and~\ref{fig:diff_xtt_cons} are colour-coded by the value of the difference in the data in units of the standard deviation of the difference from simulations.
The lower triangle shows the results over our conservative range $8 \le L \le 400$, and the upper triangle the high-$L$ range $401 \le L \le 2048$. Reading the lower triangle from the left, the reconstruction labelled on the left-hand side has an inferred lensing amplitude larger than the one labelled on the top if it has a red colour, and a smaller amplitude if it has a blue colour. Likewise, reading the upper triangle from the top, the reconstruction labelled on the top is larger for a red colour. If the amplitude shift is larger than twice the expected standard deviation, the precise value is indicated, normally in black. A small number of cells show the deviation written in white. This indicates that even though the deviation is anomalously large, it is still small in absolute terms (less than one fifth of the standard deviation of the left or upper reconstruction amplitude, whichever is larger), hence not relevant for practical purposes. The black solid lines separate for clarity the following classes of test (boldface refers to the labels used to identify the various reconstructions in Fig.~\ref{fig:diff_ptt_cons}).

\begin{figure*}
\includegraphics[width = \textwidth]{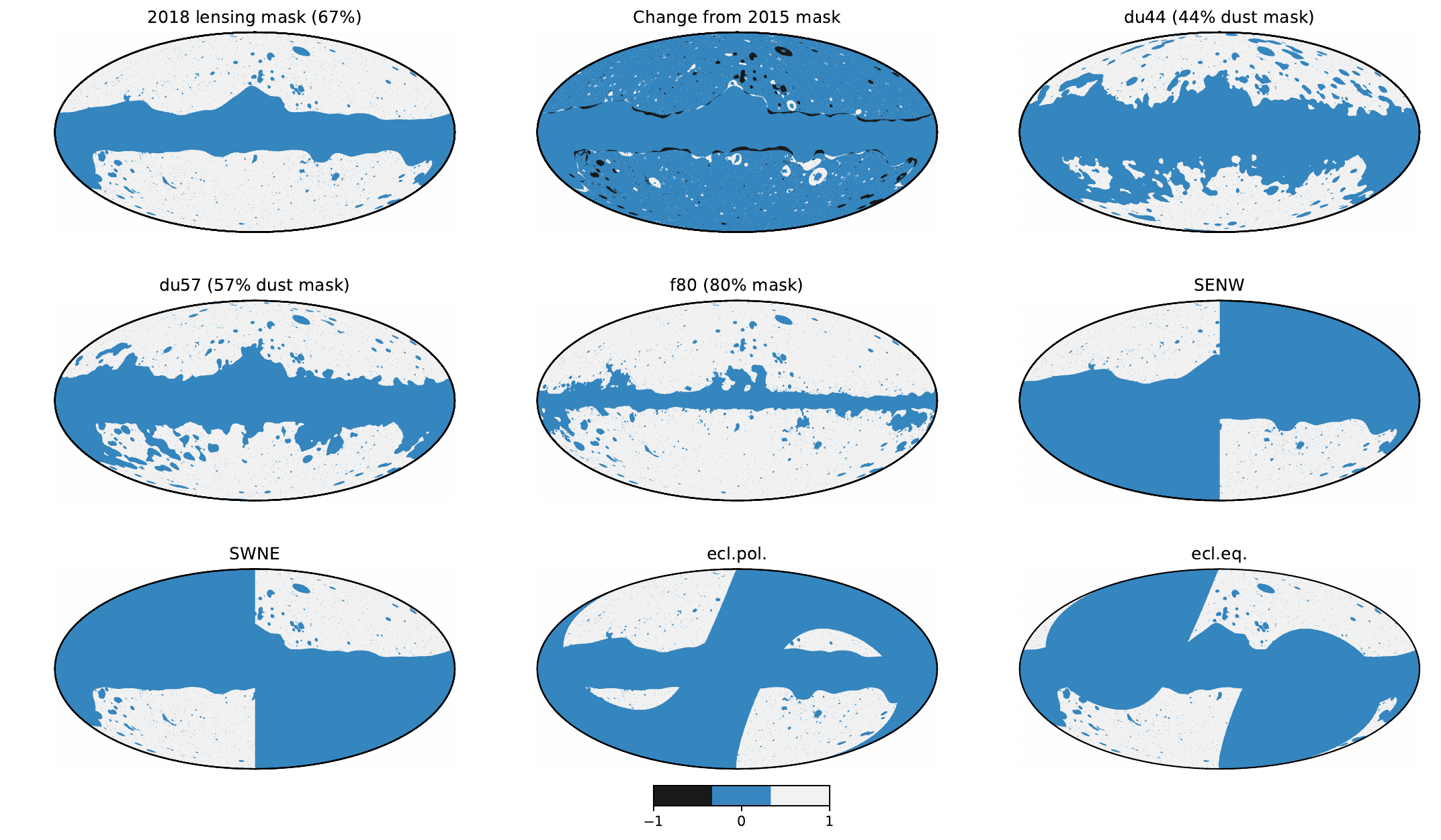}
\caption{Various masks used for the lensing analysis and consistency tests.
The top-left mask is the default lensing mask,
including cutting out point sources, resolved SZ clusters, and the Galaxy; the top-middle mask shows the difference between the new mask and the one used for the 2015 analysis. The remaining masks are used for consistency tests, as described in Sect.~\ref{subsec:matrices}. The dust masks are constructed by thresholding the smoothed pixel variance obtained from the product of the two half-mission 545-GHz \planck\ maps after filtering to include only multipoles $1000\le \ell \le 2000$.
\label{fig:masks}}
\end{figure*}

   \begin{unindentedlist}
\item \textit{Component-separation methods}. In addition to using our baseline \SMICA~map, we perform reconstructions on the CMB maps from the three other component-separation methods, \NILC, \SEVEM, and \COMMANDER, as described in \cite{planck2016-l04}, and the version of the \smica\ CMB map that deprojects the thermal SZ signal (\SMICANOSZ). In each case, the lensing reconstruction is performed on the union of the temperature and polarization confidence masks recommended by \cite{planck2016-l04}, together with the same $f_{\rm sky} = 70\,\%$ Galactic mask, CO mask, point-source mask, and SZ-cluster-targeted mask we use for our baseline reconstruction. 	
\item \textit{Sky cuts}. We test a series of sky cuts, with corresponding masks shown in Fig.~\ref{fig:masks}. Conservative Galactic dust masks with $f_{\rm sky} = 57\,\%$ (\gfiveseven) and $f_{\rm sky} = 44\,\%$ (\gfourfour) test for Galactic foreground contamination. A reconstruction using a more aggressive galactic mask with $f_{\rm sky} = 80\,\%$ is also shown (\SMICAeighty). Masking to keep only the ecliptic poles (\eclpol) or the ecliptic equator (\ecleq) tests effects related to the scanning. Masking, in Galactic coordinates, to keep only the south-east and north-west quadrants (\SENW), or the south-west and north-east quadrants (\SWNE), is relevant for effects related to {\tt HEALPix}~\citep{Gorski:2004by} pixelization. We have used further two additional masks (not shown in Fig.~\ref{fig:masks}), testing for contamination from SZ clusters. The \SMICANOSZmask\ reconstruction uses our baseline mask but without masking SZ clusters, and \SMICASZcons\ includes a more conservative SZ-cluster mask. This conservative cluster mask was constructed by extending our SZ-cluster mask to at least three times the virial radius $(\theta_{500})$ of each cluster, and enlarging the cluster catalogue: instead of using only the internally-detected \Planck\ SZ Catalogue, we added X-ray clusters listed in the Meta-Catalogue of X-ray detected Clusters of galaxies~(MCXC;~\citealt{Piffaretti:2010my}).  This contains additional objects that are not individually detected within \Planck\ data but whose SZ decrement is clearly visible after stacking. \JC{really ? did check against sz but not sz-cons}
\item \textit{CMB multipole cuts and other analysis choices}.
   	 Most of the signal in the lensing reconstruction comes from CMB multipoles centred around $\ell \simeq 1450$, with the Gaussian reconstruction noise \Nzero\ over the range $8 \le L \le 400$ taking roughly equal contributions from either side of $\ell = 1450$. The
   	 \ellmax{1450} reconstruction only includes CMB multipoles $100 \le \ell \le 1450$, and \ellmin{1451} only uses $1451 \le \ell \le 2048$. Further, \ellmax{1900} excludes the contribution from the smallest scales by considering $100 \le \ell \le 1900$. We also test consistency between two variations of our pipeline. In our baseline analysis, power spectra are evaluated from harmonic coefficients (either of the Wiener-filtered CMB maps, or the lensing potentials) using the simple estimator in Eq.~\eqref{eq:naive_phispec}. With the \MSC\ variation, we first remap the harmonic coefficients of the convergence field estimate $\hat{\kappa}$ to real space, compute
pseudo-$C_\ell$ power spectra after further masking with a 20\arcm\ apodized mask, and deconvolve the effects of the apodized mask to obtain our final power spectrum estimates.
Finally, with the \ftwofive\ reconstruction we test details of our filtering code, by reconstructing the Wiener-filtered map over a larger range of scales ($2 \leq \ell \leq 2500$), and cutting the map to our baseline range ($100 \leq \ell \leq 2048$) afterwards. This tests for leakage very close to the boundary ($\ell = 2048$ in our baseline analysis).
   \item \textit{Bias-hardened estimators}. This block tests a series of temperature-only lensing estimators orthogonalized to potential contaminants, called ``bias-hardened'' estimators, following~\citet{Namikawa:2012pe}.
  	   Sources of statistical anisotropy in the data that are not properly accounted for in the simulations can contaminate the lensing signal. Any such source, $s$, with a simple parametrization of its impact on the CMB covariance can be assigned an optimal quadratic estimator $\hat{g}^s$. We can then define a new lensing estimate (before mean-field subtraction and normalization)
  	   \begin{equation}
  	   	\hat{g}^{\phi-s}_{LM} \equiv \hat g^{\phi}_{LM} - \frac{\mathcal R^{\phi s}_{L}}{\mathcal R_L^{ss}} \hat g_{LM}^{s},
  	   \end{equation}
  	   which has vanishing response to $s$, at the cost of an increase in variance. Here, the response functions $\mathcal{R}_L^{\phi s}$ and $\mathcal{R}_L^{s s}$
describe the responses to $s$ of the lensing estimator $\hat{g}^\phi$ and source estimator $\hat{g}^s$, respectively.
We give results bias hardened against point-source ($S^2$) contamination (\citealt{Osborne:2013nna}; \SMICAbhs), \JCrev{where a spatially-varying point source signal $S^2(\hn)$ is sought in the CMB data covariance, $\av{T^{\rm dat}(\hn_i)T^{\rm dat}(\hn_j)} \ni \int d^2\hn \:S^2(\hn)  \mathcal B(\hn_i, \hn) \mathcal B^*(\hn_j,\hn) $. The corresponding unnormalized optimal estimator is }
\begin{equation}
  	   \hat g^{S^2}(\hn) \equiv \bar T ^2(\hn).
  	   \end{equation}\JCrev{The exact form  of this estimator (and of the others in this section) can be obtained easily by calculating the gradient of the CMB log-likelihood
with respect to the anisotropy source.}
We also harden against anisotropy in the variance of the instrumental noise (\SMICAbhn). \JCrev{In this case the anisotropy is parametrized as $\av{T^{\rm dat}(\hn_i)T^{\rm dat}(\hn_j)} \ni \sigma^2(\hn_i)\delta_{\hn_i \hn_j}$, with quadratic esimator}
  	   \begin{equation}
  	   \hat g^{\sigma^2}(\hn) \equiv \left(\sum_{\ell m} \frac{\bar T_{\ell m}}{b_\ell} \:_{0}Y_{\ell m} (\hn)\right)^2,
  	   \end{equation}
  	   \JCrev{where $b_\ell$ is an effective, isotropic approximation to the beam and pixel transfer function.}
  	   We further tested hardening against spatial modulation of the CMB signal by a field $m(\hn)$ (for example as a check against an effective spatial variation in calibration), which we denote by \SMICAbhf. \JCrev{The anisotropy is of the form $T^{\rm dat}(\hn_i) \ni \int d^2\hn \:\mathcal B(\hn_i, \hn) T(\hn)\:[1+ m(\hn)]$, with resulting estimator}
  	   \begin{equation}
  	   \hat g^{m}(\hn) \equiv \bar T(\hn) T^{\rm WF}(\hn).
  	   \end{equation}
  	   We have additionally attempted hardening against a residual dust amplitude (\SMICAbhd). The motivation in this case is to look for a dust component consisting of a statistically-isotropic field $d(\hn)$ with spectral shape $\mathcal D^{dd}_\ell \equiv \ell(\ell+1)C_\ell^{dd}/(2\pi)$ modulated by a varying amplitude $A_d(\hn)$:
  \begin{equation}
  	   		T^{\rm dust}(\hn) \equiv \JCrev{[1 + A_d(\hn)]} d(\hn) ; \quad \mathcal D_\ell^{d} \equiv \frac{ (100 / \ell)^{0.4}}{\left[1 + (\ell / 160)^2\right]^{0.085}}.
  	   \end{equation}
The power spectrum shape $\mathcal{D}_\ell^{dd}$ is similar to that used to model unmasked dust power in the \planck\ CMB likelihood \citep{planck2016-l05} or in \cite{Ying:2016eiz}, based on measured spectral differences between various Galactic masks. 
The quadratic estimator for the squared dust amplitude has the form
  	   \begin{equation}
  	   	\hat g^{A_d}(\hn) \equiv \bar T(\hn) \left( \sum_{\ell m} C^{dd}_\ell \bar T_{\ell m}\: _0Y_{\ell m}(\hn) \right).
  	   \end{equation}
  	   We also applied this hardened estimator to the 217-GHz channel (\twoonesevenbhd). All these contaminant estimators are orthogonal to the lensing curl mode and do not appear in Fig.~\ref{fig:diff_xtt_cons}.
\item \textit{Frequency maps}. We test temperature reconstructions using the HFI 143- and 217-GHz frequency channels. When using frequency maps, we perform a simple foreground cleaning of the maps by projecting out dust and CIB templates in the filtering step. This ensures that these two modes of the data are not used in the reconstruction, but, of course, the extent to which they remove dust and CIB depends on the quality of the templates. We take the 857-GHz channel as a dust template, and the \GNILC~CIB map at 353\,GHz for the CIB. \AC{Perhaps note that the fact that the 857\,GHz channel is dominated by CIB beyond around $\ell \simeq 1500$ (this for $40\,\%$ of sky, would be larger $\ell$ for $67\,\%$) is of no concern since just means templates not orthogonal?}Figure~\ref{fig:diff_ptt_cons} lists several reconstruction variations, one using our baseline $f_{\rm sky} = 67\,\%$ mask (\onefourthree, \twooneseven), and another using an $f_{\rm sky} = 44\,\%$ mask specifically built to minimize dust contamination (\onefourthreeg, \twooneseveng). We also list a reconstruction using the 143-GHz channel with more aggressive SZ-cluster masking (\onefourthreeszcons) or no SZ-cluster masking at all (\onefourthreeszunm).
\end{unindentedlist}
\vskip 0.25cm
The most striking inconsistencies appearing in Fig.~\ref{fig:diff_ptt_cons} are from the frequency reconstructions.  These reconstructions use a simple template-cleaning method that is cruder than that used for the main component-separated products.  On the conservative multipole range only the \twooneseven~reconstruction has possible anomalies, but on the high-$L$ range, both 143- and 217-GHz reconstructions display a striking series of inconsistencies. This demonstrates the importance of precise temperature cleaning for lensing reconstruction. Specifically, this figure strongly suggests that the \twooneseven~reconstruction contains substantial Galactic foreground contamination: the dust-masked \twooneseveng~shows very large, anomalous shifts compared to \twooneseven, on both multipole ranges displayed. While the \twooneseven~row and column are predominantly red, so this reconstruction has larger lensing amplitude than most of the others, the situation is reversed for \twooneseveng, which has slightly lower amplitudes.  This is also visible on the lower panel, where \twooneseven~shows a much larger lensing amplitude, which is substantially reduced on the $f_{\rm sky} = 44\,\%$ dust mask.

All five component-separation methods show good consistency over the conservative range. However, we do see some anomalous behaviour at high-$L$. Notably, we find
\begin{equation}
	\fidprefix\hat A_{401 \rightarrow 2048}^{\phi, \rm{TT}, \NILC} -\: \fidprefix\hat A_{401 \rightarrow 2048}^{\phi, \rm{TT}, \SEVEM} = 0.088 \pm 0.029,
\end{equation}
formally a $3\,\sigma$ deviation from zero. The actual shift in the high-$L$ amplitude between these two reconstructions is not that large however: about $0.5\,\sigma$ of our high-$L$ baseline measurement. None of the component-separation methods show large shifts in an absolute sense. Our baseline band powers using \SMICA\ appear to be stable with respect to choice of mask, giving shifts consistent with expectations for the change in sky area. The reconstruction using the aggressive galactic mask \SMICAeighty\ does not display suspicious deviations on either lensing multipole range, nor does it exacerbate the band-power features even slightly. \SMICANOSZ~relies more on the 217-GHz channel, and gives a slightly larger amplitude. While in itself not suspicious, \gfiveseven~shows a significant decrease in amplitude from \SMICANOSZ, and this could indicate a higher level of Galactic dust contamination in the \SMICANOSZ\ map. We find generally no evidence for thermal SZ (tSZ) contamination: we obtain almost identical band powers using our baseline tSZ-cluster mask, the conservative tSZ-cluster mask, or no cluster mask at all, both from the 143-GHz channel or on the \SMICA\ maps. At 143-GHz, we do however detect and subtract a much larger point-source correction without the SZ-cluster mask. This is consistent with the predictions of further tSZ-dedicated tests provided in Sect.~\ref{subsec:testforegroundscorr}.

The 143-GHz channel shows some consistency issues on the high-$L$ range significant at around the $2.5\,\sigma$ level, where it displays consistently lower amplitudes than all other reconstructions. This is true independently of sky-cuts and bias-hardening tests. The cause of these issues is currently not understood and this further motivates us to use the conservative multipole range for our baseline cosmology results.

Figure~\ref{fig:diff_xtt_cons} shows the same tests for the lensing curl spectrum estimated from temperature-only reconstructions. The curl estimator has less sensitivity to many sources of systematic effects; apart from the lensing gradient mode, it is also orthogonal (i.e., has zero \JCrev{linear} response) to residual point-source contamination, CMB signal modulation, or to the dust amplitude estimator discussed above. Hence, the curl reconstruction is a stringent test of the noise properties of the simulations versus those of the data maps. Figure~\ref{fig:diff_xtt_cons} shows that our lensing curl power is very stable across reconstructions, and that we find the same extended region of negative curl in all tests. The curl spectrum on the conservative multipole range shows sensitivity to small-scale modes of the maps used for the reconstruction (around $\ell \simeq 2000$). The shifts between \ftwofive~and \ellmax{1900}\ compared to our baseline are formally significant. As shown in the lower panel, the absolute value of the \ftwofive~shift is very small, so this probably indicates a slight mischaracterization of the covariance rather than a systematic effect that is important for the cosmological interpretation of the baseline analysis. \ellmax{1900} shifts by a standard deviation on the conservative range, giving mild evidence of a tension. However, a similar problem is not seen in our lensing reconstruction. On the high-$L$ range, the only potentially problematic test is \ellmin{1451}, which gives a lower curl. This reconstruction is noisy and only weakly constraining on the high-$L$ range, and shows no suspicious offset on the conservative $L$-range. The largest shifts overall are observed on the ecliptic poles compared to the ecliptic equators, and to a similar extent on \SWNE~versus \SENW, for both ranges. These two cuts define regions with substantial overlap, with the curl power being higher (i.e., more consistent with zero) on the ecliptic poles, where the noise is considerably lower. However, the two reconstructions being compared share no common sky area, so differences are expected, and these shifts are not anomalous compared to simulations.
The curl feature is negative, which makes it hard to explain by small inaccuracies in the simulations:
after $\RDNzero$ subtraction, which already corrects for errors in the power in the simulations to linear order, any error in the reconstructed lensing power from
inaccuracies in the simulations should be quadratic in the difference and positive, so
a negative curl power cannot easily be explained by a mismatch between the data and the simulations alone.

There is also a shift between \SWNE\ and \SENW\ (and ecliptic poles and equator) in the $\phi$ reconstruction, with the regions around the ecliptic poles (\SENW) having higher lensing amplitude. However the variance from splitting the sky in two is large, and as before these shifts are not clearly significant compared to simulations.
 It is interesting to note that the \SWNE\ region gives a larger $\Alens$ than \SENW\ when fitting for the amplitude of the lensing smoothing effect in the temperature power spectrum (see~\paramsIII\ for discussion of $\Alens$ more generally). This is the opposite to what one would expect if the difference were due to lensing, but could be consistent with a statistical fluctuations.


\AL{Two old suggestions, but no very obvious way to do it. (1) Perhaps try to give global $\chi^2$/confidence for all tests, e.g., is number of test failures OK given number of tests? (2) Can we put number on likely systematic error from known effects?}

\begin{figure*}
\includegraphics[width = \textwidth]{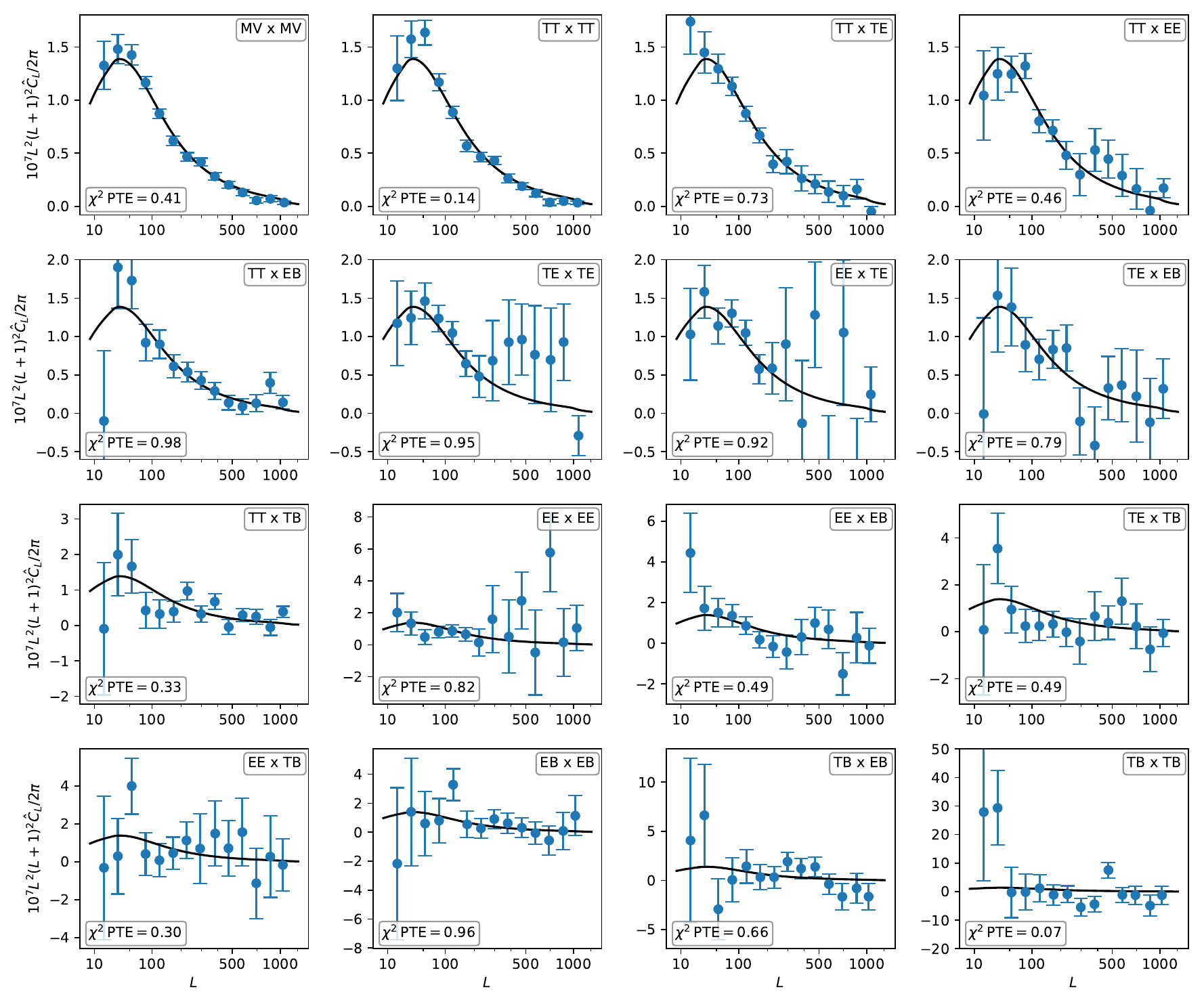}
\caption{
Lensing reconstruction band powers over the full aggressive multipole range for the minimum-variance estimator (MV), and separately for all auto- and cross-spectra of the five quadratic
lensing estimators that can be formed from the temperature ($T$) and polarization ($E$ and $B$) \smica\ CMB maps. The black line shows the \planckalllensing\ best-fit spectrum. In each lower left corner the $\chi^2$ probability to exceed (PTE) is given for the reconstruction (evaluated at the fiducial lensing spectrum), calculated across our conservative multipole range.
\label{fig:allspecs}}
\end{figure*}

\begin{figure*}
\includegraphics[width = \textwidth]{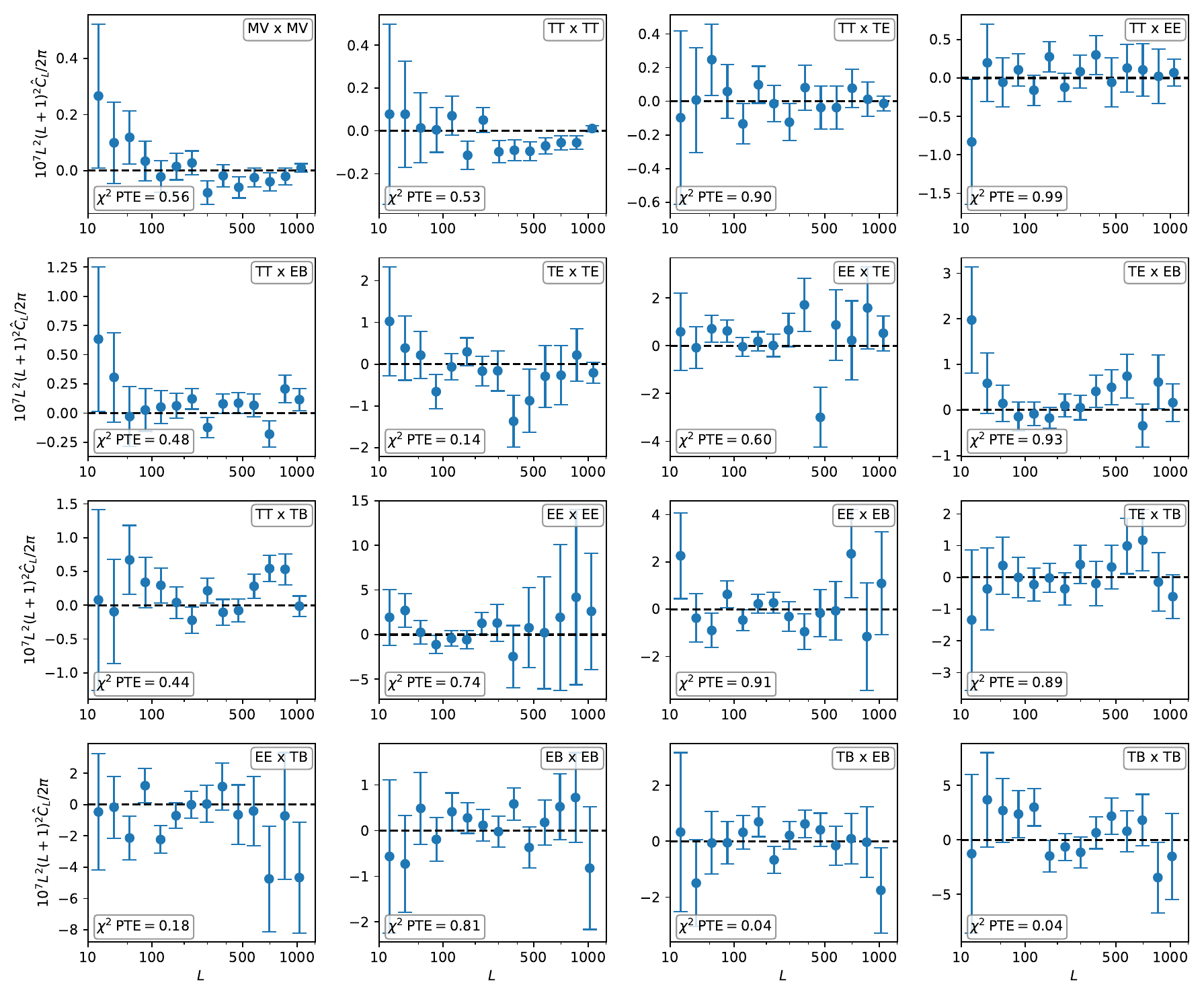}
\caption{\label{fig:allspecs_x} Same as Fig.~\ref{fig:allspecs}, but for the lensing curl reconstruction.
\AC{Some of the reported $\chi^2$ don't immediately correlate (by eye) with the scatter over the conservative range, e.g., for $TT\times EE$, which looks like it has reasonable scatter.}\JC{That is true, but the conservative BP do look less good...}}
\end{figure*}
\subsection{Individual estimator crosses and temperature-polarization consistency}
\label{subsec:polestimators}
The baseline MV reconstruction optimally combines all the information in the temperature and polarization maps (in the approximation of neglecting $T$-$E$ correlations in the filtering). We can also construct estimators separately for each pair of $T$, $E$, and $B$ maps, giving five\footnote{The $BB$ estimator has no linear response to lensing in the absence of primordial $B$-modes.} 
distinct lensing estimators. These can easily be constructed from the MV pipeline simply by zeroing the input maps that are not required for each estimator.
Figure~\ref{fig:allspecs} shows a comparison of the MV power spectrum and all 15 distinct auto- and cross-spectra of the separate reconstructions. The lensing spectrum is measured with high S/N by many of the estimators. The polarization $EB$ estimator is nearly independent of the $TT$ estimator, but their cross-spectrum shows very good agreement with the baseline reconstruction.

Figure~\ref{fig:allspecs_x} shows the reconstructed lensing curl spectra. Curl reconstruction summary amplitude statistics are very consistent on the conservative range between all pairs of estimators; however, this is less true at higher lensing multipoles. Of particular interest is the range $264 \leq L \leq 901$, built to emphasise the temperature curl feature seen in Fig.~\ref{fig:features}. Using the fiducial lensing response functions and \None subtraction, we find
            \begin{equation}
\begin{split}
	\fidprefix\hat A^{\Omega,{\rm TT}}_{264 \rightarrow 901} &= -0.072 \pm 0.017.
\end{split}
\end{equation}
The feature is specific to the temperature-only reconstruction; the next most constraining estimator over this multipole range is $TT \times EB$, with
\begin{equation}
\begin{split}
	\fidprefix\hat A^{\Omega,{\rm TT \times EB}}_{264 \rightarrow 901} &= 0.016 \pm 0.039.
\end{split}
\end{equation}
The polarization-only estimator is much noisier but does not show anomalies on this range either:
\begin{equation}
\begin{split}
	\fidprefix\hat A^{\Omega,{\rm P}}_{264 \rightarrow 901} &= 0.07 \pm 0.17.
\end{split}
\end{equation}
The amplitude of the curl feature is reduced by almost a factor of 2 in the minimum-variance reconstruction compared to the temperature-only reconstruction. This reduction is larger than expected from the FFP10 simulation behaviour. We find
\begin{equation}
\begin{split}
	\fidprefix\hat A_{264 \rightarrow 901}^{\Omega,{\rm MV}} - \:\fidprefix\hat A_{264 \rightarrow 901}^{\Omega,{\rm TT}} &= 0.0327 \pm 0.0094,
\end{split}
\end{equation}
a shift that is unexpected at the $3.5\,\sigma$ level (neglecting look-elsewhere effects). There is no similarly surprising shift in the lensing spectrum, however. 
\subsection{Noise tests}
\label{subsec:noisetests}
\begin{figure}
\centering
\includegraphics[width = \columnwidth]{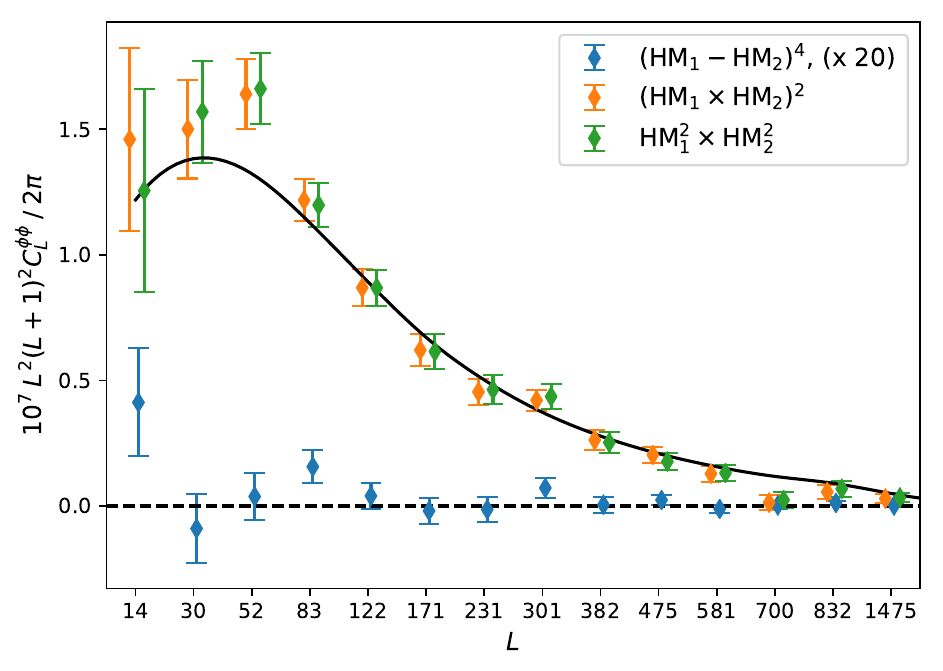}
\includegraphics[width = \columnwidth]{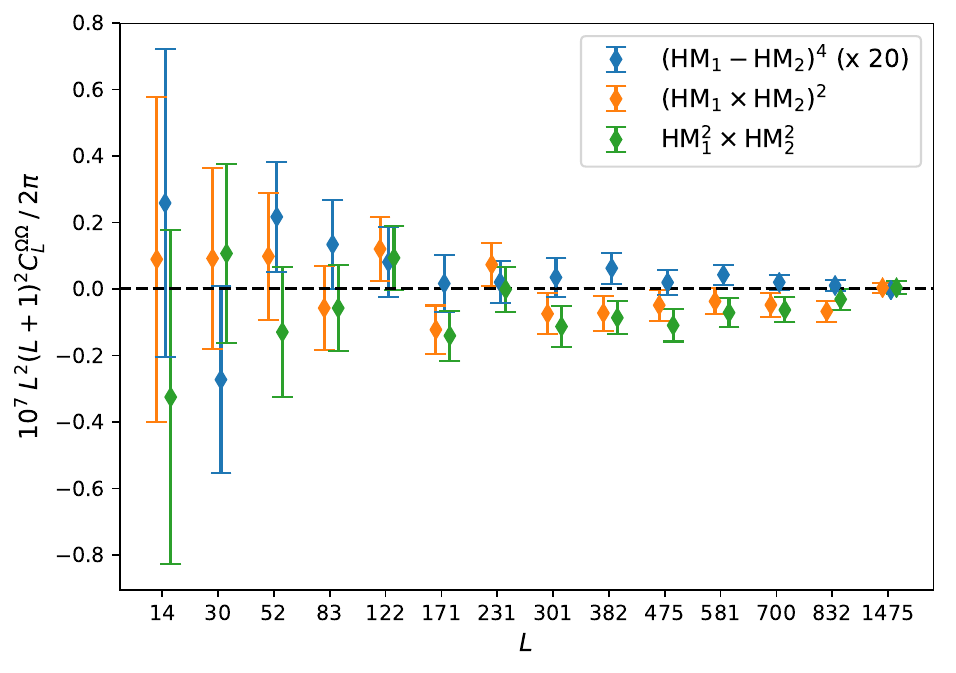}

\caption{\label{fig:halfmissions_rec.pdf}
Half-mission lensing reconstructions tests for the gradient ($\phi$; upper) and curl ($\Omega$; lower). Blue points show reconstruction power spectra from \smica\ half-mission difference maps, after multiplication by a factor of 20. The orange points show the auto-spectrum of the reconstruction built from $\rm{HM}_1$ and $\rm{HM}_2$ (with no noise mean field), and the green points the cross-spectrum of the reconstruction built from $\rm{HM}_1$ with that built from $\rm{HM}_2$ (with no noise contribution to the reconstruction noise $N^{(0)}$). The black line in the upper plot shows the \planckalllensing\ best-fit spectrum.
}

\end{figure}

We now present tests of the noise by considering reconstructions using the two \smica\ half-mission maps (HM1 and HM2) and their difference. First, we perform
gradient and curl lensing reconstructions on their half-difference, nulling the CMB signal. We use only the noise part of the simulation suite for this analysis. The result is shown in Fig.~\ref{fig:halfmissions_rec.pdf} as the blue points labelled $\left(\rm{HM}_1 - \rm{HM}_2\right)^4$ (rescaled by a factor of 20 for visibility). The band powers are consistent with zero to within $2\,\sigma$, as expected, with overall amplitude
\begin{equation}
	\begin{split}
	\fidprefix\hat A_{8 \rightarrow 2048}^{\phi,{\rm TT}} &= 0.002 \pm 0.004,\\
	\fidprefix\hat A_{8 \rightarrow 2048}^{\Omega, {\rm TT}} &= -0.00046 \pm 0.00024.
	\end{split}
\end{equation}
Second, we compare our baseline reconstruction to that obtained from half-mission maps. We consider two types of reconstruction, labelled $\left(\rm{HM}_1 \times \rm{HM}_2\right)^2$ and $\rm{HM}_1^2 \times \rm{HM}_2^2$.
In the first reconstruction, we estimate the lensing spectrum using a single lensing map, built using the first half-mission and second half-mission maps. In this way, there is no noise mean field in the estimate, and the power spectrum reconstruction noise only has one noise contraction. In the second reconstruction, we cross-correlate the lensing map built using the first half-mission data to another lensing map built using the second half-mission data. Both lensing maps now have a noise mean field, but the lensing reconstruction noise comes purely from CMB fluctuations. One caveat to these consistency tests is that there are no half-mission FFP10 CMB simulations accompanying the 2018 \planck\ release (only noise simulations). Instead, we use FFP9 half-mission CMB simulations built for the 2015 release, co-adding the frequency CMB simulations with the \smica\ weighting of the 2018 release. The FFP9 and FFP10 CMB simulations share the same random seed, and apart from the weights the main differences lie in the details of the effective beam across the sky, which is important at low lensing multipoles for the lensing mean field. Using beam-processed CMB simulations also has a slight impact on the {$\RDNzero$} bias. Hence, these half-mission reconstructions should not be considered as reliable as our baseline ones.
The gradient and curl half-mission spectra are displayed in the upper and lower plots, respectively, in Fig.~\ref{fig:halfmissions_rec.pdf}, and show overall consistency with our baseline. The curl feature is less pronounced in $\left(\rm{HM}_1 \times \rm{HM}_2\right)^2$, but only slightly so. Using the FFP10 fiducial model to build summary amplitudes over the multipole range $L=264$--$901$ covering the curl feature, we find
\begin{equation}
	\begin{split}
	\fidprefix\hat A^{\Omega,{\rm TT}}_{264 \rightarrow 901} &= -0.058 \pm 0.018 \quad  \left[({\rm HM}_1 \times {\rm HM}_2)^2\right], \\
	\fidprefix\hat A^{\Omega,{\rm TT}}_{264 \rightarrow 901} &= -0.071 \pm 0.018 \quad \left[{\rm HM}_1^2 \times {\rm HM}_2^2\right],
	\end{split}
\end{equation}
compared to $-0.072 \pm 0.017$ for our baseline. The shifts are consistent within $1\,\sigma$ with those obtained from the simulation differences. On the other hand, the ``dip'' in the gradient spectrum around $L \simeq 700$ is more pronounced. We have
\begin{equation}
	\begin{split}
	\fidprefix\hat A^{\phi,{\rm TT}}_{638 \rightarrow 762} &= 0.11 \pm 0.25 \quad \left[({\rm HM}_1 \times {\rm HM}_2)^2\right], \\
	\fidprefix	\hat A^{\phi,{\rm TT}}_{638 \rightarrow 762} &= 0.20 \pm 0.25 \quad \left[{\rm HM}_1^2 \times {\rm HM}_2^2\right],
	\end{split}
\end{equation}
compared to $0.29 \pm 0.24$ for our baseline. These shifts are again within $1\,\sigma$ of the expected differences, but lower the spectrum further with respect to the \lcdm\ prediction. The low value is driven by the first half-mission map as we now discuss. Taking amplitudes with respect to the \planck\ best-fit \planckalllensing, and marginalizing over CMB uncertainties in order to quantify better the discrepancy, we have
\begin{equation}
\begin{split}
\bfprefix\hat{A}^{\phi,\rm TT}_{638 \rightarrow 762} &= -0.51 \pm 0.35\quad \left[({\rm HM}_1)^4; \text{ CMB marginalized}\right] \\
\bfprefix\hat{A}^{\phi,\rm TT}_{638 \rightarrow 762} &= 0.43 \pm 0.32\quad \left[({\rm HM}_2)^4; \text{ CMB marginalized}\right].
\label{HM1A}
\end{split}
\end{equation}
The $({\rm HM}_1)^4$ result is formally a $4.5\,\sigma$ deviation from the best-fit lensing spectrum value. The remaining band powers of the ${\rm HM}_1$
reconstruction show no inconsistencies with the best-fit, however, and the observed shift between the two amplitudes in Eq.~\eqref{HM1A} is within $2\,\sigma$ of the difference expected from simulations. The origin of these curious dip differences is not currently understood, but the dip is not included in our conservative multipole range, so it is not directly relevant for our baseline results.

The MV baseline lensing spectrum and the half-mission spectra are still low on these multipole ranges for both curl and gradient, but less than the temperature-only spectra. We find
\begin{equation}
	\begin{split}
	\fidprefix\hat A^{\Omega,{\rm MV}}_{264 \rightarrow 901} &= -0.029 \pm 0.017\quad \left[({\rm HM}_1 \times {\rm HM}_2)^2\right], \\
	\fidprefix	\hat A^{\Omega,{\rm MV}}_{264 \rightarrow 901} &= -0.038 \pm 0.018\quad \left[{\rm HM}_1^2 \times {\rm HM}_2^2\right],
	\end{split}
\end{equation}
compared to $-0.039 \pm 0.015$ for our baseline, and
\begin{equation}
	\begin{split}
	\fidprefix\hat A^{\phi,{\rm MV}}_{638 \rightarrow 762} &= 0.43 \pm 0.24\quad \left[({\rm HM}_1 \times {\rm HM}_2)^2\right], \\
	\fidprefix	\hat A^{\phi,{\rm MV}}_{638 \rightarrow 762} &= 0.30 \pm 0.27\quad \left[{\rm HM}_1^2 \times {\rm HM}_2^2\right],
	\end{split}
\end{equation}
compared to $0.45 \pm 0.23$ for our baseline.

\subsection{Tests of CMB lensing/foreground correlations}\label{subsec:testforegroundscorr}
\begin{figure}
\centering
\includegraphics[width = \columnwidth]{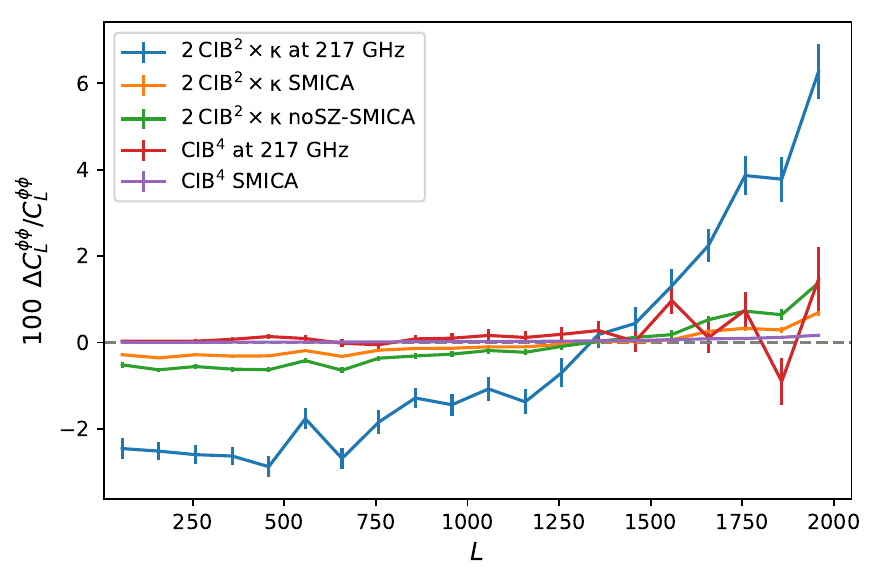}
\caption{\label{fig:psm_ata_firb} Estimates of the expected CIB-induced biases (as a fraction of the lensing potential power spectrum) to the temperature-only lensing spectrum reconstruction at 217\,GHz (without any cleaning), and for the \smica\ frequency weighting. The large-scale-structure bispectrum causes the quadratic estimator applied to the CIB map to correlate with the lensing signal, sourcing a 2\,\% negative bias at 217\,GHz (blue) on our conservative multipole range. This is reduced by an order of magnitude after \smica\ cleaning (orange). The bias remains sub-percent in the tSZ-deprojected \smica\ weighting (green). The CIB trispectrum sources a smaller contribution (red and purple at 217\,GHz and for \smica\ frequency weighting, respectively). Biases to the curl reconstruction are negligible. Error bars on this figure are estimated from the scatter of the unbinned power.}
\end{figure}
We add residual extragalactic foreground power to our simulations as an independent Gaussian component. In reality, extragalactic foregrounds are non-Gaussian and are correlated with the lensing signal since both are affected by the same matter perturbations. There is therefore a concern that residual foregrounds could lead to additional contributions to the quadratic estimator power that are not corrected for in our pipeline. In particular SZ, CIB, and clustered point-source contributions (and/or their local power) could be directly correlated with the lensing potential.
The polarization reconstruction is expected to be essentially free from these contaminants due to the much lower level of foreground power in the small-scale polarization maps. Our use of foreground-cleaned \smica\ maps is also expected to remove the bulk of the CIB foreground signal in temperature; however, SZ and unresolved extragalactic radio sources are largely unaffected,  \AC{Why?} and foreground residuals remain due to instrumental noise.
Semi-analytic estimates suggest that the possible bias on the temperature reconstruction power should not be large at the sensitivity level of \Planck~\citep{Osborne:2013nna,vanEngelen:2013rla,Ferraro:2017fac}, but it is important to test this more directly.

To assess the possible impact of correlated foregrounds we have used a small number of non-Gaussian foreground simulations that are constructed to have approximately the correct correlation structure. The Planck Sky Model (PSM; \citealt{2013A&A...553A..96D, 2016A&A...594A..12P}) software is extended to consistently generate maps of the CMB lensing potential, CIB, and SZ components, including their correlations. To do this, the Boltzmann code \CLASS\ \citep{2011JCAP...07..034B} is used to calculate the correlated angular power spectra of the unlensed CMB, lensing potential, and matter distribution (neglecting matter-CMB correlations) over 64 concentric shells between redshifts $z=0.01$ and $z=6$ up to a maximum multipole $\lmax = 4096$. The power spectra for the FFP10 fiducial model are calculated using the nonlinear \HALOFIT\ model from~\cite{Takahashi:2012em} using modes up to $k_{\rm max}=1\, h\,{\rm Mpc}^{-1}$.
The lensing potential is assumed to be Gaussian, but the matter density fields $\rho_i(\hn)$ in each shell are simulated as log-normal fields; the covariance from \CLASS\ is therefore mapped into the covariance of the log fields, so that a Gaussian realization of $s_i(\hn)$ can be exponentiated to obtain log-normal matter density fields with the correct covariance structure~\citep{Greiner:2013jea}. The CMB component is lensed with the lensing potential using \LENSPIX\ \citep{Lewis:2005tp}. We have checked that the impact of lensing of the CIB is negligible, as expected~\citep{Schaan:2018yeh}.

\begin{figure*}
\centering
\includegraphics[width = \textwidth]{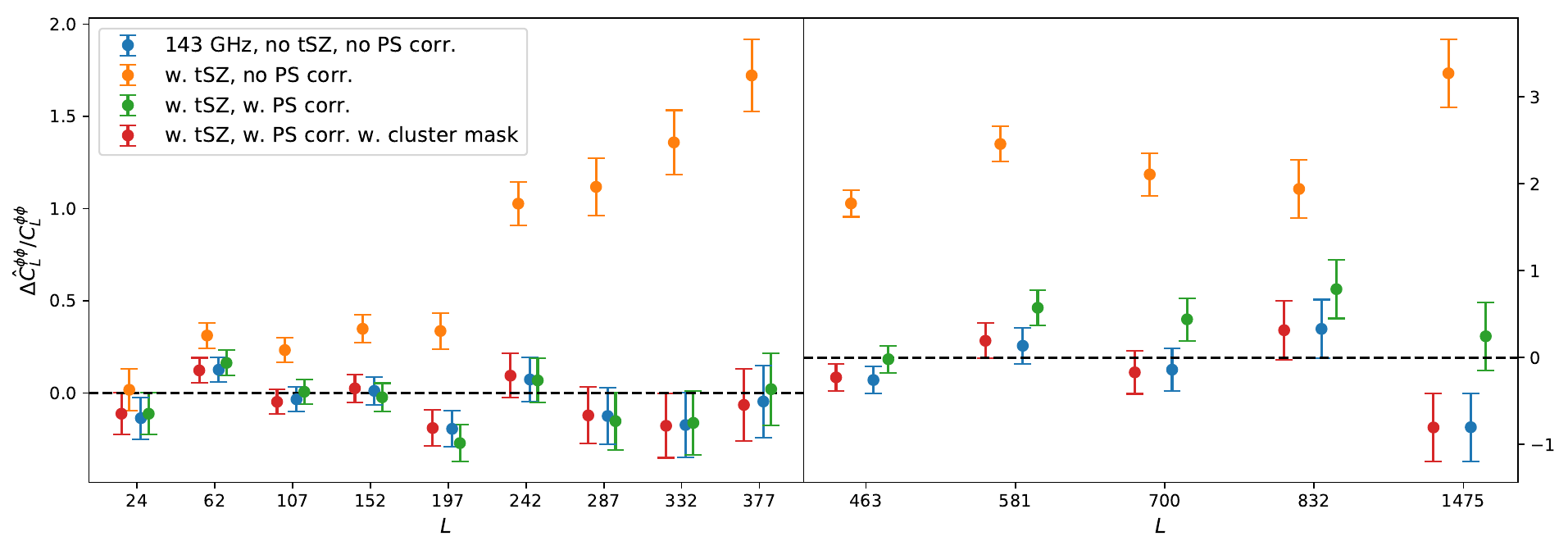}
\caption{\label{fig:psm_ata_tSZ} Impact of the thermal SZ (tSZ) on the temperature-only lensing reconstruction spectrum at 143\,GHz. We show the fractional difference between several reconstruction power spectra based on a 143\,GHz full-sky simulation and the input lensing potential power spectrum (where the simulation includes tSZ and CMB, and Gaussian noise generated according to the 143-GHz channel pixel variance map). Without any masking, the contamination appears predominantly as a large point-source trispectrum contribution (orange). The point-source correction is quite effective at reducing this term below $L \simeq 400 $, with visible residual contamination at higher $L$ (green). After applying a $f_{\rm sky} = 0.997$ cluster mask (red), the differences from the power reconstructed in the reference CMB-only simulation with no masking (blue) are further reduced to a small fraction of a standard deviation in all bin. The left panel shows the conservative multipole range and the right panel the higher lensing multipoles.
\JC{Difference to CMB-only rec only ?}}
\end{figure*}

The CIB galaxies are grouped into three different populations according to their spectral energy distributions (SEDs): proto-spheroid, spiral, and starburst. The flux density in each population is randomly distributed according to the redshift-dependent number counts from \Planck\ ERCSC \citep{2013MNRAS.429.1309N}, JCMT/SCUBA-2 \citep{2013ApJ...762...81C}, AzTEC/ASTE \citep{2012MNRAS.423..575S}, and \textit{Herschel}/SPIRE \citep{2013A&A...557A..66B} observations. Each redshift shell is populated with CIB galaxies and galaxy clusters with probabilities proportional to the density contrast distribution, with a population-dependent bias. 
\JCrrev{The CIB maps constructed in this way have power spectra that agree with measurements from \planck\ data at high frequencies using several different methods for dealing with Galactic dust and CMB contamination~\citep{planck2013-pip56, Ying:2016eiz, Lenz:2019ugy}.}

Galaxy clusters are simulated by drawing a catalogue of halos from a Poisson distribution of the Tinker mass function \citep{2008ApJ...688..709T}. The thermal SZ emission of each halo at a given redshift and mass is modelled according to the temperature \citep{2005A&A...441..893A} and the pressure profiles given by \citet{2010A&A...517A..92A}. The simulated SZ clusters are then distributed over the redshift shells with probabilities proportional to matter density fields to have consistent correlations between SZ, CMB, and CIB.  The resulting cluster counts are consistent with those observed by \planck. The kinetic SZ component is simulated by using the electron density of the clusters and a Gaussian realization of the 3D cluster velocity field derived from the density fluctuations \citep[following][]{Peebles:93}.

We first use the simulations to assess the expected impact of the CIB component on our lensing reconstruction. We test the 217-GHz channel component (without any cleaning), and a cleaned map constructed by combining the different frequency simulations using the \smica\ frequency weights. We use the same multipole cuts as our baseline analysis, but include the full sky for simplicity. \JCrrev{We see a CIB signature in the lensing spectrum consistent with previous work~\citep{vanEngelen:2013rla}, with two principal biases}: a positive bias from the CIB trispectrum (labelled $\rm CIB^{4}$ in the following); and a larger contribution from the bispectrum ($\rm CIB^2 \times \kappa $). We isolate the first contribution by performing direct lensing reconstruction on the simulated CIB map, then estimating the power spectrum subtracting (disconnected) biases with Gaussian isotropic realizations from the CIB power spectrum. To estimate the bispectrum contribution, we cross-correlate the lensing estimator $\hat \phi$ as applied on the CIB map to the input lensing potential. The resulting bias is twice this cross-spectrum.
 Figure~\ref{fig:psm_ata_firb} shows the biases that are expected. For the uncleaned 217-GHz channel (blue) there is at most about a $2\,\%$ bias over the conservative multipole range. This dominant (bispectrum) contribution is effectively reduced by one order of magnitude after \smica\ cleaning (orange), making it negligible. It is larger, but still sub-percent level, in the case of the tSZ-deprojected \smica\ weighting (green). The trispectrum biases are even smaller (red and purple).
 A similar analysis for the lensing curl shows that all these terms are expected to be completely negligible for the curl power.
 \AL{Long list: foreground-projected 217 would be nice, to know how much we expect to see it the consistency tests section}
\AC{Perhaps note that masking bright dusty galaxies will have little impact on these results (except, perhaps, for the trispectrum contribution)?}

 We now turn to the thermal SZ component, focussing on the 143-GHz channel since it is small in the 217-GHz channel where the detector bandpasses are centred on the null of the tSZ spectrum. We perform several reconstructions on the same CMB simulation, with and without thermal SZ, and with and without masking to emulate the effect of the cluster mask that we apply in our baseline analysis. \JCrev{These tests allow us both to demonstrate the good performance of our point-source correction procedure, even in the absence of masking, as well as the expected robustness of our band-powers.} To build the simulation cluster mask, we mask the same number of objects as our baseline mask (rescaled by the respective observed area), starting from the objects with the strongest integrated Compton $Y_{500}$ emission, and using the same criteria to define the masking radii for each object. This procedure results in a total of 1\,148 objects masked (out of 377\,563) for a resulting masked sky fraction of roughly $0.3\,\%$. Some effectively pure point-source signal remains in the map, which would be detected by the \planck\ source-detection methodology and also masked in our actual data lensing mask. However, these residual signals have no impact on these tests and we perform no further cleaning.

  Figure~\ref{fig:psm_ata_tSZ} collects the different reconstructions, showing for clarity the conservative multipole range on the left panel, and the remaining high-$L$ multipoles on the right panel. We show deviations of the reconstruction from the fiducial input lensing power spectrum, with the blue points showing the reconstruction for the reference simulation without the SZ mask. This full-sky reference simulation includes instrumental noise in the form of Gaussian noise, independent between pixels but with amplitude in each pixel following the 143-GHz channel variance map. Owing to the larger sky area, the blue error bars are slightly tighter than those on our 143-GHz channel data reconstruction. The tSZ signal appears largely as a point-source-like contamination, and we distinguish results before (orange) and after (green) point-source correction, calculated and subtracted from the band powers in the same way as in the main data analysis. The correction is very effective at correcting the power over the conservative range, but residual additional power is clearly visible at high $L$. Finally, the red points show the result after application of the cluster mask. At this point, the contamination (i.e., the difference from the reconstructed power for the reference simulation) becomes a negligible fraction of a standard deviation. We also perform the same tests for the curl reconstruction band powers; the curl estimator does not respond to point-source signals, and we find a negligible impact, even without any masking.

  Applying a cluster mask systematically removes small regions of the sky with strong lensing convergence power, and could bias the lensing spectrum estimate. We assess the size of this effect on simulations by comparing lensing spectra before and after masking. The cluster masks are produced for each simulation in the way described above, and we obtain spectra after deconvolving from pseudo-$C^{\kappa\kappa}_L$ power spectra the effects of the (slightly enlarged and then apodized) mask. We find this bias to be at most $0.25\,\%$ in amplitude in all bins (limited by sampling noise), and we do not consider it further. 

Finally, we perform test reconstructions using the entire set of foregrounds available (CIB, thermal and kinetic SZ, and point sources) using the \smica\ frequency weighting, and incorporating the cluster mask described above. The foregrounds contribute an additional $5\,\%$ power around $\ell = 2000$ (from the CIB and point sources), in good agreement with the mismatch observed in the data compared to the (clean) FFP10 simulation suite. We perform a full lensing reconstruction on this map, adding Gaussian isotropic power to the Monte Carlo simulations to account for the difference in power, in the same way as for our analysis of the actual data, and compare to a reference reconstruction that includes the CMB component only. We find shifts in all bins that are typically $0.1(0.13)\,\sigma$ of our baseline reconstructions in the conservative (aggressive) multipole range. All deviations seem consistent with the scatter expected from the addition of the extra power at high CMB multipoles.

\AL{Comment on size of ksz in sims, and cite ~\citep{Ferraro:2017fac}?}\JC{Sent expl. request on ksz size}
\subsection{Test of dependence on fiducial model}\label{subsec:testfid}

\begin{figure}
\centering
\includegraphics[width = \columnwidth]{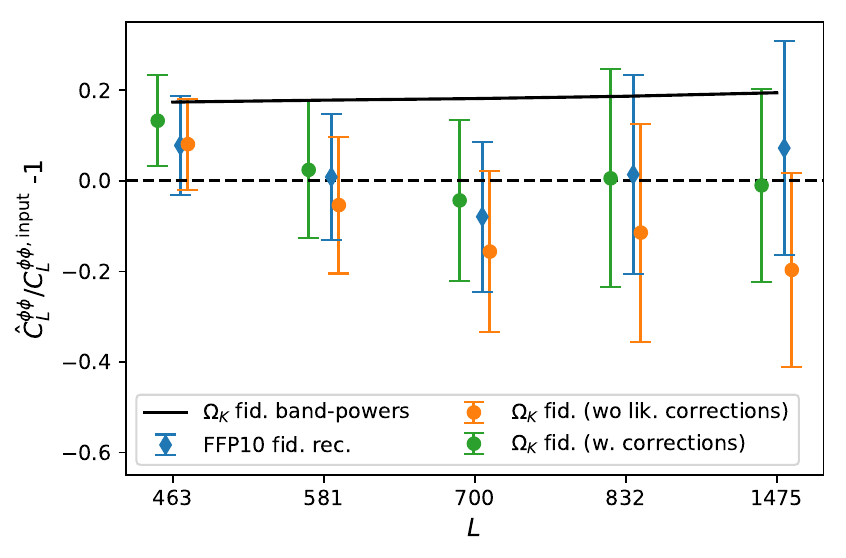}
\caption{\label{fig:omk_test} 
Comparison of reconstructions over the same high-$L$ multipole range on the same simulated map, using different fiducial model assumptions. The blue points use a fiducial model, set of simulations, and estimator ingredients based on the FFP10 input cosmology. The orange points use a cosmology with $\Omega_K\simeq -0.04$, where the lensing power spectrum is higher by 10--20\,\% (black line), and the CMB spectra differ by roughly 2\,\% at high $\ell$. The linear corrections to the likelihood account for the differences in fiducial model, and are mostly important at high-$L$ in this case; they are effective (green) at bringing the reconstructions with the different fiducial models into good agreement.}
\end{figure}

The lensing pipeline assumes a fiducial model for the true CMB and lensing power spectra, with the likelihood perturbatively correcting for differences between the fiducial model and the actual model spectra at each point in parameter space.
For our results to be robust, they should be almost independent of which fiducial model was assumed. To test this, we carry out a simplified analysis using a different fiducial model: a non-flat \lcdm\ model best-fit to the CMB power spectra with $\Omega_K\simeq -0.04$. The test model has 15--20\,\% more lensing power than the FFP10 fiducial model, and also has CMB power spectra that differ by approximately 2\,\% at high $\ell$. Note that we do not need to test large deviations of the CMB spectra because they are now empirically measured to good accuracy: the main variations still allowed in the spectral shape are due to residual cosmic variance and foreground uncertainties, both of which are a small fraction of the total CMB signal on scales relevant for lensing reconstruction.

Running a full set of FFP10 simulations with a different fiducial model would be numerically very expensive, so we instead generate a set of simpler idealized isotropic-beam simulations with the FFP10 fiducial model and the test $\Omega_K \ne 0$ model, and check for differences in the lensing reconstruction based on these two sets of simulations and theoretical spectra. The spectra enter into the filtering of the data, the analytic estimate of the estimator response, and via the simulations used for evaluation of the mean field, $\RDNzero$ and MC corrections, and via the likelihood in the perturbative correction functions and fiducial value of $\None$. The simpler set of simulations we use here lacks the detailed beam shape and scanning model that affects the signal at low-$L$. For this reason we compare the reconstructions not on the actual \Planck\ data, but on one of the simulations, generated with the FFP10 CMB spectra as input. We show in Sect.~\ref{subsec:N1tests} below that the \None\ contribution on the cut sky is accurately modelled by full-sky analytic results, so we use the full sky for convenience in this test.

We test robustness to the choice of fiducial model by analysing a
single FFP10 ($\Omega_K = 0$) simulation
with the consistent fiducial simulation set, and then by analysing it using the
inconsistent ($\Omega_K \ne 0$) fiducial simulation set. 
With the consistent fiducial model, using the full lensing multipole range to fit for a lensing amplitude we find
\begin{equation}
	\fidprefix \hat A_{8 \rightarrow 2048}^{\hat \phi,{\rm TT}} = 1.032 \pm 0.025 \textrm{ (FFP10 fiducial sim.).}
\label{eq:fidAfidsim}
\end{equation}
Over this same multipole range, fitting the same reconstructed power with
a lensing amplitude relative to the $\Omega_K$ fiducial model results in $0.88 \pm 0.021$, consistent with the increased lensing power in this model and
excluding the model at more than $5\,\sigma$. If we now perform lensing reconstruction on the same data, but with the $\Omega_K$ model replacing the fiducial FFP10 model throughout the analysis, and then fit a lensing amplitude relative to the FFP10 model in the likelihood, we get
\begin{equation}
   \begin{split}
	&\fidprefix A_{8 \rightarrow 2048}^{\hat \phi,{\rm TT}} = 1.009 \pm 0.026 \textrm{ (no lin. corrections)} \\
	&\fidprefix A_{8 \rightarrow 2048}^{\hat \phi,{\rm TT}} = 1.026 \pm 0.026 \textrm{ (with lin. corrections)}\\
	&\textrm{($\Omega_K$ fiducial simulations)}.
	\end{split}
\end{equation}
While the estimate without linear corrections differs by one standard deviation from that in Eq.~\eqref{eq:fidAfidsim}, the linear corrections are effective at reducing the discrepancy to a small fraction of a standard devitation, demonstrating consistency of our estimator and likelihood methodology. Note that we do not expect perfect agreement, since the estimator's filtering and quadratic estimator weights do differ slightly. For the most part, the linear correction in the likelihood occurs at high lensing multipoles for the $\None$ bias subtraction, where the lensing power spectrum is up to $20\,\%$ higher in the $\Omega_K \ne 0$ model than the flat model (see Fig.~\ref{fig:omk_test}).

\subsection{Lensing Gaussianity assumption}\label{subsec:testnl}
The quadratic estimator formalism and \FFP\ simulations assume that the lensing potential is Gaussian, but it is expected to be non-Gaussian at some level due to nonlinear growth of large-scale structure (LSS) and post-Born lensing~\citep{Pratten:2016dsm}. A non-vanishing LSS bispectrum will source contributions to the lensing reconstruction spectrum involving three powers of $\phi$, potentially giving rise to an additional $N^{(3/2)}_L$ lensing bias as well as the \Nzero\ and {\None}\ biases that we already model~\citep{Bohm:2016gzt}. A subset of the contractions leading to this bias were first studied by \citet{Bohm:2016gzt} assuming tree-level perturbation theory for the LSS bispectrum and neglecting post-Born contributions, finding that the terms are negligible at \Planck~noise levels. \citet{Beck:2018wud} and \citet{Bohm:2018omn} have shown on simulations that including post-Born effects reduces the signal even further, as expected from the opposite sign of post-Born and LSS contributions to most configurations of the lensing bispectrum~\citep{Pratten:2016dsm}. We therefore consider the bias on the lensing reconstruction power spectrum to be negligible and neglect it. The bias on cross-correlations between the lensing spectrum and low-redshift large-scale structure tracers could, however, be larger due to larger low-redshift LSS non-Gaussianity, and smaller opposite-sign post-Born contributions at low redshift.
\ALrev{However, for \planck\ noise levels, following \citet{Fabbian:2019tik} we estimate that the cross-correlation bias remains below 1\,\% for tracers at $z\gtrsim 0.2$ where there is a useful cross-correlation signal, and hence should also be negligible compared to errors.}

Note that although the lensing signal is expected to be close to Gaussian, the lensing reconstruction noise is expected to be non-Gaussian, since it is a nonlinear function of the maps, for example the reconstruction 1-point function is skewed~\citep{Liu:2016nfs}. However, this non-Gaussianity does not affect our analysis, and the power spectrum band powers are well approximated as Gaussian to the required level of accuracy.


\subsection{Tests of the \None\ lensing bias}
\label{subsec:N1tests}

\begin{figure}
\centering
\includegraphics[width = \columnwidth]{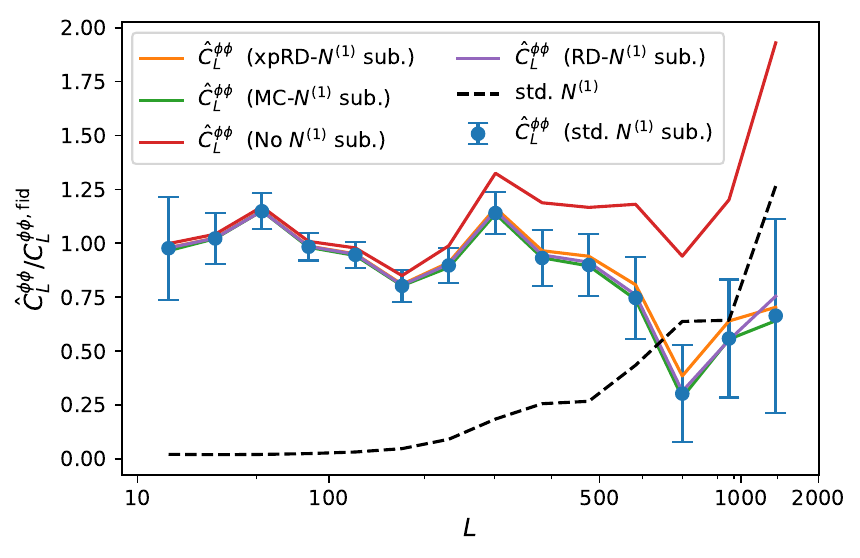}
\includegraphics[width = \columnwidth]{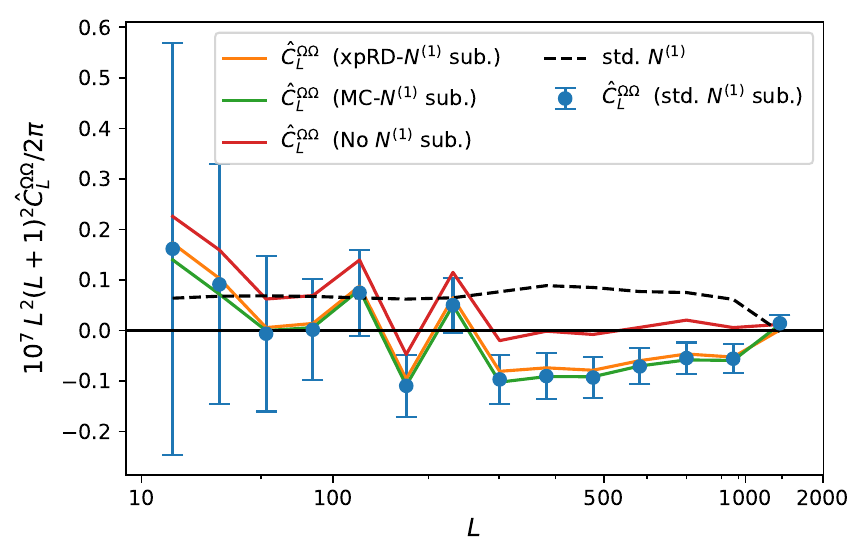}
\caption{\label{fig:RDN1} Comparison of reconstruction band powers for various methods to remove the {\None} lensing bias. Our baseline band powers are built by subtracting the bias calculated analytically with the FFP10 fiducial model, and are shown as the blue points with error bars. The curves labelled ``$\RDNone$'' and ``$\xpRDNone$'' invert Eqs.~\eqref{eq:RDN1} and \eqref{eq:xpRDN1}, respectively, and make no assumption about the theory lensing spectrum. The curves labelled ``$\MCNone$'' subtract a simulation-based estimate of {\None}. For comparison the red lines show the band powers without any {\None} subtraction, and the dashed black lines the {\None} bias itself. The top plot shows the lensing reconstruction from temperature only, and is normalized to the FFP10 fiducial $C_L^{\phi\phi, \rm fid}$. The lower plot shows the temperature-only curl reconstruction (which on small scales appears to be coincidentally close to the result one would expect with no $\None$ at all).} 
\end{figure}

The {{\None}} bias originates from secondary contractions of the lensed CMB trispectrum \citep{Kesden:2003cc,Hanson:2010rp}, and affects both gradient and curl deflection reconstructions. We follow \PlanckLensTwo\ in that our baseline band powers simply subtract the {\None} bias evaluated in a fiducial model, with the likelihood correcting for the model dependence perturbatively. This section outlines some tests to show that this is not significantly suboptimal, nor a source of bias. We first discuss two ways to make {\None}-subtracted band powers without assuming a fiducial lensing spectrum. We then test a simulation-based {\None} estimator, assessing the impact of sky cuts and sky curvature; this also forms our baseline {\None} calculation in the case of inhomogeneous filtering.

The {{\None}} correction captures the estimator's linear response to $C^{\phi\phi}$ from non-primary contractions. As such, we can write (neglecting point-source and MC corrections, and complications due to masking in the following discussion)
\begin{equation}\label{eq:RDN1}
\left\langle {|\hat \phi_{LM}|^2} \right\rangle = N^{(0)}_L + \sum_{L'}\left( \delta_{LL'} + N^{(1)}_{LL'}\right) C_{L'}^{\phi\phi}.
\end{equation}
Instead of subtracting the right-most term using a fiducial $C_L^{\phi\phi}$, we can invert this matrix relation to obtain the lensing spectrum. This can also be generalized to calculate the curl \None, which is used for the curl null test.
With $X,Y$ each standing for the gradient or curl deflection, let $N^{(1),XY}$ be the $Y$-induced {{\None}} contribution to the $X$-spectrum estimate, we can  write
\begin{multline}
\label{eq:xpRDN1}
\begin{pmatrix} \left\langle {|\hat \phi_{LM}|^2} \right\rangle \\ \left\langle {|\hat \Omega_{LM}|^2} \right\rangle \end{pmatrix} =
\begin{pmatrix} N_L^{(0),\phi\phi} \\ N_L^{(0),\Omega\Omega} \end{pmatrix} \\+ \sum_{L'}\begin{pmatrix} \delta_{LL'} + N^{(1),\phi\phi}_{LL'} & N^{(1),\phi\Omega}_{LL'} \\ N^{(1),\Omega\phi}_{LL'} & \delta_{LL'} + N^{(1),\Omega\Omega}_{LL'} \end{pmatrix} \begin{pmatrix} C_{L'}^{\phi\phi} \\ C_{L'}^{\Omega\Omega} \end{pmatrix}.
\end{multline}
In this equation $C_L^{\Omega\Omega}$ represents a hypothetical curl-like deflection caused, for example, by an instrumental systematic effect that we assume is uncorrelated with the lensing signal. Sub-pixel effects or pointing errors could create such a signal (see Appendix~\ref{subsec:subpix}).

Inverting Eq.~\eqref{eq:xpRDN1} provides slightly different lensing gradient and curl band powers, labelled ``$\xpRDNone$'' in Fig.~\ref{fig:RDN1}, while inverting Eq.~\eqref{eq:RDN1} provides yet another set of lensing band powers, labelled ``$\RDNone$.'' The results for the temperature-only reconstructions are shown in Fig.~\ref{fig:RDN1}. Lensing band powers are shown in the top panel, after taking the ratio to the FFP10 fiducial spectrum; the lower panel shows the curl band powers. In both cases, the points with error bars show our baseline reconstructions, using the fiducial {\None} subtraction. For comparison, the red curves display the case of no {\None} subtraction, and the black dashed curves the {\None} bias itself. In implementing the direct inversion of the linear relations above, we have truncated $L'$ at 3000, and the matrix elements are evaluated numerically according to the flat-sky isotropic analytic form given in Appendix~\ref{app:biases}. These different methods for dealing with the {\None} bias have very little impact on the band powers, with a change of at most a small fraction of $1\,\sigma$ on the smallest scales, and are consistent with results from our fiducial model. The {\None} subtraction method does not substantially affect our curl null-test results either. A deconvolution of the {\None} bias might also be desirable in order to reduce off diagonal contributions to the band powers' covariance \citep{Peloton:2016kbw}, but we found the gain to be negligible.

We also test a simulation-based {\None} estimation, labelled ``$\MCNone$'' in Fig.~\ref{fig:RDN1}. We use pairs of simulations that share the same lensing deflections but have different CMB realizations, to build an estimate of the bias. The precise way to do this follows \cite{SPTPOLLensing}, and is also reproduced for completeness in Appendix~\ref{app:biases}. In contrast to the other methods described above, this way of estimating the bias takes into account our exact sky cuts, and also includes sky curvature (which is neglected in our analytic {\None} calculations). We use simplified, effectively isotropic CMB simulations that do not contain the full anisotropic \Planck\ beam model or \smica~processing, since a full re-simulation would be too numerically expensive. The agreement between the simulation-based \None\ estimator and the other methods is good, to the point that it does not visibly affect the band powers.

\begin{figure}
\centering
\includegraphics[width = \columnwidth]{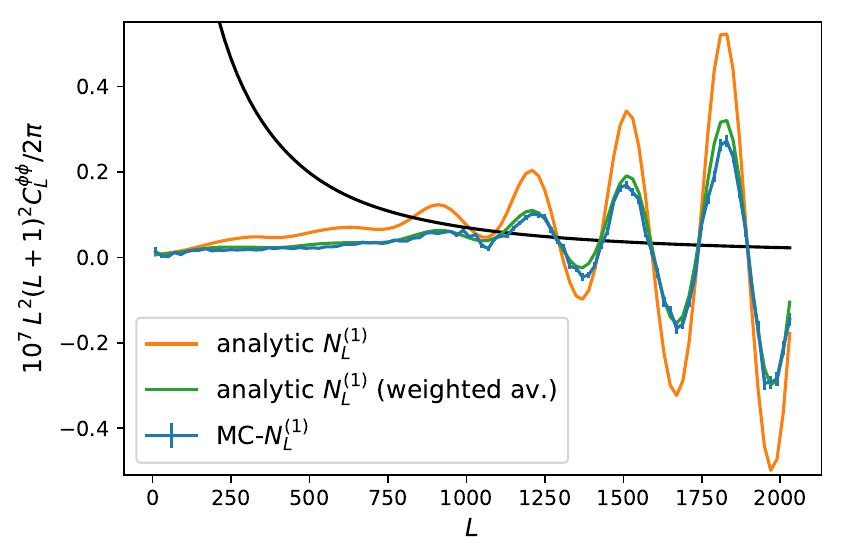}
\caption{\label{fig:MCN1}Comparison of the simulation-based {\None} lensing bias estimates (blue points) to analytic estimates in the case of our inhomogeneously-filtered polarization reconstruction. The orange curve shows the naive prediction using a single map-averaged noise level, which does not capture the variation of the {\None} bias across the sky. The green curve shows the weighted average Eq.~\eqref{eq:N1wei}. The black line shows the fiducial lensing power spectrum.
}
\end{figure}

We use the simulation-based $\MCNone$ as the baseline for our results using inhomogeneous filtering. Figure~\ref{fig:MCN1} shows $\MCNone$ (blue points) for the polarization reconstruction using the inhomogeneously-filtered maps. The large dynamic range of the noise-variance map used for filtering makes the naive analytic prediction (the orange curve) fail to reproduce the $\MCNone$ shape. However, it is straightforward to understand the simulation result analytically by using the same toy model discussed in Sect.~\ref{subsec:inhogfilt}. Let $\None_L(\hn) $ be the analytic (full-sky) {\None} bias calculated using the noise levels at angular position $\hn$, as specified by the variance map used in the filtering. The green curve in Fig.~\ref{fig:MCN1} shows the weighted average
\begin{equation}\label{eq:N1wei}
\begin{split}
	\None_L &\simeq \int \frac{d\hn}{4\pi} \left(\frac{\mathcal{R}_L(\hn)}{\mathcal{R}_L^{\rm fid}}\right)^2\:\None_L(\hn). \\
\end{split}
\end{equation}
This is evaluated by splitting the sky into 15 patches with roughly constant noise levels, and successfully reproduces the simulation-based {\None}.

\section{Data products}
\label{sec:products}
The final \planck\ lensing data products are available on the Planck Legacy Archive,\footnote{\PLA} and described in more detail in the Explanatory Supplement~\citep{planck2016-ES}. The lensing analysis data release consists of the following products:
\begin{itemize}
\item the baseline MV convergence ($\kappa$) reconstruction map up to $L_{\rm max} = 4096$ based on the \smica\ CMB map, along with the corresponding simulations, mask, mean field, lensing biases and response functions;
\item variations using only temperature information, only polarization information, and with no SZ-cluster mask;
\item temperature, polarization and MV lensing maps using optimal filtering for the noise inhomogeneity, together with simulations.
\item temperature lensing maps and simulations built from the \smica\ tSZ-deprojected maps (up to $L_{\rm max} = 2048$);
\item joint MV+CIB and $TT$+$TE$+$EE$+CIB reconstruction $\kappa$ maps, primarily for use with delensing and for plotting the current best-estimate lensing potential, together with the corresponding simulation suite.
\item lensing $B$-mode templates and simulations built from our best $E$-mode estimate and the $TT$+$TE$+$EE$ lensing reconstruction, and in combination with the CIB;
\item power spectrum band powers, covariance, and linear correction matrices for the various analyses;
\item likelihood code using the band powers for the conservative and aggressive multipole ranges from the baseline analysis; and
\item cosmological parameter tables and MCMC chains.
\end{itemize}

\section{Conclusions}
\label{sec:conclude}
We have presented the final official \planck\ lensing analysis, and described in detail
the limits of our understanding of the data. The baseline lensing reconstruction, over nearly 70\,\% of the sky and using lensing multipoles $8\le L\le 400$, is robust to a wide variety of tests. The reconstruction S/N is of order 1 at $L\alt 100$, but retains significant statistical power to smaller scales. It gives interesting constraints on cosmological parameters on its own, yielding percent-level estimates of $\sigma_8\Omm^{0.25}$, and tight constraints on individual parameters when combined with BAO and a baryon density prior. The \planck\ lensing results are currently competitive with galaxy survey constraints, and in the case of $\sigma_8$ substantially more powerful due to the weaker degeneracy with $\Omm$. In combination with lower-redshift tracers, degeneracies can be further broken, and CMB lensing provides a powerful high-redshift baseline for joint constraints.

We showed that a joint analysis with the \planck\ CIB map can be used to provide lensing estimates out to much smaller scales than with lensing reconstruction alone.  The combined map provides our current best estimate of the integrated mass in the Universe between today and recombination, and can be used for delensing analyses. We demonstrated that delensing can already be achieved by \planck, giving peak-sharpening and a reduction in $B$-mode power in line with expectations. For future $B$-mode polarization observations, delensing will probably be essential.

We studied in detail the limits of our understanding of the data. In particular, the deficit of curl power on small scales persists and appears to be very robust to analysis choices. The roughly $3\,\sigma $ significance of this signal is at a level where it starts to be concerning, and may be correlated with sky direction in a way that suggests an unknown systematic issue. However, the changes could also be purely statistical, and we cannot at this point give any likely origin of a systematic signal
that could explain it. The reconstructed power on smaller scales, $L > 400$,
also shows less stability to changes in foreground modelling and map choices, so we restrict attention to the conservative multipole range $8 \le L \le 400$ for our main cosmology results. Our likelihoods are made publicly available for both the conservative and aggressive ($8 \le L \le 2048$) ranges, and users may opt to use the full multipole range at their discretion.
For cross-correlation studies, where the detailed modelling of the auto-spectrum bias is not required, the full multipole range may be more reliable, but we have not performed detailed consistency checks for that application.

In addition to the baseline analysis, we have also provided a number of substantial analysis improvements. In particular, we gave the first demonstration that anisotropic filtering of the polarization can significantly improve the S/N in the reconstruction, giving our best polarization-only band power lensing estimates.

Our baseline lensing reconstruction maps and simulations are made publicly available, along with the joint CIB map and variations with and without SZ-cluster masking. No planned experiments will be able to provide comparable quality maps over the full sky for many years. Ongoing and forthcoming ground-based observations are expected to improve greatly the reconstruction over patches of the sky,
but the \planck\ reconstructions are likely to remain the only tracers of the very largest-scale lensing modes for the immediate future.

\section*{Acknowledgements}
We thank Duncan Hanson for all his work on the previous \planck\ analyses without which the current work would not have been possible. We thank the DES collaboration for sharing their data. Some of the results in this paper have been derived using the {\tt HEALPix} package. Support is acknowledged from the European Research Council under the European Union's Seventh Framework Programme (FP/2007-2013) / ERC Grant Agreement No. [616170],and from the Science and Technology Facilities Council {[}grant numbers ST/L000652/1 and ST/N000927/1, respectively{]}. The Planck Collaboration acknowledges the support of: ESA; CNES, and CNRS/INSU-IN2P3-INP (France); ASI, CNR, and INAF (Italy); NASA and DoE (USA); STFC and UKSA (UK); CSIC, MINECO, JA, and RES (Spain); Tekes, AoF, and CSC (Finland); DLR and MPG (Germany); CSA (Canada); DTU Space (Denmark); SER/SSO (Switzerland); RCN (Norway); SFI (Ireland); FCT/MCTES (Portugal); ERC and PRACE (EU). A description of the Planck Collaboration and a list of its members, indicating which technical or scientific activities they have been involved in, can be found at \href{https://www.cosmos.esa.int/web/planck/planck-collaboration}{https://www.cosmos.esa.int/web/planck/planck-collaboration}.

\bibliography{Planck_bib,texbase/cosmomc,lensing,texbase/antony,julien}{}
\bibliographystyle{aat}

\appendix
\newcommand{\MFerr}{\delta \phi^{\rm MF}}
\newcommand{\MF}{\phi^{\rm MF}}
\newcommand{\ds} { {\hat{C}_L^{di}}}
\newcommand{\sij}{ {\hat{C}_L^{ij}}}
\newcommand{\sst}{ {\hat{C}_L^{ii'}}}
\newcommand{\dd} { {\hat{C}_L^{ii }}}

\newcommand{\sone}{s}
\newcommand{\stwo}{\tilde s}
\newcommand{\Nbias}{N_{\rm bias}}
\newcommand{\NMF}{N_{\rm MF}}
\newcommand{\MCbias}{ {N_{\rm bias}}}

\section{Power spectrum biases}\label{app:biases}

\begin{figure}
\centering
\includegraphics[width = \columnwidth]{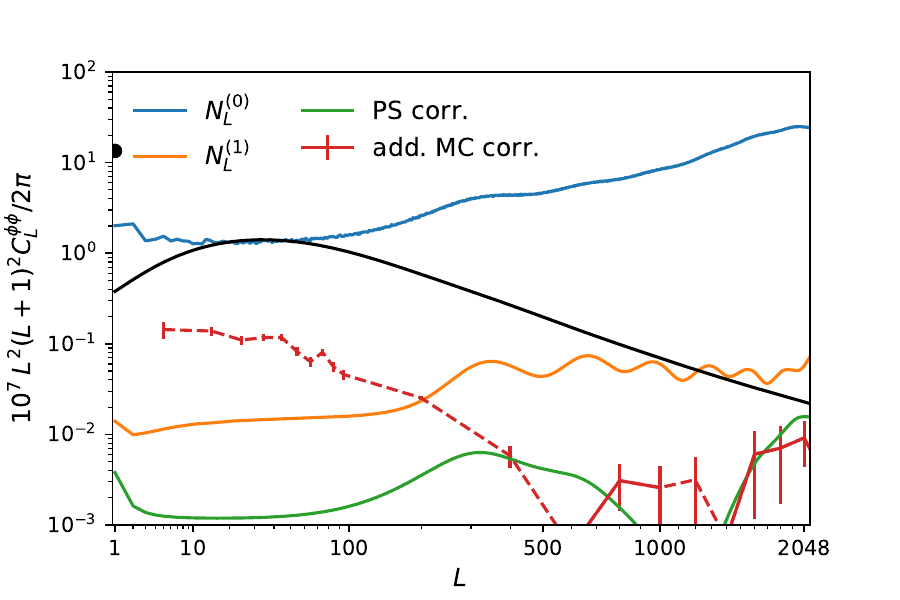}
\caption{\label{fig:biases_app} Compilation of the power spectrum reconstruction biases for our baseline MV analysis. The blue and orange lines show $\RDNzero$ and the analytic $\None$ estimate. The green curve shows the point-source correction, and the red points the additive Monte Carlo correction (shown here after a coarse binning was applied) defined by the difference of Eq.~(\ref{eq:MCcorr_app}) to the simulation input fiducial lensing spectrum. This correction is most prominent at low lensing multipoles, where it is negative (dashed) and due to masking. The error bars are obtained from the 240 simulations used to calculate the correction.
The black line shows the fiducial lensing power spectrum. The black dot at $L=1$ shows the almost pure dipole effective deflection caused by our motion with respect to the CMB frame (aberration). This dipole is included in our simulations, hence subtracted from our lensing reconstructions together with other sources of anisotropy through the mean field.}
\end{figure}

This appendix describes in more detail the calculation of the lensing power spectrum biases. For brevity, here we only present results for our baseline analysis, where all four input maps entering the power spectrum estimator come from the same data and simulation sets. \PlanckLensTwo\ contains a thorough description of the most general case with mixtures of input maps. Figure~\ref{fig:biases_app} shows the size of the various biases for our baseline analysis.

We introduce three types of lensing power spectrum estimators built using the simulations:
$\dd$ is the power spectrum of the quadratic estimator with both legs\footnote{By ``leg,'' we mean one of the two fields entering the quadratic estimator.} being simulation $i$; $\ds$ is the analogous spectrum where one leg of the quadratic estimator is always the data, and the second leg is simulation $i$; and $\sij$ is where simulation $i$ is the first leg and simulation $j$ the second leg and $i\neq j$.
We note that
$\dd$ contains a mean-field subtraction, but the other two spectra do not (the mean field would vanish, since the different maps are independent to a very good approximation). The Gaussian lensing reconstruction noise biases $\MCNzero$ and $\RDNzero$ are then defined as
\begin{equation}\label{eq:N0_app}
\begin{split}
	\MCNzero_L &\equiv \av{2 \sij}_{\MCbias}, \\
	\RDNzero_L &\equiv \av{4 \ds - 2 \sij}_{\MCbias},
\end{split}
\end{equation}
where angle brackets denote an average over simulations, with index $i$, and in our implementation $j$ is always equal to $i + 1$ (cyclically). \JCrev{The numerical factors in front of $\sij$ (or $\ds$) in Eqs.~\eqref{eq:N0_app} account for the fact that the quadratic estimators entering these spectra are built with two independent CMB maps, hence only half of the Gaussian contractions are captured.} In principle, many more pairs could be used to estimate $\av{2\sij}$; this would require considerably more resources, and the Monte Carlo error on this term is already a small correction, as demonstrated in Appendix~\ref{app:cov}.
The Monte Carlo reconstruction $\av{\hat C^{\phi\phi}_L}_{\rm MC}$ (used to build the Monte Carlo correction, Eq.~\ref{Binning}, after binning) is
\begin{equation}\label{eq:MCcorr_app}
 \av{\hat C^{\phi\phi}_L}_{\rm MC} \equiv \av{ \dd- 2 \sij - \None_L}_{\MCbias}.
\end{equation}
In the last definition, the analytic expression for $\None$ is used, with the exception of our inhomogeneously-filtered reconstructions, for which a Monte Carlo estimate $\MCNone$ is used instead. In general, for a source of anisotropy $s$,
the analytic $s$-induced $\None$ is calculated according to the flat-sky approximation using~\citep{Kesden:2003cc}
\begin{multline}
\None_L{}^{XYIJ, s} = \frac{1}{\mathcal{R}_L^{XY}}\frac 1 {\mathcal{R}_L^{IJ}}\int \frac{d^2\vl_1 d^2\vl_1'}{(2\pi)^4} F^X_{l_1}F^Y_{l_2}F^I_{l_1'}F^J_{l_2'}\\ \times W^{XY}(\vl_1,\vl_2)W^{IJ}(\vl_1',\vl_2')\\
\times \left[C^{ss}_{\left|\vl_1 +\vl_1'\right|} f^{XI,s}(\vl_1,\vl_1') f^{YJ,s}(\vl_2,\vl_2') \right. \\
      + \left. C^{ss}_{\left|\vl_1 + \vl_2'\right|} f^{XJ,s}(\vl_1,\vl_2') f^{YI,s}(\vl_2,\vl_1') \right],
        \label{Nonedef}
\end{multline}
with $\vl_1 +\vl_2 = \bf{L} = -(\vl_1' +\vl_2')$,
and the labels $X,Y,I,J$ stand for the $T$, $E$, or $B$ map. Further, $F^X_l$ is the analytic inverse-variance filter $\transfmultipolestopix^\dagger \Cov^{-1}\transfmultipolestopix$ in the isotropic limit for field $X$ (diagonal in $T,E,B$ space since we perform temperature and polarization filtering separately), $W^{XY}$ are the flat-sky estimator weights as applied to the inverse-variance filtered $X$ and $Y$ fields, $f^{XY, s}$ is the CMB $XY$ flat-sky covariance response to anisotropy source $s$, and $C_L^{ss}$ is the spectrum of that source of anisotropy. Results in this paper use lensing as source of anisotropy $s$ almost exclusively, but we have also built curl-induced and point-source induced \None\ biases for robustness tests.

The Monte Carlo $\MCNone$ \citep{SPTPOLLensing} is built as follows. We generate additional noiseless simulations in pairs, where each pair shares the same input deflection field, but have independent unlensed CMB. These simulations are not propagated through the \planck\ beam processing, since this would be computationally too expensive, but are generated with the effective, isotropic transfer function instead (the accuracy requirements on $\None$ are much lower than on $\Nzero$, and Sect.~\ref{subsec:N1tests} demonstrates that the impact of the CMB non-idealities on $\None$ are negligible). These simulations also neglect $C_L^{T \phi}$ and $C_L^{E\phi}$, which are negligible on the relevant scales. 
We again form spectra mixing quadratic estimates $\sij$ and $\sst$ from these simulations, where $\sij$ is defined as above on the first pair member of these new simulations, and $\sst$ takes the quadratic estimate built from the first pair member on one leg and the second pair member on the second leg. Then

\begin{equation}
\MCNone_L \equiv \av{2 \sst - 2 \sij}.
\end{equation}

\section{Mean fields}\label{subsec:subpix}

\begin{figure*}[h!]
\centering
\includegraphics[width = \textwidth]{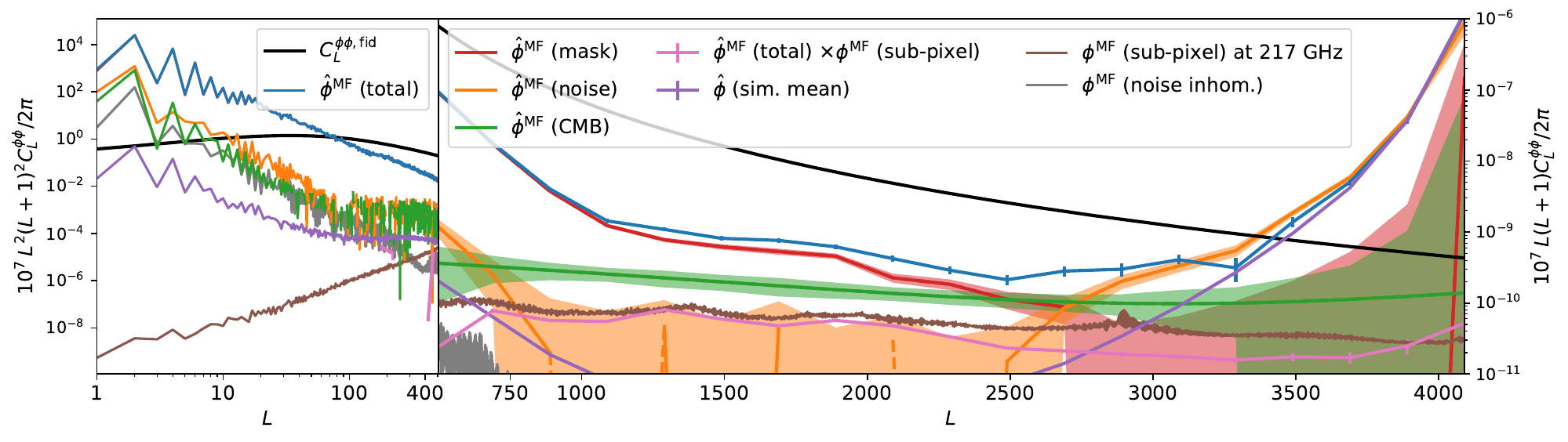}
\caption{\label{fig:MF}
Characterization of the mean-field power spectrum and its components in our temperature-only reconstruction. The left-hand part shows the convergence-like power spectrum $L^2(L+1)^2 C_L^{\phi\phi}$ on large scales, while the right-hand part shows the deflection power spectrum $L(L+1)C_L^{\phi\phi}$ on smaller scales. The blue line shows the total mean field as obtained from the \smica\ FFP10 simulation set. Lines showing the contributions from statistical anisotropy of the mask (red), noise (orange), and the beam-convolved and pixelized CMB anisotropies (green) are plotted with rough $68\,\%$ confidence regions shown shaded on the right panel (the uncertainty arises from the finite number of simulations used to estimate the mean fields). The mask mean field dominates at low multipoles and is barely distinguishable from the total on the left-hand plot.
The orange and green curves were obtained from full-sky lensing reconstructions using idealized, statistically-isotropic CMB or noise components, respectively. The red curve was obtained by differencing the mean-field spectra on the full anisotropic simulations, with and without masking. Pixelization effects due to sub-pixel pointing offsets are expected to appear as an approximately white-noise lensing deflection component of amplitude $0\parcm05$ (the brown line shows the prediction for the 217-GHz channel), and is clearly detected by cross-correlating the mean field to the sub-pixel deflection prediction (pink curve). The rise of the mean field at $L \simeq 3000$ originating in the noise maps is mostly a simulation artefact, sourced by nonlinearities in the data processing that cause slight correlations between the simulations (do to the fixed fiducial CMB and foregrounds used to make the noise simulations); the purple curve shows the lensing quadratic estimator applied to the empirical average of the noise simulations. The predicted mean field due to inhomogeneity of the variance of idealized uncorrelated pixel noise (grey curve, as derived from Eq.~\ref{eq:noisemf}) is much weaker on smaller scales and is only visible at low lensing multipoles in this figure.}
\end{figure*}

In this appendix we characterize the main components of the mean field, including very high lensing multipoles $L > 2048$ that are not used in this paper, but are included in the lensing maps that we release. The mean field is obtained by averaging lensing reconstruction maps constructed from the FFP10 simulations, and for many effects we rely on the fidelity of these simulations for accurate subtraction of the mean field from the data.

Figure~\ref{fig:MF} shows the different contributions to the mean field of the temperature-only reconstruction. The left-hand panel focuses on the low-$L$ range ($L \leq 400$) and the right-hand panel on higher multipoles. The total mean field is shown as the blue curve. The orange curve shows the contribution to the mean field from the noise simulations. This captures various effects, including the large inhomogeneity of the pixel noise variance (see below and Fig.~\ref{fig:vmapsmica}), and residual noise correlations from the mapmaking procedure. The green curve shows the contribution from statistical anisotropy of the observed CMB, through the effects of beam anisotropies and pixelization.
The orange and green curves were obtained from full-sky reconstructions to remove the effect of masking, and used idealized statistically-isotropic CMB and noise components, respectively (which do not contribute to the mean field). The mask contribution itself is shown as the red curve, which we obtain by differencing the mean-field spectra obtained from the full, anisotropic, and masked simulations to the mean field from the same simulations without any masking.

As in previous releases, the mean field is very large at low multipoles, where it is mainly due to masking, with the noise and beam-anisotropy contribution being the next most important. The pink curve in Fig.~\ref{fig:MF} shows the expected contribution from the inhomogeneities in the pixel noise variance only (using the variance map in Fig.~\ref{fig:vmapsmica} and assuming there are no pixel-pixel correlations). The expected noise mean-field contribution is, from Eq.~(\ref{eq:lensing2pt}),
\begin{equation}\label{eq:noisemf}
	\av{ _1\hat d(\hn)} \textrm{(noise inhom. only)} = \sum_{LM} h_L \sigma^{2}_{T, L M}\: _1 Y_{L M}(\hn) ,
\end{equation}
and hence, according to Eq.~(\ref{eq:gcLM}), is a pure gradient field with $g_{LM} = - \sqrt{L (L + 1)} h_L \sigma^{2}_{T, LM}$.
Here, $\sigma^{2}_{T, L M}$ is the spin-0 harmonic transform of the temperature noise variance map, and
\begin{equation}
\begin{split}
		h_L &\equiv 2\pi \int_{-1}^{1}d\mu \: \xi^{\bar T}_{0, 0}(\mu)\:\xi^{ T^{\rm WF}}_{0, 1}(\mu)\: d^L_{-1,0}(\mu) , \\
		 \xi^{\bar T}_{0, 0}(\mu) &\equiv \sum_{\ell = 100}^{2048} \left( \frac{2 \ell + 1}{4\pi}\right) \left(\frac{b_\ell }{b_\ell^2C_\ell^{TT} + N_\ell} \right)_{\rm fid} d^\ell_{0,0}(\mu), \\
		 		 \xi^{T^{\rm WF}}_{0, 1}(\mu) &\equiv \sum_{\ell = 100}^{2048}\left( \frac{2 \ell + 1}{4\pi}\right) \left(\frac{b_\ell C_\ell^{TT} }{b_\ell^2C_\ell^{TT} + N_\ell} \right)_{\rm fid}
\sqrt{\ell(\ell+1)}
d^\ell_{1, 0}(\mu).
\end{split}
\end{equation}
Here, $b_\ell$, $C_\ell^{TT}$, and $N_\ell$ are the fiducial combined beam and pixel window function, fiducial CMB spectrum, and noise spectrum used in the filtering (we use a flat noise level of $35\muKarcmin$).
The contribution to the mean field from inhomogeneous pixel noise is only visible at low lensing multipoles.

At higher multipoles, $500 \leq L \leq 2500$, the noise mean field drops sharply, but the CMB mean field remains clearly visible, at least partly because of pixelization effects. To some approximation, the mapmaking procedure simply bins individual time-ordered data into the assigned pixels. Approximating the temperature field locally as a gradient, this gives rise to an effective deflection field, which can be predicted given the full pointing information for each frequency channel. Since this sub-pixel mis-centring is the same in all simulations, it appears rather directly in the mean-field estimate (with a possible bias, since, unlike lensing, the sub-pixel deflections act on the sky signal \emph{after} beam convolution). The prediction for the 217-GHz channel sub-pixel deflection spectrum is shown as the purple curve in Fig.~\ref{fig:MF},
and has the approximate power spectrum of a white-noise displacement component $C_L^{\kappa\kappa, \rm{sub-pixel}} \propto L^2$ (as expected for a large number of independent hits) with an amplitude of around $0\parcm05$.
The cross-spectrum of the mean-field estimate for the \smica\ map with the predicted sub-pixel displacements for the 217\,GHz channel recovers the expected amplitude in the range $500 \leq L \leq 2500$ fairly well.

At yet higher multipoles, $L \geq 2500$, while the CMB mean field can still be seen in cross-correlation, the most striking feature is a very sharp rise in the mean field sourced by the noise simulations.  We caution that this rise is likely to be largely an artefact of the way that the simulations were constructed: 
as discussed in Sect.~\ref{sec:pipe}, the noise simulations are not exactly independent, due to small nonlinearities in the data processing causing correlations after subtraction of the common fiducial CMB and foreground realizations used to make the simulations.
This introduces a contribution to the mean field given by the result of applying the lensing quadratic estimator to the common component of the noise simulations. We estimate this common component from empirical averages of the first 150 noise simulations and the last 150 noise simulations. By applying the lensing quadratic estimator to each of these simulation averages in turn, and cross-correlating the results, we attempt to isolate the mean-field power sourced by the common signal. The result is shown as the red curve in
Fig.~\ref{fig:MF}; this agrees well with the sharp increase in the mean field (blue curve) seen at $L \geq 2500$. The red curve contains a residual Monte Carlo noise level of $150^{-2}$ 
in the mean field, which is visible at low $L$. On these scales the mean field of the common component is too small to be separated from the noise. In polarization, the mean field of the common component is slightly larger than in temperature; however, its contribution to the spectrum is a negligibly small fraction of the standard deviation of the band powers.

\section{Covariance matrix corrections}
\label{app:cov}
As described in Sect.~\ref{subsec:cov}, our covariance matrix takes into account Monte Carlo errors in the various quantities obtained from simulations, increasing the covariance by roughly $10\,\%$ over our conservative multipole range. These errors come mainly from the mean-field estimate and the lensing $N^{(0)}$ bias (see Appendix~\ref{app:biases}). We use $\NMF = 60$ simulations for the mean field and $\Nbias = 240$ independent simulations for $\Nzero$. This section describes how the impact of the Monte Carlo error is calculated. For simplicity we use the following form of the Monte Carlo-corrected final power spectrum estimator, where the Monte Carlo correction is applied additively instead of multiplicatively:
\begin{equation}
\label{eq:app_recspec}
\begin{split}
	\hat C_L^{\phi\phi} &= \hat C^{dd}_L - \RDNzero_L - \None_L - \av{\hat C^{\phi\phi}_L}_{\rm MC} + \clppfid \\
	&= \hat C^{dd}_L- \av{4 \ds - 4 \sij + \dd}_{\MCbias} + \clppfid \\
\end{split}
\end{equation}
(and neglecting the point-source correction), obtained combining Eqs.~(\ref{eq:N0_app}) and (\ref{eq:MCcorr_app}). Here, $\hat{C}_L^{dd}$ is the lensing reconstruction power spectrum estimate from the data, including mean-field subtraction (i.e., it is equivalent to
$\hat{C}_L^{\hat{\phi}\hat{\phi}}$ defined in the main text in Eq.~\ref{eq:naive_phispec}, but here we wish to emphasise that it is constructed from the data).

\subsection{Monte Carlo errors on the mean field}

The mean field is defined as an ensemble average $\MF_{LM} \equiv \av{\hat{g}_{LM}}/\mathcal{R}_L^\phi$,
but evaluated with a finite number of simulations $\NMF$. The mean field only enters two terms in the reconstructed spectrum of Eq.~(\ref{eq:app_recspec}): the data lensing reconstruction spectrum $\hat C^{dd}_L$; and similarly $\dd$ on the simulations used for the Monte Carlo correction calculation. To avoid Monte Carlo noise bias in these spectrum estimates, we use two independent sets of $\NMF / 2$ simulations to subtract the mean field from each quadratic estimator before forming their cross-spectrum.
Denoting the (small and independent) error on the estimated mean field for each quadratic estimate by
$\MFerr_{1,2}$, all relevant terms are of the form
\begin{equation}
\hat C_L^{\phi\phi} \supset
\left(\frac{\hat{g}_{LM}}{\mathcal{R}_L^\phi}- \MF_{LM} - \MFerr_{1,LM} \right) \left(
\frac{\hat{g}_{LM}}{\mathcal{R}_L^\phi}- \MF_{LM}- \MFerr_{2,LM}\right)^*,
\end{equation}
where
$\hat{g}_{LM}$ is either the data lensing map $\hat{g}^d_{LM}$ (for $\hat C^{dd}_L $) or that of simulation $i$ ($\hat{g}^i_{LM}$ for $\hat C^{ii}_L $). Summing the terms from $\hat C^{dd}_L$ and $\dd$ at fixed $L$ and $M$, the contribution to the error on the spectrum is
\begin{multline}
\delta \hat{C}_L^{\phi\phi} \supset -\MFerr_{1, LM} \left[\left(\frac{\hat{g}^d_{LM}}{\mathcal{R}_L^\phi}-\MF_{LM} \right) - \av{
\frac{\hat{g}^i_{LM}}{\mathcal{R}_L^\phi}
-\MF_{LM} }_{\MCbias}\right]^* \\ + (1 \leftrightarrow 2)^\ast.
\end{multline}
We note that terms of the form $\MFerr_{1, LM} (\MFerr_{2, LM})^\ast$ cancel between
$\delta \hat C^{dd}_L$ and $\delta \dd$.
We proceed 
to calculate the averaged squared error on the spectrum. The spectrum of the mean-field error is
\begin{equation}
\left\langle{|\MFerr_{LM}|^2} \right\rangle \simeq \frac{2}{N_{\rm MF}} \left(C_L^{\phi\phi} + N_L^{(0)}\right),
\end{equation}
neglecting the $\None$ contribution.
Note that the factor of two is present since we use only half of the $N_{\rm MF}$ simulations on each leg. We also have
\begin{multline}
\av{\left|\left(\frac{\hat{g}^d_{LM}}{\mathcal{R}_L^\phi}
-\MF_{LM} \right) - \av{ \frac{\hat{g}^i_{LM}}{\mathcal{R}_L^\phi}
-\MF_{LM} }_{\MCbias}\right|^2} \simeq \left(C_L^{\phi\phi} + N_L^{(0)}\right)\\
\times \left( 1+ \frac 1 {\Nbias}\right).
\end{multline}
Finally, averaging over $M$, allowing crudely for an effective number of modes $(2L + 1)f_{\rm sky}$, and neglecting the correction scaling with $1/\Nbias$, the total
error induced in the reconstruction spectrum due to Monte Carlo errors in the mean-field evaluation has variance:
\begin{equation} \label{eq:MFerr}
\var\left(\left.\delta \hat C_L^{\phi\phi} \right|_{\rm MF}\right) \simeq 	\frac{4}{N_{\rm MF}} \frac{\left(C_L^{\phi\phi} + N_L^{(0)} \right)^2}{\left(2L + 1\right)f_{\rm sky}}.
\end{equation}

\subsection{$\Nzero$ Monte Carlo errors}
We now consider the error on the spectrum due to the finite number $\Nbias$ of simulations in the average in Eq.~(\ref{eq:app_recspec}),
\begin{equation}
\left.\delta \hat C_L^{\phi\phi} \right|_{\rm Biases} \equiv 	-\delta \av{ 4\ds - 4\sij +\dd}_{\MCbias}.
\end{equation}
We proceed by approximating the reconstructions $\hat{g}_{LM}$ as Gaussian, in which case correlations between the errors in $\dd$, $\ds$, and $\sij$ vanish. Empirical verification confirms that to a good approximation, errors in these spectra are independent, with approximate Gaussian variance
\begin{eqnarray}
	\var \left(\delta\av{\dd}_{\MCbias}\right) &\simeq& \frac 1 {\Nbias}\frac{2\left( C_L^{\phi\phi} + N_L^{(0)}\right)^2}{(2L + 1)f_{\rm sky}}, 	
\label{eq:N0errsim}
\\
	\var\left(\delta\av{ 2\ds}_{\MCbias}\right) \!&=& 	\!
\var\left(\delta\av{ 2\sij}_{\MCbias}\right)
\simeq \frac 1 {\Nbias}\frac{2\left(N_L^{(0)}\right)^2}{(2L + 1)f_{\rm sky}}.
\label{eq:N0err}
\end{eqnarray}
\subsection{Total Monte Carlo errors}

\begin{figure}
\centering
\includegraphics[width = \columnwidth]{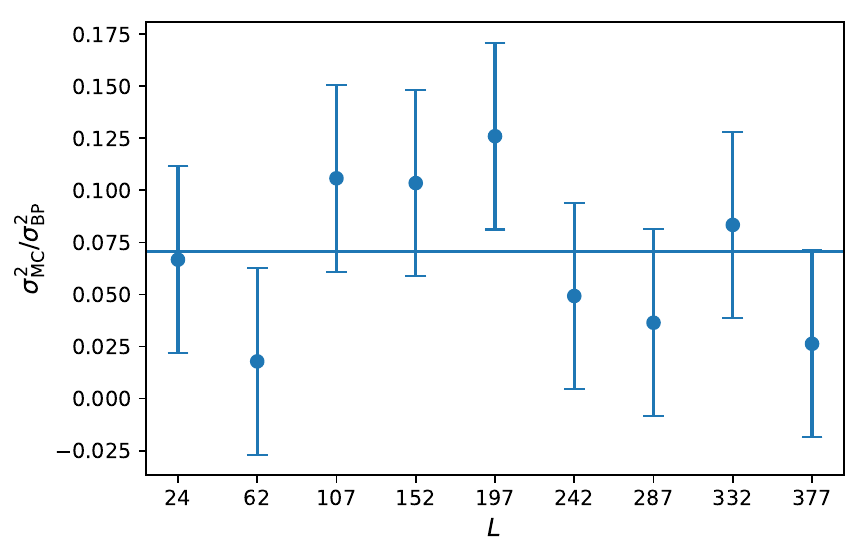}
\caption{\label{fig:MCerr_app}
Crude empirical estimates of the additional variance
on the MV reconstruction band powers (on the conservative multipole range) due to the finite number of simulations used for mean-field and bias correction. The additional variance is normalized by the baseline statistical variance of the band powers.
These empirical estimates were obtained by splitting the simulations into five independent subsets, building reconstruction band powers using each subset of the simulations, and measuring the scatter between the five sets of band powers. The results are scaled by $1/5$ to approximate the full simulation set used in the main analysis.
The horizontal line shows the simple constant relative correction applied to the covariance matrix, Eq.~\eqref{eq:cov_corr}, as obtained in Appendix~\ref{app:cov}. The error bars on the empirical variance (identical for each bin, after rescaling by $
\sigma^2_{\rm BP}$) are those expected in the fiducial model of Eq.~\eqref{eq:cov_corr}, assuming Gaussian statistics of the lensing maps.
}
\end{figure}

Combining Eqs.~(\ref{eq:MFerr}--\ref{eq:N0err}) gives the total variance of the Monte Carlo error on the power spectrum of the reconstruction.
Neglecting the absence of $C_L^{\phi\phi}$ in Eq.~\eqref{eq:N0err}, which is accurate on almost all scales,
we obtain
\begin{equation}
\label{eq:cov_corrtwo}
\sigma^2_{{\rm MC},L} \simeq \left(\frac{2}{N_{\rm MF}} + \frac 9 {N_{\rm Bias}} \right) \frac{2\left(C_L^{\phi\phi}+N_L^{(0)}\right)^2}{(2L + 1)f_{\rm sky}}.
\end{equation}
This gives Eq.~(\ref{eq:cov_corr}) in the main text, after identifying the diagonal band-power Gaussian variance with $\sigma^2_{\rm BP}$.
Figure~\ref{fig:MCerr_app} shows a rough empirical estimate of the Monte Carlo errors on our MV band powers compared with the analytic estimate of Eq.~(\ref{eq:cov_corrtwo}).
The points were estimated from the scatter of five data band powers, each constructed using 60 independent simulations (with $\NMF = 12$ and $\Nbias = 48$) out of the full set of 300 simulations. The figure shows the empirical variance (divided by 5) normalized to the statistical variance of the band powers, $\sigma^2_{\rm BP}$; this agrees reasonably well with the analytic estimate (the correction is small, so it does not need to be calculated accurately).

\raggedright
\end{document}